%% file: supplementary202607.tex
\documentclass[lettersize,journal]{IEEEtran}

\usepackage{cite,balance}
\usepackage{amsmath,amssymb,amsfonts,nicematrix}
\usepackage{bbm}
\usepackage{xcolor}
\usepackage{booktabs}

\usepackage{algorithm}

\usepackage{algcompatible}
\usepackage{algpseudocode}

\def\BibTeX{{\rm B\kern-.05em{\sc i\kern-.025em b}\kern-.08em
    T\kern-.1667em\lower.7ex\hbox{E}\kern-.125emX}}
\usepackage{nicematrix,tikz}
\usepackage{graphicx}
\usepackage{textcomp}
\usepackage{array,booktabs,makecell}
\usepackage[font=footnotesize]{caption}
\usepackage{float}
\usepackage{tikz}
\usepackage{cite,balance}
\usetikzlibrary{shapes.geometric, arrows}
\usepackage{pgfplots}
\usepackage{caption}
\usepackage{mathtools}
\usepackage{subcaption}
\usepackage{comment}
\usepackage{enumitem}
\usepackage{stackengine}
\usepackage{fix-cm}    
\usepackage[most]{tcolorbox}
\newtcolorbox[auto counter, number within=section]{problem}[2][]{
  colback=gray!5!white,
  colframe=gray!80!black,
  fonttitle=\bfseries,
  title=Question~\thetcbcounter:
}

\usepackage{hyperref}
\usepackage{makecell}

\newcommand\numberthis{\addtocounter{equation}{1}\tag{\theequation}}
\usepackage{xcolor}
\definecolor{OliveGreen}{rgb}{0,0.6,0}
\def\BibTeX{{\rm B\keri-.05em{\sc i\keri-.025em b}\keri-.08em
    T\keri-.1667em\lower.7ex\hbox{E}\keri-.125emX}}
\usepackage{multirow}

\usepackage{amsthm}
\newtheorem{theorem}{Theorem}
\newtheorem{condition}{Condition}
\theoremstyle{definition}
\newtheorem{example}{Example}
\newtheorem{lemma}{Lemma}
\newtheorem{construction}{Construction}
\newtheorem{assumption}{Assumption}
\newtheorem{definition}{Definition}
\newtheorem{remark}{Remark}

\usepackage{booktabs}
\usepackage{multirow}
\usepackage{pifont}

\allowdisplaybreaks
\usepackage{mathtools}
\usepackage{pifont}

\setlist[enumerate]{wide=0pt, leftmargin=15pt, labelwidth=15pt, align=left}
\usepackage{tabstackengine}

\usepackage{bm}
\usetikzlibrary{arrows.meta}
\tikzset{every picture/.style={line width=0.6pt}}
\usetikzlibrary{calc,positioning}
\usetikzlibrary{arrows.meta,
                backgrounds,
                chains,
                fit,
                quotes}

\begin{document}
\title{Coding-enforced Robust Secure Aggregation for Federated Learning\\
Under Unreliable Communication}

\author{Shudi Weng,~\IEEEmembership{Graduate Student Member,~IEEE,}
Chao Ren,~\IEEEmembership{Member,~IEEE,}
Yizhou Zhao,~\IEEEmembership{Member,~IEEE,}\\
Ming Xiao,~\IEEEmembership{Senior Member,~IEEE,}
and Mikael Skoglund,~\IEEEmembership{Fellow,~IEEE}\vspace{-1em}
\thanks{Shudi Weng, Ming Xiao, Chao Ren, and Mikael Skoglund are with the Department of Information Science and Engineering (ISE), KTH Royal Institute of Technology, Stockholm, Sweden. Email: \{shudiw, mingx, chaor, skoglund\}@kth.se. }
\thanks{Yizhou Zhao is with the College of Electronic and Information Engineering, Southwest University, Chongqing, China. Email: \{onezhou\}@swu.edu.cn.}
\thanks{\textit{Corresponding Author: Shudi Weng}.}
}
\markboth{Journal of \LaTeX\ Class Files,~Vol.~14, No.~8, August~2021}%
{Shell \MakeLowercase{\textit{et al.}}: A Sample paper Using IEEEtran.cls for IEEE Journals}

\maketitle

\begin{abstract}
{This work studies privacy-preserving federated learning (ppFL) under unreliable communication. Zero-sum privacy noise can protect model privacy without sacrificing model accuracy, effectively overcoming the privacy-utility trade-off.    
However, unreliable communications can randomly disrupt the coordination of zero-sum noise, resulting in unpredictable client participation and aggregation errors, which can severely impair training effectiveness.}
To address this issue, we exploit the linearity of the coding scheme and the target function to develop a coding-enforced structured secure aggregation method, termed \underline{Sec}ure \underline{Co}operative \underline{G}radient \underline{C}oding (SecCoGC), which leads to an all-or-nothing outcome under unreliable communication: either exact reconstruction of the global model or a non-meaningful result.  
We evaluate the local differential privacy (LDP) across all protocol layers in SecCoGC, which accounts for the deterministic network coding scheme, correlations among privacy noise, and random realization of communication networks. 
Additionally, we present a complete problem formulation and constructions of real-field zero-sum privacy noise, and introduce fairness as a privacy metric among clients.  
Finally, we provide a distinct non-convex convergence analysis for FL algorithms with binary global model reconstruction.
Experimental results demonstrate the superiority of SecCoGC under unreliable communication while maintaining varying levels of privacy preservation, yielding significant performance improvements over benchmarks. 
\end{abstract}
\section{Introduction}
\subsection{Backgrounds}
\IEEEPARstart{F}{ederated learning} (FL) enables numerous clients to collaboratively train models based on their local datasets. By only sharing local models (updates), FL enhances user privacy and reduces the volume of data exchanged during training \cite{ren2025advances}. 
However, the adversary may still infer sensitive information captured by local models, making FL vulnerable to various model-based attacks, e.g., \textit{inversion attacks} (IA) and \textit{membership inference attacks} (MIA), etc. 

To defend against these and other attacks, a widely adopted strategy in \textit{privacy-preserving FL} (ppFL) is privacy noise injection to obscure sensitive information, through, e.g., the well-known \textit{Gaussian mechanism} and \textit{Laplacian mechanism}. 
Massive studies investigate the use of independent Gaussian privacy noise with provable convergence guarantee \cite{saha2024privacy}. 
However, in these settings, the privacy noise is fully preserved in the server aggregation.
Strong privacy inevitably leads to significant degradation in model accuracy. 
This inherent tension between privacy and model accuracy is referred to as \textit{privacy-utility trade-off}. 
To alleviate this tension, recent works employ correlated privacy noise \cite{zhang2025locally,11240322,nordlund2025secureovertheaircomputationmultiple}, which enables partial cancellation of randomness and mitigates the privacy noise accumulation at the server.
Despite these improvements, the privacy-utility trade-off remains a fundamental limitation of these approaches. 

\textit{Secure aggregation} (SA), which enables the server to learn the accurate aggregate without observing each individual, offers a promising solution to overcome the privacy-utility trade-off. 
Existing SA methods primarily build upon techniques such as \textit{homomorphic encryption} (HE), \textit{Shamir's Secret Sharing} (SSS), multi-party computation, zero-sum discrete and analog noise, and combinations of these techniques \cite{bonawitz2017practical,8241854,zhao2022information,zhang2025fundamentallimitshierarchicalsecure,9929413} \footnote{Zero-sum noise is considered a SA method, as it permits the server to obtain an accurate global model without revealing the individuals. }.  
In SA, global model reconstruction relies on a coordinated encryption structure, which in turn depends on reliable communication. 
However, reliable communication is often unrealistic in practical deployments due to physical-layer impairments such as fading and shadowing. 
As a result, communication uncertainty gives rise to unpredictable client participation and coordination patterns, posing a fundamental challenge to the design and reliability of SA schemes.

\begin{table*}[ht]
\centering
\caption{{Comparison between existing SA protocols and our proposed method \textbf{under unreliable communication}.}}
\label{tab:comparison}
\renewcommand{\arraystretch}{1.15}
\setlength{\tabcolsep}{4pt}
\small
\resizebox{\linewidth}{!}{
\begin{tabular}{c|cccccc}
\toprule
Methods
&
Robust Privacy Mechanism
&
Model Accuracy
&
Objective Consistency
&
No Partial Recovery
&
No Multi-round Exposure
&
Low Complexity
\\

\midrule

Independent privacy noise
\cite{saha2024privacy}
 & \checkmark & $\times$ & $\times$ & \checkmark
& \checkmark & \checkmark
\\

Correlated privacy noise
\cite{zhang2025locally,11240322,nordlund2025secureovertheaircomputationmultiple}
 & $\times$ & $\times$ & $\times$ & \checkmark
& \checkmark & \checkmark
\\

Pairwise masking
\cite{bonawitz2017practical}
& \checkmark & \checkmark & $\times$ & $\times$
& $\times$ & \checkmark
\\

HE-based SA
\cite{8241854}
 & -- & \checkmark & $\times$ & --
& -- & $\times$
\\

2-round Zero-sum noise
\cite{zhao2022information,li2026informationtheoreticdecentralizedsecureaggregation}
& \checkmark & \checkmark & $\times$ & $\times$
& $\times$ & \checkmark
\\


SSS + Coded datasets
\cite{schlegel2023codedpaddedfl}
& \checkmark & -- & \checkmark & \checkmark
& \checkmark & $\times$ 
\\

\midrule

\textbf{Coding-enforced Robust SA (Ours)}
& \checkmark
& \checkmark
& \checkmark
& \checkmark
& \checkmark
& \checkmark
\\

\bottomrule
\end{tabular}
}
\vskip.2\baselineskip
{\begin{tabular}{p{.98\textwidth}}
$^*$
\begin{footnotesize}    
Note that $\checkmark$, $\times$, and -- denote support, lack of support, and design-dependent support, respectively.
\end{footnotesize}
\end{tabular}}
\vspace{-7mm} 
\end{table*}

{
Another critical challenge arising from unreliable communication is the persistent, strict sub-optimality of the FL training process. 
The data generated at the network edge is inherently uncontrollable and typically \textit{non-independent and identically distributed} (non-IID) across clients.
For this reason, the local models could deviate from the aggregate, referred to as \textit{client drift} \cite{karimireddy2019scaffold}. 
Unreliable communication further intensifies this issue by introducing statistically uneven participation of local models in the global aggregation, resulting in convergence to a stationary point that is misaligned with the true global optimum, known as \textit{objective inconsistency} \cite{wangTackling,wang2021quantized}. 
}

{
In ppFL, the negative effects of unreliable communication become even more pronounced because communication and privacy mechanisms are tightly coupled.
For HE-based methods, unreliable communication may cause decryption failure. 
For noise-injection-based methods, the combined effect can severely degrade global model accuracy and may even cause the training process to diverge. More specifically, the injected privacy noise can randomly perturb both the convergence direction and the effective step size in each training round, resulting in unstable optimization trajectory. When this instability interacts with communication uncertainty, the resulting trajectory becomes highly unpredictable.
Moreover, the random network topology can disrupt the intended coordination of privacy noise. For example, correlated noise components that are designed to cancel each other during aggregation may fail to do so because of communication failure. 
As a result, the intended noise-cancellation mechanism becomes ineffective, leading to excessive residual noise and limiting the overall performance and reliability of the method.
}

Although coding strategies were originally developed to combat uncertainty rather than privacy-related challenges, their underlying structural properties offer a unique advantage for ppFL.
In essence, coding strategy exploits deterministic algebraic structures to introduce redundancy and enable reliable recovery to combat uncertainty. 
This structural certainty it provides, rather than mere statistical unbiasedness, makes coding-enforced strategy particularly a valuable future direction for robust SA under unreliable communication. 

\vspace{-0.3em}
\subsection{Related Works}
{
While extensive efforts have been devoted to \textbf{enhancing robustness against unreliable communication}, these methods are generally not readily compatible with the zero-sum mechanism.
For instance, a line of research focuses on balancing communication uncertainty, e.g., adaptive resource allocation \cite{wang2021quantized}, heterogeneity-aware aggregation \cite{zheng2023federated,perazzone2022communication}, collaborative relaying \cite{yemini2023robust}, heterogeneity-aware client sampling \cite{weng2026heterogeneityawareclientsamplingoptimal}, etc. 
Although these approaches can preserve the statistical unbiasedness of model aggregation, they do not guarantee the reliable coordination and cancellation of privacy noise under unreliable communication conditions.
Other methods \cite{wang2022unified,wang2024lightweight,xiang2024efficient} address this issue through carefully designed accumulation mechanisms. However, in ppFL, improper accumulation of privacy noise can severely degrade convergence and may even lead to training divergence.
Consequently, bridging the gap between straggler-resilient design and robust privacy protection remains a critical challenge. 
}

{
Despite the extensive efforts on SA, most existing studies overlook the challenges posed by unreliable communication. 
To achieve \textbf{robust SA under unreliable communication}, recent studies have investigated integrating secret sharing schemes (SSS) with \textit{gradient coding} (GC)-based replicated coded datasets \cite{schlegel2023codedpaddedfl}, pairwise masking \cite{bonawitz2017practical}, and two-round strategic secure communication protocols \cite{zhao2022information,li2026informationtheoreticdecentralizedsecureaggregation}.
Despite their significant contributions, these approaches exhibit several limitations. 
Specifically, \cite{schlegel2023codedpaddedfl} renders high encoding and decoding complexity, while \cite{bonawitz2017practical, zhao2022information,li2026informationtheoreticdecentralizedsecureaggregation} allow the server to recover partial sums of local models under partial participation, leaking additional information beyond the intended aggregate, and may even expose individual client updates at the server across multiple rounds \cite{so2023securing}. Table \ref{tab:comparison} summarizes the key differences between our work and existing SA schemes. 
}

{Notably, parallel to our work, another research direction has emerged on the \textbf{fundamental limits of coded SA}. In particular, Zhang \emph{et al.} \cite{zhang2025fundamentallimitshierarchicalsecure} characterized the fundamental limits of coding-enabled efficient SA and established a trade-off between communication efficiency and security under reliable communication. 
More closely related to the present work, \cite{weng2026resilientefficientlinearsecure} investigated coding-enforced SA under unreliable communication, establishing fundamental limits and providing theoretical insights complementary to the protocol design herein. 
Another related line of research considers secure computation of linearly separable functions \cite{9678313}. Although it shares some conceptual connections with our approach, its objectives and distributed learning setting differ substantially from those considered here.
}

\vspace{-0.5em}
\subsection{Our Contributions} 
{
In this work, we propose a robust coding-enforced structured SA method,  which simultaneously guarantees accurate computation of the global model and complete cancellation of privacy noise under unreliable communication. 
In addition, the binary decoding property of the coding scheme provides an additional privacy safeguard beyond noise injection, ensuring that the server either correctly recovers the intended aggregate or gains no meaningful information.
Both rigorous theoretical analysis and rich experimental validations are provided to support our findings. 
The key contributions are outlined below.
}
\vspace{0.1em}\\
\noindent(\textbf{C1}) We present a complete problem formulation 
of real-field zero-sum noise, covering its design principles, existence conditions, and constructions. 
Moreover, we introduce fairness as a metric in privacy preservation, which is defined by allocating privacy protection levels in proportion to the clients' learning weights. \footnote{Fairness in privacy preservation warrants further investigation from both algorithmic and system-design perspectives. Furture directions include fairness-aware privacy budget allocation and adaptive scheduling.}
\vspace{0.5em}\\
\noindent(\textbf{C2})  {We propose a robust coding-enforced SA method, termed SecCoGC, which exploits the matched structure between the target function, privacy mechanisms, and the coding schemes to achieve reliable, private and accurate computation.  Despite the structural perturbations caused by unreliable communication, SecCoGC exhibits robustness in the following respects: }
\begin{itemize}
    \item Whenever the global model can be correctly formed, the privacy noise is enforced to fully cancel out through the predetermined coded aggregation structure, irrespective of the privacy levels. This ensures the optimality of FL algorithms and avoids any negative effect arising from the unpredictable partial participation patterns of clients.
    \item Whenever the global model cannot be correctly formed, the privacy noise remains present in any aggregated result. Consequently, no information beyond the intended global model can be accurately revealed to the server. This guarantees privacy preservation for all communication patterns that does not support successful recovery of the global model. 
    \vspace{0.2em}
\end{itemize}
\noindent(\textbf{C3}) 
{A novel non-convex convergence analysis is developed for FL algorithms with binary global model reconstruction, accommodating arbitrary choice of local solvers, arbitrary privacy levels, and unreliable communication. }
Notably, the resulting convergence theorem applies not only to SecCoGC but also to a broad class of SA protocols in ppFL, 
where global reconstruction failure carries no meaningful information about the training process (e.g., decryption failures).
Distinctively, both the number of consecutive training rounds between two successful global model reconstructions and the total number of training rounds are r.v.s, and they are statistically dependent. This dependence significantly complicates the convergence analysis. 
{Our theoretical analysis establishes a linear convergence rate with probability 99.74\%, given the failure probability of global model recovery.}
\vspace{0.2em}\\
\noindent(\textbf{C4}) {Multi-view and in-depth privacy analyses are provided to examine several key aspects of privacy in SecCoGC. Our analysis addresses four distinctive challenges that, to the best of our knowledge, have not been tackled yet.}
\begin{itemize}
    \item  In cooperative networks, a local update may arrive at the server both directly and indirectly through different neighboring nodes. This multi-path delivery introduces a substantial conceptual challenge. 
    \item The correlation among privacy noise, coupled with unreliable communication, poses significant challenges to LDP analysis, as the variance of the aggregated noise varies randomly with the communication pattern. 
    \item Correlations among secret keys enables a client to infer other clients' local models from its own secret key and multiple correlated received messages.
    \item Unreliable communication causes the local updates and privacy noise to be simultaneously present or absent, rendering the sensitivity function unbounded.
\end{itemize}
{The analytical framework lays a foundation for future ppFL studies under correlated noise, unreliable communication, and cooperative networks.}


\section{System Model and Preliminaries}\label{sec:system model}
\subsection{Federated Learning (FL)}\label{sec: ml model}
Consider a FL system consisting of $K$ clients, each holding local datasets $\mathcal{D}_k$, coordinated by a central server. The goal is to learn a global model $\boldsymbol{\Theta}\in\mathbb{R}^D$ that minimizes the \textit{global objective function (GOF)} defined in (\ref{eq: goal}) through iterative training and communication: 
\begin{align}
\min_{\boldsymbol{\Theta}\in\mathbb{R}^D}
\;\;\left[ \mathcal{L}(\boldsymbol{\Theta})
\triangleq
\sum_{k=1}^{K} w_k \mathcal{L}_k(\boldsymbol{\Theta}; \mathcal{D}_k)\right].
\label{eq: goal}
\end{align}
The learning weight $w_k$ measures the importance of clients, satisfying $\sum_{k=1}^{K} w_k=1$. In this work, we assume $w_k=\frac{1}{K}$. The \textit{local objective function (LOF)} $\mathcal{L}_k(\boldsymbol{\Theta}, \mathcal{D}_k)$ that measures the accuracy of the model $\boldsymbol{\Theta}$ on local datasets $\mathcal{D}_k$ is defined by the average loss on all training samples in $\mathcal{D}_k$,
\begin{align}
    \mathcal{L}_k(\boldsymbol{\Theta}, \mathcal{D}_k)\triangleq \frac{1}{\lvert \mathcal{D}_k \rvert}\sum_{\xi\in \mathcal{D}_k} l(\boldsymbol{\Theta}, \xi),
    \label{eq: LOF}
\end{align}
where $l(\boldsymbol{\Theta}, \xi)$ is the sample loss of the model $\boldsymbol{\Theta}$ evaluated on a single training sample $\xi$. 

In the $t$-th training round, where $t\in[T]$, all clients are initialized with the latest global model, i.e., $\boldsymbol{\Theta}_k^{t,0}=\boldsymbol{\Theta}^{t-1}$. The local training on client $k$ at the $i$-th iteration is given by   
\begin{align}
    \boldsymbol{\Theta}_k^{t,i}=\boldsymbol{\Theta}_k^{t,i-1}-\eta a_{i} \nabla \mathcal{L}_k(\boldsymbol{\Theta}_k^{t, i-1}, \boldsymbol{\xi}_k^{t, i}), i\in [I],
    \label{eq: local iteration}
\end{align}
where $\eta$ is the learning rate, $I$ is the number of local iterations, $a_{i}$ characterizes the type of local solver, $\boldsymbol{\xi}_k^{t, i}$ is the data patch containing $n_p$ data samples extracted from $\mathcal{D}_k$ with $\lvert \boldsymbol{\xi}_k^{t, i} \rvert=n_p$. The gradient $\nabla\mathcal{L}_k(\boldsymbol{\Theta}_k^{t, i-1}, \boldsymbol{\xi}_k^{t, i})$ computed on the data patch $\boldsymbol{\xi}_k^{t, i}$ is defined by the averaged gradients over all data samples $\xi\in\boldsymbol{\xi}_k^{t, i}$. In a matrix form, the local model update $ \Delta\boldsymbol{\Theta}_k^{t,I}=\boldsymbol{\Theta}_k^{t,I}-\boldsymbol{\Theta}^{t}$ can be expressed by
\begin{align}
    \Delta\boldsymbol{\Theta}_k^{t,I}=-\eta\boldsymbol{a}\cdot \nabla\boldsymbol{\mathcal{L}}_k^{t,I}, 
    \label{eq: local training}
\end{align}
where $ \boldsymbol{a}=\left[a_{1}, \cdots, a_{I} \right]$ describes how the gradients accumulate over iterations \cite{wangTackling}, and the gradients obtained at all iterations are stored in $\nabla\boldsymbol{\mathcal{L}}_k^{t,I}$, defined as $\nabla\boldsymbol{\mathcal{L}}_k^{t,I}=\left[ \nabla \mathcal{L}_k(\boldsymbol{\Theta}_k^{t, 0}, \boldsymbol{\xi}_k^{t, 1}), \cdots, \nabla \mathcal{L}_k(\boldsymbol{\Theta}_k^{t, I-1}, \boldsymbol{\xi}_k^{t, I}) \right]^\top $. 

With full participation of clients, the server updates the global model $\boldsymbol{\Theta}^t$ based on the local model updates as follows, 
\begin{align}
\boldsymbol{\Theta}^{t}=\boldsymbol{\Theta}^{t-1}+\frac{1}{K}\sum_{k=1}^K \Delta\boldsymbol{\Theta}_k^{t,I}.
    \label{eq: sever_agg}
\end{align}
\subsection{Threat Model}


Our threat model follows a standard assumption adopted in multi-party computation (MPC) \cite{bonawitz2017practical,goldreich2004foundations,saha2024privacy,zhang2025informationtheoreticdecentralizedsecureaggregation}. The server and clients are assumed honest-but-curious: they faithfully execute the protocol and but attempt to infer other local models based on observable information and their local knowledge. 
To be more specific, the clients may try to learn the local model or the participation of a client by exploiting e.g., the received messages from its neighbors or the computed partial sums, and correlations among secret keys. 
{The server may try to infer the local model, the participation of a client at another relaying client, or the participation of a client based on the combined partial sum by applying the predefined combinator in the protocol}. 

\subsection{Communication and Network Model}
The inter-user and uplink communication channels are assumed to be orthogonal and modeled as independent Bernoulli r.v.s. The link $\tau_k$, from client $k$ to the sever, is modeled by $\tau_k\sim \mathrm{Ber}(1-p_k)$. The link $\tau_{k,m}$, from client $m$ to client $k$, is modeled by $\tau_{k,m}\sim \mathrm{Ber}(1-p_{k,m})$, where $p_k$ and $p_{k,m}$ are the probabilities when the channel is in outage. $\tau_k=1$ and $\tau_{k,m}=1$ imply a successful communication and perfect recovery, and $\tau_k=0$ and $\tau_{k,m}=0$ imply a complete failure. For $\forall k$, $p_{k,k}=0$, $\tau_{k,k}=1$, since client $k$ has access to its own local model and requires no transmission. For $k\neq m$, if no transmission occurs from client $m$ to client $k$, then $p_{k,m}=1$, $\tau_{k,m}=0$, as clients do not have direct access to each other’s local models. Moreover, any link is assumed to be statistically independent from each other, i.e., $\mathrm{(i)}\; \forall k_1\neq k_2: \tau_{k_1}\perp\tau_{k_2}$; $\mathrm{(ii)}\; \forall (k_1,m_1) \neq (k_2,m_2):\tau_{k_1,m_1}\perp\tau_{k_2,m_2}$; and $\mathrm{(iii)}\; \forall k_1, k_2, m: \tau_{k_1}\perp\tau_{k_2,m}$.   
{These assumptions are widely adopted in existing literature \cite{saha2024privacy,yemini2023robust,zheng2023federated}. Its correspondence with the underlying communication process was stated in our prior work \cite[Section II-3)]{10901785}.}

\subsection{Cyclic Gradient Codes}
{The cyclic gradient code is formed by a pair of matrices, the allocation matrix $\boldsymbol{G}=[g_{k,m}] \in \mathbb{R}^{K\times K}$ and the combination matrix $\boldsymbol{C}=[c_{f_t,k}] \in \mathbb{R}^{f\times K}$  such that $\boldsymbol{C}\boldsymbol{G}=\mathbf{1}_{f\times K}$, where $f={K \choose s}$.}
In particular, $\boldsymbol{G}$ is the allocation matrix of size $K\times K$ in a cyclic pattern with $s$ nonzero entries per row, and $\boldsymbol{C}$ is the combination matrix of size $f\times K$, where each row represents a unique pattern with exactly $s$ zeros per row, encompassing all such possible patterns. The structure ensures that, even if up to $s$ rows are missing from $\boldsymbol{G}$, its resulting product with $\boldsymbol{C}$ still includes at least one all-one row. The construction for GC codes can be found in \cite[Algo. II]{tandon2017gradient}.

\section{Real-Field Zero-Sum Noise} \label{Sec: formulation_zero_sum}
This section begins by discussing general real-field zero-sum noise. Then, fairness is introduced as a performance metric to ensure equitable privacy protection among clients.
\subsection{General Generation} \label{sec: code_construct_SecCoGC}
Define $L$ independent r.v.s $\boldsymbol{Z}_l^t\sim \mathcal{N}(\mathbf{0}, \boldsymbol{I}_D)$,
where $l\in[L]$ and $\boldsymbol{I}_D$ denotes $D\times D$ identity matrix. 
The secret key\footnote{In this paper, secret key and privacy noise are used interchangeably.} on each client $k$ is generated from a linear combination of these r.v.s, i.e., $\boldsymbol{N}_k^t=\sum_{l=1}^L\alpha_{k,l}\boldsymbol{Z}_l^t$. Moreover, the secret keys $\{\boldsymbol{N}_k^t\}_{k=1}^K$ are designed to sum to zero,
\begin{align}
    \sum_{k=1}^K \boldsymbol{N}_k^t=0.
    \label{eq: design_SecCoGC}
\end{align}
Let $\boldsymbol{A}$ be the generator matrix comprising all $\alpha_{k,l}$. 
In matrix form, the secret key construction can be expressed as \footnote{{The construction in \eqref{eq: key_generation} remains valid regardless of the distribution of the secret keys.}}
\begin{align}
    \begin{bmatrix}
    \boldsymbol{N}_1^t\\
    \vdots\\
    \boldsymbol{N}_K^t
    \end{bmatrix}
    =\boldsymbol{A}\cdot
    \begin{bmatrix}
        \boldsymbol{Z}_1^t\\
        \vdots\\
        \boldsymbol{Z}_L^t
    \end{bmatrix}.
    \label{eq: key_generation}
\end{align}
{To ensure the effectiveness of the randomized privacy mechanism, the key generation is designed to prevent unexpected complete cancellation of the injected randomness such that the underlying privacy mechanism remains effective. This structural requirement is independent of the adopted privacy metric and is therefore applicable to both LDP and mutual information privacy (MIP) \cite{zhao2022information}.}
To ensure effective privacy preservation while maintaining the desired zero-sum property, $\boldsymbol{A}$ must satisfy the following constraints. 
\begin{condition}\label{con: secure} 
First, to ensure that the secret keys $\{\boldsymbol{N}_k^t\}_{k\in[K]}$ sum to zero in \eqref{eq: design_SecCoGC}, each column of the generator matrix $\boldsymbol{A}$ must sum to zero, 
\begin{align}
[\text{Correctness}]\;\;\;\;\;\;  \forall l\in [L]: \sum_{k=1}^K\alpha_{k,l}=0.
\label{eq: correctness}
\end{align}
Second, any linear combination involving fewer than $K$ secret keys must remain non-trivial. That is, no combination of fewer than $K$ secret keys may sum to the all-zero vector, i.e.,
\begin{align}
[\text{Inter-user\;Security}]\;\;\;\;\;
\mathrm{rank}\left(\boldsymbol{A}\right)=K-1. 
\label{eq: security}
\end{align}
Third, to ensure that every secret key contributes to privacy preservation, each row of $\boldsymbol{A}$ must contain at least one non-zero entry, i.e.,
\begin{align}
[\text{Per-user\;Security}]\;\;\;\;\;\;
\forall k, \exists l: \alpha_{k,l}\neq 0. 
\label{eq: per-user security}
\end{align}
\end{condition}
\begin{remark}[Necessity of \eqref{eq: security}]\label{remark: neccessity}
{The rank of $\boldsymbol{A}$ cannot exceed $K-1$ while maintaining correctness. Achieving rank $K-1$ guarantees security under partial participation, as complete cancellation of the secret keys requires all $K$ keys.
In the context of this work, client-to-client communication outages may cause the randomly perturbed coefficient matrix observed by the server to become full-rank, thereby enabling the recovery of individual updates through alternative decoding techniques \cite{11175173}. Therefore, the security condition in \eqref{eq: security} is essential under unreliable communication, as it prevents complete cancellation of the injected randomness and thereby preserves the effectiveness of the underlying randomized privacy mechanism, independent of the adopted privacy measure.}
\end{remark}
\begin{remark}\label{remark: general_num_rv}
The minimum number of independent r.v.s required to construct zero-sum noise that satisfies the correctness and security constraints is $L \geq K-1$, as a result of \eqref{eq: security}. 
\end{remark}

\subsection{Fairness-aware Generation}
The correctness and security requirements in Condition \ref{con: secure} often result in uneven privacy noise power among clients, which directly causes unequal privacy preservation among clients. 
Consider that the predetermined learning weights are independent of communication and privacy, such a disparity lacks a principled justification and may compromise fairness.
Moreover, excessively low- or high-power signals are more susceptible to distortion and clipping by nonlinear power amplifiers, which impairs aggregation accuracy. 
To ensure that all privacy noise terms have identical power $\lambda^2$ assuming equal client importance 
i.e., $\boldsymbol{N}_k^t\sim \mathcal{N}(\mathbf{0}, \lambda^2\boldsymbol{I}_D)$, $\forall k$. This leads to the following condition. 
\begin{condition}[Fairness condition]\label{con: fair}
To ensure equitable privacy allocation, the sums of squared $\ell_2$-norm of rows in $\boldsymbol{A}$ should be identical,
    \begin{align}[\text{Inter-user Fairness}]\;\;\;\;\forall k\in [K]: \lVert \underline{\boldsymbol{\alpha}}_k \rVert_2^2=\lambda^2, \label{eq:fairness}
    \end{align}
    where $\underline{\boldsymbol{\alpha}}_k$ is the $k$-th row in $\boldsymbol{A}$ and $\lambda>0$ is a constant. 
\end{condition}
\begin{remark}
The fairness condition incurs no additional randomness requirement beyond that imposed by correctness and security. Specifically, the minimum number of independent r.v.s required to construct zero-sum noise satisfying all three constraints remains $L \geq K-1$, $ L\geq 2$, $K\geq 2$. 
The fairness condition in \eqref{eq:fairness} imposes $K$ additional constraints. A non-trivial solution for $\boldsymbol{A}$ satisfying both \eqref{eq: correctness} and \eqref{eq:fairness} exists if $KL\geq K+L$, i.e., $K\geq 2$, $L\geq 2$. Combining this observation with Remark~\ref{remark: general_num_rv} establishes the claim. 
\end{remark}
\begin{remark}
If $\boldsymbol{A}$ is of cyclic structure, each row must contains at least two zeros, i.e., $\forall k:\;\lVert \boldsymbol{\alpha}_k\rVert_0 \geq 2$.
\end{remark}

To generate fair zero-sum privacy noise, equations \eqref{eq: correctness}, \eqref{eq: security}, and \eqref{eq: per-user security} must be simultaneously satisfied. These constraints includes include the column-wise linear constraints for correctness and security, as well as the row-wise $\ell_2$-norm constraints required for fairness. The simultaneous row- and column-wise requirements complicate code construction, as they require the joint design of all rows and columns rather than allowing them to be treated independently. 
A direct approach is to solve these equations using methods such as substitution or Gröbner basis. However, such methods become computationally prohibitive in large-scale systems. 

To this end, \cite{9929413} proposes a covariance-based construction of zero-sum noise. A covariance matrix satisfying the zero-sum and per-user power constraints is first designed through an optimization problem, and correlated Gaussian perturbations are then generated by multiplying independent Gaussian noise with its matrix square root. 
This approach enables efficient construction of the fair zero-sum noise while significantly reducing computational complexity. Moreover, the privacy power can be flexibly allocated, which is potentially useful for heterogeneous learning weights.  
However, the constraints in \cite[(8b), (8c)]{9929413} alone are insufficient to guarantee the desired privacy preservation. The inter-user and per-user security requirements in \eqref{eq: security} and \eqref{eq: per-user security} must also be satisfied.

Consider multi-source inference, in which each client receives multiple correlated masked messages. Stronger correlation among secret keys provides more side information to an adversary and therefore leads to weaker privacy guarantees. Conversely, weaker correlation among secret keys reveals less side information and thus offers stronger privacy protection. 
Therefore, rather than seeking the optimal covariance matrix, this paper focuses on symmetric privacy preservation to avoid the unjustified imbalance in information leakage induced by secret-key correlations. Accordingly, we provide a simplified covariance-based generator-matrix construction to achieve a symmetric secret-key design. 
{For a given privacy-noise power $\lambda^2$, the correlation matrix is uniquely determined by the symmetry requirement and the zero-sum constraint,
\begin{align}
\boldsymbol{R}
=
\frac{K\lambda^2}{K-1}
\left(
\mathbf{I}_K-\frac{1}{K}\mathbf{1}_K\mathbf{1}_K^{\top}
\right). 
\label{eq: symmetric_covariance}
\end{align}
Equivalently,
\begin{align}
\boldsymbol{R}_{km}
=
\begin{cases}
\lambda^2, & k=m,\\[4pt]
-\frac{\lambda^2}{K-1}, & k\neq m.
\end{cases}
\end{align}
Then, the generator matrix $\mathbf{A}$ can be chosen uniquely as the principal square root of the correlation matrix \cite{9929413}, 
\begin{align}
\mathbf{A}=\boldsymbol{R}^{\frac{1}{2}}
=
\sqrt{\frac{K\lambda^2}{K-1}}
\left(
\mathbf{I}_K-\frac{1}{K}\mathbf{1}_K\mathbf{1}_K^{\top}
\right).
\label{eq:generator_matrix}
\end{align}
It can be verified that the generator matrix in \eqref{eq:generator_matrix} satisfies the correctness, security, fairness, and symmetry constraints.
}
\begin{remark}
    Symmetric privacy preservation requires exactly $K$ independent randomness. 
\end{remark}


\section{Secure Cooperative Gradient Coding (SecCoGC)}
\label{sec: Secure CoGC}

Each client $k\in\{1, \cdots, K\}\triangleq [K]$ holds a raw local update $\Delta\boldsymbol{\Theta}_k^{t,I}$ and a secret key $\boldsymbol{N}_k^t$ \footnote{
{Following prior studies \cite{zhao2022information,zhang2025fundamentallimitshierarchicalsecure,li2026informationtheoreticdecentralizedsecureaggregation,9929413,10932699}, it is assumed that the secret keys are generated and securely distributed by a trusted third party prior to the execution of the learning protocol.}}, as depicted in Fig. \ref{fig:SecCoGC}. 
To protect the local models, each client applies additive masking to its raw local model update prior to transmission,
\begin{align}
\boldsymbol{Y}_k^t=\Delta\boldsymbol{\Theta}_k^{t,I}+\boldsymbol{N}_k^t.
\label{eq: mask_input}
\end{align}
Then, each client $k$ transmits $\boldsymbol{Y}_k^t$ to its neighbors $m\in\mathcal{V}_k$, where $\mathcal{V}_k\triangleq\{m: g_{m,k}\neq 0\}$ (including client $k$ itself), through unreliable link $\tau_{m,k}^t$.   

Meanwhile, client $k$ also acts as a relay, listening from its neighboring client $m\in\mathcal{U}_k$, where $\mathcal{U}_k\triangleq\{m: g_{k,m}\neq 0\}$ (including client $k$ itself). Client $k$ may or may not be able to receive information from client $m$, depending on the realization of link $\tau_{k,m}^t$, where 
\begin{align}
   \tau_{k,m}^t=
   \begin{cases}
       1,\;\;\; \text{client $k$ can decode $\boldsymbol{Y}_m^t$ perfectly}, \\
       0,\;\;\; \text{client $k$ cannot hear from client $m$}.   
   \end{cases}
\end{align}
Let $\mathcal{U}_k^t$ denote the set of updates that client $k$ can successfully learn (including its own updates) at the $t$-th round, i.e., $\mathcal{U}_k^t=\{m: \tau_{k,m}^t=1\}$.
Then, client $k$ computes the product $g_{k,m}\boldsymbol{Y}_m^t$ for each $m\in \mathcal{U}_k^t$ according to the encoding matrix $\boldsymbol{G}$, and sums these products to compute the partial sum,
\begin{align}
     \boldsymbol{S}_k^t=\sum_{m\in \mathcal{U}_k^t} g_{k,m}\boldsymbol{Y}_m^t
     =\sum_{m=1}^K\tau_{k,m}^t\cdot g_{k,m}\boldsymbol{Y}_m^t.
     \label{eq: partial sum}
\end{align}
If $\mathcal{U}_k^t=\mathcal{U}_k$, i.e., client $k$ can successfully decode the message from all of its neighbors, then $\boldsymbol{S}_k^t$ is referred to as a complete partial sum, since the coefficients in the partial sum $\boldsymbol{S}_k^t$ fully matches those in $\boldsymbol{G}$. 

Then, client $k$ sends $\boldsymbol{S}_k^t$ (not necessarily complete) to the server through link $\tau_k^t$, where 
\begin{align}
   \tau_k^t=
   \begin{cases}
       1,\;\;\; \text{the server can recover $\boldsymbol{S}_k^t$ perfectly}, \\
       0,\;\;\; \text{the server hears nothing from client $k$}.   
   \end{cases}
\end{align}
\begin{figure}[t]
\centering
\scalebox{1}{\input{SecCoGC}}
\vspace{1mm}
\vspace{-2pt}
\caption{The proposed SecCoGC with simplified notations ($K=3$, $s=1$).
} 
\label{fig:SecCoGC}
\vspace{-0.3cm}
\end{figure}
Let $\mathcal{U}^t$ denote the set of complete partial sums arrived at the server, i.e., $\mathcal{U}^t=\{k: \mathcal{U}_k^t=\mathcal{U}_k \}\cap \{k: \tau_k^t=1 \}$. The server combines these complete partial sums to obtain the expected aggregation as follows. 
\begin{itemize}
\item \textit{Match the Combinator}: Let $\boldsymbol{c}_{f_t}$ denote the $f_t$-th row in the combination matrix $\boldsymbol{C}$. The combinator $\boldsymbol{c}_{f_t}$ corresponds to $\mathcal{U}^t$ should satisfy
\begin{align}
\forall k:\; c_{f_t,k} \neq 0 \Rightarrow k \in \mathcal{U}^t.
\label{eq: combinator}
\end{align}
Such $\boldsymbol{c}_{f_t}$ exists iff $\lvert\mathcal{U}^t\rvert\geq K-s$.
\item \textit{Server Aggregation}: If such $\boldsymbol{c}_{f_t}$ exists, the server can compute the exact global model by combining these complete partial sums accordingly,
\begin{align}
   \boldsymbol{\Theta}^{t}
   &=\boldsymbol{\Theta}^{t-1}+\frac{1}{K}\sum_{k=1}^K c_{f_t,k}\boldsymbol{S}_k^t.
   \label{eq: sever_agg_CoGC}
\end{align}
\end{itemize}
If the server receives at least $K-s$ complete partial sums, the server aggregation in \eqref{eq: sever_agg_CoGC} gives exactly the global model pursued by \eqref{eq: sever_agg} without introducing any residual privacy noise,
\begin{align}
    \frac{1}{K}\sum_{k=1}^K c_{f_t,k}\boldsymbol{S}_k^t
    &=\frac{1}{K}\sum_{k=1}^K \left(\Delta\boldsymbol{\Theta}_k^{t,I}+\boldsymbol{N}_k^t\right)
    =\frac{1}{K}\sum_{k=1}^K \Delta\boldsymbol{\Theta}_k^{t,I}.
\end{align}    
If the server receives less than $K-s$ complete partial sums, the global model cannot be formed correctly. In this case, the clients proceed to the next round by initializing with their most recent local models, i.e., by setting $\boldsymbol{\Theta}_k^{t+1,0}=\boldsymbol{\Theta}_k^{t, I}$ in \eqref{eq: local iteration}.

\textit{Stopping Rule:} The entire training process terminates when the total number of executed training rounds reaches $T$ and the global model is successfully formed in the final round.
The complete training procedure is detailed in Algorithm \ref{alg:seccogc}.

Notably, the privacy noise can be canceled only under the linear transformation defined in \eqref{eq: design_SecCoGC}. This transformation is tied to the same combinators that govern the server-side aggregation. Consequently, the server can recover either the noiseless global model or a noisy linear combination that still contains residual privacy noise. {As a result, successful global computation guarantees the accuracy of the recovered global model, while failed global computation guarantees privacy preservation. Moreover, the coding structure naturally precludes the negative bias induced by heterogeneous networks.}

\begin{remark}[Encoding and decoding complexity]
{During client-to-client communication, each client encodes its local update for transmission and decodes up to $s$ messages received from its $s$ neighboring clients, depending on the communication link conditions. During client-to-server transmission, $K$ clients encode their complete partial sums. In total, each client performs $2$ encoding operations and up to $s$ decoding operations, and the server performs at most $K$ decoding operations.
The total communication cost and storage cost in terms of $K$ and $s$ is the same as \cite[Remark 3]{11175173}.}\footnote{{The reported communication and decoding costs exclude the overhead of secret key allocation, which incurs additional communication costs between the clients and the key management service.}}
\end{remark}
\begin{remark}[Compatibility with communication systems]
{The SecCoGC can be readily integrated with practical communication-system models under limited transmission resources for communication and privacy. For instance, the communication system framework and theoretical analyses developed in our prior work \cite{10901785} can be immediately applied to the proposed SecCoGC framework.}\footnote{{Future work may extend the SecCoGC framework to address practical communication issues, including communication delays, multi-packet transmissions, and heterogeneous client behaviors.}}
\end{remark}


\begin{algorithm}[t]
\caption{Pseudocode of the SecCoGC Protocol}
\label{alg:seccogc}
\begin{algorithmic}[1]

\State \textbf{Input:} $T, K, \eta$, $\boldsymbol{\Theta}_k^{0}$, $t \gets 0$, $T_s$, $R_{t_s} \gets 1$, $t_s \gets 0$

\While{$\sum_{j=1}^{t_s} \le T$}
    \State $t\gets t+1$
    \State \texttt{Client side}
    \For{ client $k\in [K]$ }
    \If {$\iota=1$}
    \State $\boldsymbol{\Theta}_k^{t,0}=\boldsymbol{\Theta}^{t-1}$
    \Else
    \State $\boldsymbol{\Theta}_k^{t+1,0}=\boldsymbol{\Theta}_k^{t, I}$
    \EndIf
    \For{iteration $i \in [I]$}
        \State $\boldsymbol{\Theta}_k^{t,i}\gets \boldsymbol{\Theta}_k^{t,i-1}-\eta a_{i} \nabla \mathcal{L}_k(\boldsymbol{\Theta}_k^{t, i-1}, \boldsymbol{\xi}_k^{t, i})$
    \EndFor
    \State Update $\Delta\boldsymbol{\Theta}_k^{t,I}=\boldsymbol{\Theta}_k^{t,I}-\boldsymbol{\Theta}_k^{t,0}$
    \State Mask $\boldsymbol{Y}_k^t=\Delta\boldsymbol{\Theta}_k^{t,I}+\boldsymbol{N}_k^t$
    \State Send $\boldsymbol{Y}_k^t$ to neighbors in $k\in\mathcal{V}_k$
    \State Hear $\boldsymbol{Y}_m^t$ from neighbors in $m\in\mathcal{U}_k$
    \State Compute $\boldsymbol{S}_k^t=\sum_{m\in \mathcal{U}_k^t} g_{k,m}\boldsymbol{Y}_m^t$
    \State Send $\boldsymbol{S}_k^t$ to the server
    \EndFor
    \State \texttt{Server side}
    \State Detect the combinator $\boldsymbol{c}_{f_t}$ 
    \If{$\boldsymbol{c}_{f_t}$ exists}
    \State $\iota=1$  \Comment{Indicator for GC decoding}
    \State $t_s\gets t_s+1$ \Comment{Index for successful recovery}
    \State Update $\boldsymbol{\Theta}^{t}=\boldsymbol{\Theta}^{t-1}+\frac{1}{K}\sum_{k=1}^K c_{f_t,k}\boldsymbol{S}_k^t$ 
    \State $R_{t_s}\gets 1$
    \Else
    \State $\iota=0$
    \State $R_{t_s}\gets R_{t_s}+1$ \Comment{Consecutive training rounds}
    \EndIf
    \State $r \gets r + 1$
\EndWhile
\State Update $T_s\gets t_s$  \Comment{Total number of global updates}
\end{algorithmic}
\end{algorithm}
\vspace{-3mm}

\section{Binary Convergence Analysis}\label{sec:convergence}
A novel convergence analysis is developed for FL algorithms with binary global model reconstruction, where each communication round results in either an exact reconstruction of the global model or a non-meaningful outcome. 
This convergence framework applies not only to SecCoGC but also to a broader class of SA protocols and coded distributed learning (DL) schemes.
\begin{figure}[t]
\centering
\includegraphics[width=1\linewidth]{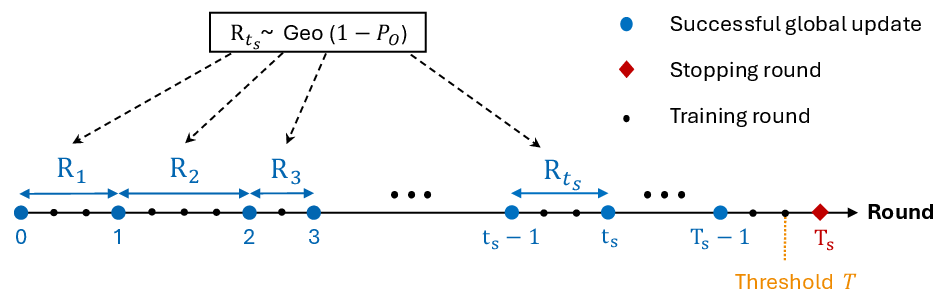}
\vspace{-5mm}
\caption{
{Illustration of the preset training-round threshold $T$, the number of consecutive training rounds between two successful global recoveries $R_{t_s}$, and the total number of successful global updates $T_s$.}
}
\label{fig: global_update_round}
\vspace{-5mm}
\end{figure}
{
\subsection{Statistical Characterization of the Training Process}
A distinctive feature of this class of algorithms is that both the number of consecutive training rounds between two successive successful global model recoveries and the total number of training rounds are r.v.s. 
In addition, these r.v.s are dependent, as the latter is determined by the realizations of the former and the preset threshold of the total number of training rounds $T$.}

{Specifically, by the stopping rule, training terminates at the first successful global model recovery for which the cumulative number of training rounds reaches or exceeds the threshold $T$.  
Let $t_s$ index successful global model recoveries, $R_{t_s}$ denote the number of consecutive training rounds (each training round contains $I$ local iterations) between the $(t_s-1)$-th and $t_s$-th successful recoveries, and $T_s$ denote the total number of successful global updates. Then, it follows that
\begin{align}
    T_s=\min\left\{n: \sum_{t_s=1}^n R_{t_s}\geq T\right\}. 
    \label{eq: Ts expression}
\end{align}
Fig. \ref{fig: global_update_round} provides a visual illustration of \eqref{eq: Ts expression}.
}

{
After each successful global model recovery, the recovery process restarts. 
Since the global model reconstruction succeeds independently with probability $1-P_O$ 
\footnote{ 
{The derivation of $P_O$ for SecCoGC is identical to that for CoGC, as both schemes employ the same $(K,s)$ coding structure. Consequently, the analyses in \cite{10901785,11175173} apply directly and are not repeated here. Notably, $P_O$ is not necessarily a monotonic function of the redundancy parameter $s$. Specifically, \cite{11175173} characterizes $P_O$ in heterogeneous networks and provides a detailed analysis of its dependence on $s$ and network conditions (see Section IV and Remark 5), while \cite{10901785} studies $P_O$ in homogeneous networks and establishes its relationship with the signal-to-noise ratio (SNR).
}
}
in each round, $R_{t_s}$ represents the waiting time until the next successful recovery. Therefore, 
\begin{align}
    R_{t_s}\overset{\mathrm{i.i.d.}}{\sim}\mathrm{Geo}(1-P_O), \qquad \forall\, t_s.
\end{align}
Consequently,
\begin{subequations}
\begin{align}
&e_R \triangleq \mathbb{E}[R_{t_s}] = \frac{1}{1-P_O},\\
&\nu_R \triangleq \mathrm{var}(R_{t_s}) = \frac{P_O}{(1-P_O)^2}.
\end{align}
\end{subequations}
}
{
In the convergence analysis, $R_{t_s}$ and $T_s$ are key yet statistically dependent parameters. The optimization error accumulates over $R_{t_s}$ consecutive training rounds between successive successful global updates, whereas the convergence bound is established by averaging the accumulated error across the $T_s$ successful global updates. 
Due to the dependence between them, establishing a valid convergence bound requires a joint characterization of the r.v.s $T_s$ and $\frac{1}{T_s}\sum_{t_s=1}^{T_s} R_{t_s}^2$.
To this end, the mean and variance of $R_{t_s}^2$ are required. 
\begin{subequations}
\begin{align*}
    &e_{R^2}\triangleq\mathbb{E}[R_{t_s}^2]=\sum_{R_{t_s}=1}^{+\infty}\hspace{-1mm} R_{t_s}^2 P_O^{R_{t_s}-1}(1-P_O)\\
    &\hspace{2cm}=\hspace{-1mm}\frac{1-P_O}{P_O}\mathrm{Li}_{-2}(P_O),\numberthis\\
    &\nu_{R^2}\triangleq\mathrm{var}(R_{t_s}^2)=\mathbb{E}[R_{t_s}^4]-e_{R^2}^2\\
    &\hspace{5mm}=\frac{1-P_O}{P_O}\mathrm{Li}_{-4}(P_O)-\left(\frac{1-P_O}{P_O}\mathrm{Li}_{-2}(P_O) \right)^2.
    \numberthis
\end{align*}
\end{subequations}
The polylogarithm function $\mathrm{Li}_{-v}(z)=\sum_{k=1}^\infty k^v z^k$ can be evaluated via $\mathrm{Li}_{-v}(z)= \left(z\frac{\partial (-\ln{(1-z)})}{\partial z}\right)^v \frac{z}{1-z}$. }

{
We next characterize the joint behavior of $T_s$ and $\frac{1}{T_s}\sum_{t_s=1}^{T_s} R_{t_s}^2$ with high probability. Since \eqref{eq: Ts expression} defines a renewal counting process with i.i.d. inter-renewal times $R_{t_s}$, classical results from renewal theory apply. In particular, by the strong law of large numbers (SLLN),
\begin{align}
    \frac{T_s}{T}
    \overset{\mathrm{a.s.}}{\rightarrow}
    \frac{1}{e_R},
    \qquad T\rightarrow\infty,
    \label{eq: Ts_slln}
\end{align}
where $\overset{\mathrm{a.s.}}{\rightarrow}$ denotes almost sure convergence. Then, by the central limit theorem (CLT),
\begin{align}
    \frac{T_s-\frac{T}{e_R}}{\sqrt{T}}
    \overset{d}{\rightarrow}
    \mathcal{N}\!\left(0,\frac{\nu_R}{e_R^3}\right).
    \label{eq: Ts_dist}
\end{align}
where $\overset{d}{\rightarrow}$ denotes convergence in distribution.
}

{
Since $T_s\to\infty$ almost surely as $T\to\infty$, the SLLN yields that
$\frac{1}{T_s}\sum_{t_s=1}^{T_s}R_{t_s}^2
\overset{\mathrm{a.s.}}{\longrightarrow}
e_{R^2}$. 
According to CLT, 
\begin{align}
    \sqrt{T_s}\left(\frac{1}{T_s}\sum_{t_s=1}^{T_s} R_{t_s}^2-e_{R^2}\right)\overset{d}{\rightarrow}\mathcal{N}(0, \nu_{R^2}).
    \label{eq: R^2_dist}
\end{align}
Applying the $3\sigma$-rule then yields the following high-probability characterization.
}

{
\begin{lemma}
Define events $\mathcal{C}_1\triangleq\left\{ T_s>\frac{T}{e_R}-3\sqrt{\frac{\nu_R T}{e_R^3}} \right\}$, and  
$\mathcal{C}_2\triangleq \left\{  \frac{1}{T_s}\sum_{t_s=1}^{T_s} R_{t_s}^2 <e_{R^2}+3\sqrt{\frac{\nu_{R^2}}{T_s} } \right\}$. Then, the following result holds: 
\begin{subequations}
\begin{align}
    \mathrm{P_r}\left(\mathcal{C}_1 \cap \mathcal{C}_2  \right)&\ge
1-\Pr(\mathcal C_1^\mathrm{c})-\Pr(\mathcal C_2^\mathrm{c}) \\
&\approx
1-2\times 0.0013
=
0.9974.
    \label{eq: bound_Ts}
\end{align}
\end{subequations}
\end{lemma}
}

\subsection{Non-convex Convergence Analysis}
The following assumptions on LOFs are adopted throughout the convergence analysis.
{The combination of Assumption 1--3 is the least restrictive \cite{weng2026heterogeneityawareclientsamplingoptimal} among commonly used assumption sets for convergence analysis \cite{wang2021quantized,wangTackling}, enabling broad applicability to practical deep neural network (DNN) training.}
\begin{assumption}[{Smoothness}]\label{assump:1}
    For $\forall k \in [K]$, each local objective function is lower bounded by $\mathcal{L}_k(x)\geq \mathcal{L}^\star$ and is Lipschitz differentiable and its gradient $\nabla \mathcal{L}_k(x)$ is $G$-smooth, i.e., $\lVert \nabla \mathcal{L}_k(x)-\nabla \mathcal{L}_k(y) \rVert\leq G\lVert x-y \rVert$, $\forall x, y\in \mathbb{R}^D$.
\end{assumption}
\begin{assumption}[Unbiasedness and Bounded Variance]\label{assump:2}
   For $\forall k \in [K]$, the local stochastic gradient is an unbiased estimator of the true local gradient, i.e., $\mathbb{E}_\xi[\nabla \mathcal{L}_k(x\vert \xi)]=\nabla \mathcal{L}_k(x)$, where $\nabla \mathcal{L}_k(x)$ has bounded data variance $\mathbb{E}_\xi[\lVert \nabla \mathcal{L}_k(x\vert \xi)-\nabla \mathcal{L}_k(x)\rVert^2]\leq \sigma^2 $, $\sigma^2>0$.  
\end{assumption}
\begin{assumption}[Bounded Dissimilarity]\label{assump:3}
    For any set of $\forall\{w_k\}_{k=1}^{K}: \sum_{k=1}^{K} w_k=1$, the dissimilarity between the local objective functions $\nabla \mathcal{L}_k(x)$ and the global objective function $\nabla \mathcal{L}(x)$ is bounded by $\sum_{k=1}^K w_k\lVert \nabla \mathcal{L}_k(x) \rVert^2\leq \beta^2\lVert  \nabla \mathcal{L}(x) \rVert^2+\kappa^2$, where $\beta^2\geq 1, \kappa^2\geq 0$.
\end{assumption}

Another lemma concerning the accumulation of local updates over $R_{t_s}$ consecutive training rounds under an arbitrary local solver is presented below.
\begin{lemma}\label{lemma: local accum}
    With an arbitrary choice of local solver $\boldsymbol{a}$, the accumulated local updates at client $k$ after $R_{t_s}$ consecutive training rounds is given by 
    \begin{align}
         \Delta\boldsymbol{\Theta}_k^{t_s,R_{t_s} I}=-\eta\cdot
         [\underbrace{\boldsymbol{a}\;\cdots\;\boldsymbol{a}}_{R_{t_s} \;\mathrm{copies}}]\cdot \nabla\boldsymbol{\mathcal{L}}_k^{t_s,R_{t_s} I},
         \label{eq:local_accum}
    \end{align}
    where $\Delta\boldsymbol{\Theta}_k^{t_s,R_{t_s} I}$ and $\nabla\boldsymbol{\mathcal{L}}_k^{t_s,R_{t_s} I}$ follow a definition similar to that in Section \ref{sec: ml model}. 
    Moreover, the following inequality holds 
    \begin{align}
        &\sum_{i_s=1}^{R_{t_s}I} a_{i_s}\mathbb{E}\left[ \left\lVert \boldsymbol{\Theta}_k^{t_s,i_s-1} -\boldsymbol{\Theta}^{t_s-1} \right\rVert^2 \right]\notag\\
        &\leq \frac{2\eta^2R_{t_s}^3\lVert \boldsymbol{a} \rVert_1^3}{1-2\eta^2G^2R_{t_s}^2\lVert \boldsymbol{a} \rVert_1^2} \mathbb{E}\left[\lVert \nabla \mathcal{L}_k\left( \boldsymbol{\Theta}^{t_s-1}\right) \rVert^2\right],
        \label{eq:local_accum_var}
    \end{align}
    where $a_{i_s}=a_{i_s\bmod I}, i_s=1, \cdots, R_{t_s}I$, and $a_{i_s\bmod I}$ corresponds to entries in $\boldsymbol{a}$, defined in \eqref{eq: local training}. 
\end{lemma}
\begin{proof}[Sketch Proof]
    By replacing $I$ with $R_{t_s}I$ and substituting \eqref{eq:local_accum} into \cite[Lemma 5]{weng2026heterogeneityawareclientsamplingoptimal}, Lemma \ref{lemma: local accum} is obtained.
\end{proof}

\begin{theorem}\label{theo:Converge}
Consider any FL algorithm with binary global model recovery, where the recovery failure probability is $P_O$.
Given an arbitrary local solver $\boldsymbol{a}$,
Let $\eta=\frac{1}{G}\sqrt{\frac{M}{T}}$. Under Assumptions~\ref{assump:1}--\ref{assump:3}, if the threshold $T$ for the number of training rounds is sufficiently large, then, {although the optimization error depends on the coupled random variables $R_{t_s}$ and $T_s$, the following deterministic convergence bound holds with probability $99.74\%$:}
\begin{align*}
        &\frac{1}{T_s}\sum_{t_s=1}^{T_s} \hspace{-0.5mm}\mathbb{E}\left[ \lVert \Delta \mathcal{L}(\boldsymbol{\Theta}^{t_s-1})\rVert^2 
        \right]
        \leq \frac{1}{T_s}\hspace{-0.5mm}\sum_{t_s=1}^{T_s} \hspace{-0.5mm} R_{t_s} \mathbb{E}\left[ \lVert \Delta \mathcal{L}(\boldsymbol{\Theta}^{t_s-1})\rVert^2 
        \right]\\
        &\leq \frac{2G(\mathcal{L}(\boldsymbol{\Theta}^0)-\mathcal{L}^\star)}{\lVert \boldsymbol{a}\rVert_1 (1-P_O)\sqrt{KT}}
        + \frac{2\lVert \boldsymbol{a}\rVert_1 (1+P_O)}{\left(1-P_O \right)^2} \sqrt{\frac{K}{T}} (\sigma^2+2\kappa^2), 
        \numberthis 
        \label{eq: theorem 1}
    \end{align*}
    where $T_s=\min\{n: \sum_{t_s=1}^n R_{t_s}\geq T\}$ is the total number of successful global updates. 
\end{theorem}
\begin{proof}
    Proofs are provided in Appendix \ref{Appx: theo:Converge}.
\end{proof}
\begin{remark}
If the learning rate $\eta$ is chosen independently of $T$, a linear convergence rate of $\mathcal{O}\!\left(\frac{1}{T}\right)$ exhibits. 
\end{remark}
\begin{remark}[Role of privacy]
{The convergence behavior of SecCoGC is unaffected by the privacy parameters, guaranteeing convergence to the optimal solution regardless of the privacy level or privacy budget.}
\end{remark}
\begin{remark}[Role of communication]
{The impact of communication links on convergence is entirely captured by the global model recovery failure probability $P_O$. Poor communication conditions lead to a large overall outage probability $P_O$. Consequently, the convergence rate is degraded in two ways: first, the update frequency is reduced by a factor of $1-P_O$; second, an additional slowdown proportional to $(1-P_O)^2$ is incurred due to data heterogeneity and local data variance. 
Due to the binary global model recovery mechanism, the objective inconsistency arising from network heterogeneity is naturally avoided.}
\end{remark}

\section{Privacy Analysis}\label{sec: LDP}
{This section examines several types of model privacy mainly under local differential privacy (LDP) framework in the proposed SecCoGC protocol.} 
The relevant definitions, adapted to our context, are presented in
\cite[Appendix A]{weng2025codingenforcedrobustsecureaggregation}. 
{Following LDP analysis in \cite{saha2024privacy}, the objective of the protocol is that no party can accurately infer any client's local model or participation, while the global model is intended to be publicly available.}
{It is worth noting that LDP is defined to characterize privacy preservation for a specific randomized mechanism, instead of every possible protocol transcript. Unlike MIP, which models the sensitive data as random variables and quantifies statistical information leakage, the randomness in the LDP framework arises solely from the privacy mechanism, while the local models are treated as fixed throughout the analysis. 
Consequently, the updated global model is uniquely determined once the local models are treated as fixed and therefore cannot be used as additional side information for removing randomness in the LDP framework. 
Moreover, the privacy analysis assumes $s<K-1$, i.e., each client is not connected with all other clients.}
{For convenience, we summarize the notations used throughout the privacy analysis in Table \ref{tab:notations} in Appendix \ref{appx: notation}. }
\vspace{-0.5em}
\subsection{Peer-to-Peer LDP}
The local model update is masked by \eqref{eq: mask_input} before transmission. 
Even if client $m$ can decode client $k$'s transmission, it should not confidently learn about $\Delta\boldsymbol{\Theta}_k^t$. However, unlike \cite{9929413,saha2024privacy}, client $m$ has access to $\boldsymbol{N}_m^t=\boldsymbol{n}_m$ as auxiliary information when inferring the $k$-th local model, and this auxiliary information is correlated with $\boldsymbol{N}_k^t$.
This is equivalent to adding privacy noise 
$\boldsymbol{N}_k^t\vert\boldsymbol{N}_m^t\sim\mathcal{N}\left(
\frac{\boldsymbol{n}_m}{ \lambda^2}\boldsymbol{R}_{m,k} ,\;
 (\lambda^2 - \frac{1}{ \lambda^2}\boldsymbol{R}_{m,k}\boldsymbol{R}_{k,m})\cdot\boldsymbol{I}_D
\right)$ that is independent of the local models.
\begin{theorem}[Peer-to-Peer LDP]\label{theo: p2p LDP}
    Given the per-dimension covariance matrix $\boldsymbol{R}$ of secret keys $\{\boldsymbol{N}_k^t\}_{k=1}^K$, for $\forall \Delta\boldsymbol{\Theta}_k^t, \Delta\boldsymbol{\Theta}_k^{t'}\in\mathbb{B}_D(R)$, and $\forall \delta_{m, k}^{(1)}\in(0,1]$, the masked transmission from client $k$ to client $m$ via unreliable link $\mathrm{Ber}(1-p_{m,k})$ is $(\epsilon_{m, k}^{(1)},(1-p_{m,k})\delta_{m, k}^{(1)})$-differentially private, where
\begin{align}
    \epsilon_{m, k}^{(1)}= \frac{2R}{\sqrt{\lambda^2 - \frac{1}{ \lambda^2}\boldsymbol{R}_{m,k}\boldsymbol{R}_{k,m}}}\left[2\log\left( \frac{1.25}{\delta_{m, k}^{(1)}} \right)\right]^{\frac{1}{2}}. 
\end{align}
    In other words, for any measurable set $\mathcal{S}$, it holds that
    \begin{align*}
         \mathrm{P_r}(\boldsymbol{Y}_k^{t}\in \mathcal{S} \vert \Delta\boldsymbol{\Theta}_k^t\in\mathcal{O}_k)
         \leq& e^{\epsilon_{m,k}^{(1)}}\mathrm{P_r}(\boldsymbol{Y}_k^{t}\in \mathcal{S} \vert \Delta\boldsymbol{\Theta}_k^{t'}\in\mathcal{O}_k)\\ 
         &\hspace{5mm}+(1-p_{m,k})\delta_{m,k}^{(1)}.
         \numberthis
    \end{align*}
\end{theorem}
\begin{proof}
    Proofs are provided in \cite[Appendix A]{weng2025codingenforcedrobustsecureaggregation}. 
\end{proof}
\begin{remark}[{Multi-source information leakage}]
\label{remark: Privacy leakage in global model}
{Under the LDP framework, the multi-source inference gain arising from jointly exploiting multiple correlated observations cannot be explicitly quantified, as LDP is defined for a specific mechanism. Thereby, we additionally examine the resulting mutual information (MI) leakage to capture multi-source information leakage as follows.}

{Assuming independent local training process across clients and multiple superposed sources of randomness (e.g., data sampling and data noise), by the Central Limit Theorem (CLT), the local updates are modeled as Gaussian r.v.s with $ \operatorname{Cov}(\Delta\boldsymbol{\Theta}_k^t) = \mathrm{Diag}(\boldsymbol{I}_D\zeta^2)$. 
Assume that client $m$ successfully received local updates from client $k'\in \mathcal{U}_k^t$, the information leakage of a local update $\boldsymbol{\Theta}_{k}^{t}$, $k\in \mathcal{U}_k^t$, based on client $m$'s received messages from its neighbors in $\mathcal{U}_m^{t,-m}\triangleq \mathcal{U}_k^t\setminus\{m\}$ and local information $\{\boldsymbol{\Theta}_m^{t}, \boldsymbol{S}_m^t\}$ and the global model $\frac{1}{K}\sum_{k=1}^K \Delta\boldsymbol{\Theta}_k^{t,I}$ can be quantified by}
\begin{subequations}
\begin{align}
    &\mathrm{I}\left(  \boldsymbol{\Theta}_{k}^{t}; \{ \boldsymbol{Y}_{k'}^t \}_{k'\in \mathcal{U}_k^t}  \middle\vert \frac{1}{K}\sum_{k=1}^K \Delta\boldsymbol{\Theta}_k^{t,I}, \boldsymbol{\Theta}_m^{t}, \boldsymbol{S}_m^t \right) \notag\\
    =&\mathrm{I}\left(  \boldsymbol{\Theta}_{k}^{t}; \{ \boldsymbol{Y}_{k'}^t \}_{k'\in \mathcal{U}_m^{t,-m}}  \middle\vert \frac{1}{K}\sum_{k=1}^K \Delta\boldsymbol{\Theta}_k^{t,I}, \boldsymbol{S}_m^t \right)\\
    =&\mathrm{h}\left(  \{ \boldsymbol{Y}_{k'}^t \}_{k'\in \mathcal{U}_m^{t,-m}}  \middle\vert \boldsymbol{S}_m^t \right) \notag\\
    &\hspace{2mm}-\mathrm{h}\left(  \{ \boldsymbol{Y}_{k'}^t \}_{k'\in \mathcal{U}_m^{t,-m}}  \middle\vert \frac{1}{K}\sum_{k=1}^K \Delta\boldsymbol{\Theta}_k^{t,I}, \boldsymbol{\Theta}_k^{t}, \boldsymbol{S}_m^t  \right)\\
    =& \resizebox{0.95\linewidth}{!}{$%
    \begin{aligned}[t]
    \frac{D}{2}\log\!\left(
    \frac{
    \mathrm{det}\left(\zeta^2 \boldsymbol{I}_{\lvert \mathcal{U}_m^{t,-m} \rvert }+\operatorname{Cov}\left(\{ \boldsymbol{S}_{k'}^t(d) \}_{k'\in \mathcal{U}_m^{t,-m}} \middle\vert \boldsymbol{S}_m^t(d)\right)\right)
    }{
    \mathrm{det}\left(\operatorname{Cov}\left( \{ \boldsymbol{Y}_{k'}^t(d) \}_{k'\in \mathcal{U}_m^{t,-m}} \middle\vert \boldsymbol{\Theta}_k^{t}(d), \boldsymbol{S}_m^t(d) \right)\right)
    }
    \right),
    \end{aligned}%
    $}
    \label{dist: MI_multi_leakage}
\end{align}
\end{subequations}
{where \eqref{dist: MI_multi_leakage} applies the independence across dimensions, and the conditional variance $\operatorname{Cov}(\{ \boldsymbol{S}_{k'}^t \}_{k'\in \mathcal{U}_m^{t,-m}} \vert \boldsymbol{S}_m^t)$ and $\operatorname{Cov}( \{ \boldsymbol{Y}_{k'}^t \}_{k'\in \mathcal{U}_m^{t,-m}} \vert \boldsymbol{\Theta}_m^{t}, \boldsymbol{S}_m^t )$ are given in \eqref{eq: distr_key} and \eqref{dist: Y|theta,S}.}
\begin{proof}
{
In the $d$-th dimension, where $d\in [D]$, due to \eqref{eq: mask_input} and the independence between the local models and secret keys, it holds that}
\begin{align}
&\operatorname{Cov}\left(\{ \boldsymbol{S}_{k'}^t(d) \}_{k'\in \mathcal{U}_m^{t,-m}} \middle\vert \boldsymbol{S}_m^t(d)\right) \notag \\
=&\boldsymbol{R}_{\mathcal{U}_m^{t,-m},\mathcal{U}_m^{t,-m}}-\boldsymbol{R}_{\mathcal{U}_m^{t,-m}, m} \boldsymbol{R}_{mm}^{-1}\boldsymbol{R}_{m, \mathcal{U}_m^{t,-m}}, \label{eq: distr_key}
\end{align}
{where $\boldsymbol{R}_{\mathcal{U}_m^{t,-m},\mathcal{U}_m^{t,-m}}$, $\boldsymbol{R}_{\mathcal{U}_m^{t,-m},m}$ and $\boldsymbol{R}_{m,m}$ are corresponding submatrices in the covariance matrix of secret keys in \eqref{eq: symmetric_covariance}. }
\begin{align}
&\{ \boldsymbol{Y}_{k'}^t(d) \}_{k'\in \mathcal{U}_m^{t,-m}}  \vert \boldsymbol{S}_m^t(d) \sim \notag\\
&\mathcal N\!\left(\hspace{-0.5mm} 0,\zeta^2 \boldsymbol{I}_{\lvert \mathcal{U}_m^{t,-m} \rvert }\hspace{-0.5mm}+\hspace{-0.5mm}\operatorname{Cov}\left( \{ \boldsymbol{S}_{k'}^t(d) \}_{k'\in \mathcal{U}_m^{t,-m}} \middle\vert \boldsymbol{S}_m^t(d) \right)\hspace{-0.5mm}\right), 
\label{dist: Y|S}
\end{align}
{where \eqref{eq: distr_key} is the Schur complement. Moreover, we have}
\begin{align}
&\operatorname{Cov}\left( \{ \boldsymbol{Y}_{k'}^t(d) \}_{k'\in \mathcal{U}_m^{t,-m}} \middle\vert \frac{1}{K}\sum_{k=1}^K \Delta\boldsymbol{\Theta}_k^{t,I}(d), \boldsymbol{\Theta}_k^{t}(d), \boldsymbol{S}_m^t(d) \right) \notag\\
=&\operatorname{Cov}\left(\{ \boldsymbol{\Theta}_{k'}^t(d) \}_{k'\in \mathcal{U}_m^{t,-m}} \middle\vert \frac{1}{K}\sum_{k=1}^K \Delta\boldsymbol{\Theta}_k^{t,I}(d), \boldsymbol{\Theta}_{k}^{t}(d)\right) \notag\\
&\hspace{3cm}+\operatorname{Cov}\left(\{ \boldsymbol{S}_{k'}^t(d) \}_{k'\in \mathcal{U}_m^{t,-m}} \middle\vert \boldsymbol{S}_m^t(d)\right), \notag\\
=&\zeta^2 \boldsymbol{I}_{\lvert \mathcal{U}_m^{t,-m} \rvert }^{-o_k} \left(\boldsymbol{I}_{\lvert \mathcal{U}_m^{t,-m} \rvert }-\frac{1}{K-1}\mathbf{1}_{\lvert \mathcal{U}_m^{t,-m} \rvert }\mathbf{1}_{\lvert \mathcal{U}_m^{t,-m} \rvert }^\top\right)\boldsymbol{I}_{\lvert \mathcal{U}_m^{t,-m} \rvert }^{-o_k} \notag\\
&\hspace{2cm}+\operatorname{Cov}\left(\{ \boldsymbol{S}_{k'}^t(d) \}_{k'\in \mathcal{U}_m^{t,-m}} \middle\vert \boldsymbol{S}_m^t(d) \right),
\end{align}
{where $\boldsymbol{I}_{\lvert \mathcal{U}_m^{t,-m} \rvert }^{-o_k}$ denote the identity matrix, whose $o_k$-th diagonal element is $0$, and $o_k$ is the index of $k$ in $\mathcal{U}_m^{t,-m}$. Moreover, } 
\vspace{-5mm}
\begin{align}
& \{ \boldsymbol{Y}_{k'}^t(d) \}_{k'\in \mathcal{U}_m^{t,-m}} \bigg\vert \frac{1}{K}\sum_{k=1}^K \Delta\boldsymbol{\Theta}_k^{t,I}(d), \boldsymbol{\Theta}_m^{t}(d), \boldsymbol{S}_m^t(d)\sim \notag\\
&\hspace{0.7cm} \mathcal N\!\left(\hspace{-0.5mm} 0,\operatorname{Cov}\left( \{ \boldsymbol{Y}_{k'}^t(d) \}_{k'\in \mathcal{U}_m^{t,-m}} \middle\vert \boldsymbol{\Theta}_m^{t}(d), \boldsymbol{S}_m^t(d) \right)\hspace{-0.5mm}\right). 
\label{dist: Y|theta,S}
\end{align}
{Substituting \eqref{dist: Y|S} and \eqref{dist: Y|theta,S} into \eqref{dist: MI_multi_leakage} completes the proof.}
\end{proof}

\end{remark}
\subsection{LDP in Partial Sum Relaying} 
{This section examines the model privacy in the partial sum $\boldsymbol{S}_k^t$ relayed by client $k$.}
{After communication among clients, the relaying client $k$ computes the partial sum $\boldsymbol{S}_k^t$ in \eqref{eq: partial sum}, based on the received masked local updates in $\mathcal{U}_k^t$. 
Let $\boldsymbol{S}_k^{t, -j}\triangleq \sum_{m\in\mathcal{U}_k^t\setminus \{j\}} g_{k,m}(\Delta\boldsymbol{\Theta}_m^{t}+\boldsymbol{N}_m^t)$ denote the partial sum excluding client $j$’s contribution.
Define the auxiliary vector $\boldsymbol{\Lambda}_k^t=[\boldsymbol{\Lambda}_k^t(1),\cdots,\boldsymbol{\Lambda}_k^t(K)]$ with $\boldsymbol{\Lambda}_k^t(m)=\tau_{k,m}g_{k,m}$, and $\boldsymbol{\Lambda}_k^{t,-k_1}$ denote $\boldsymbol{\Lambda}_k^t$ with its $k_1$-th entry set to zero. 
Let $\boldsymbol{M}_k^t=\sum_{m\in\mathcal{U}_k^t} g_{k,m}\boldsymbol{N}_m^t$ be the aggregated privacy noise in the partial sum $\boldsymbol{S}_k^t$.
Similarly, define $\boldsymbol{M}_k^{t,-j}=\sum_{m\in\mathcal{U}_k^t\setminus \{j\}} g_{k,m}\boldsymbol{N}_m^t$. 
Then, their covariance matrices are r.v.s respective to the network topology,  $\mathrm{var}(\boldsymbol{M}_k^t)=\nu_k^t \boldsymbol{I}_D=\lVert \boldsymbol{\Lambda}_k^t\boldsymbol{A}\rVert^2\boldsymbol{I}_D$, and $\mathrm{var}(\boldsymbol{M}_k^{t,-j})=\nu_k^{t,-j} \boldsymbol{I}_D=\lVert \boldsymbol{\Lambda}_k^{t,-j} \boldsymbol{A}\rVert^2 \boldsymbol{I}_D$.
Then, the expectation of $\nu_k^t$ and $\nu_k^{t,-j}$ are given by 
$\Bar{\nu}_k=\mathbb{E}[\nu_k^{t}]=\lambda^2\sum_{m=1}^K(1-p_{k,m})g_{m,k}^2+\boldsymbol{\Lambda}_k^{\mathrm{E}}\boldsymbol{A}\boldsymbol{A}^\top(\boldsymbol{\Lambda}_k^{\mathrm{E}})^\top$ and
$\Bar{\nu}_k^{-j}=\mathbb{E}[\nu_k^{t,-j}]=\lambda^2\sum_{m=1}^K(1-p_{k,m})g_{m,k}^2+\boldsymbol{\Lambda}_k^{\mathrm{E},-j}\boldsymbol{A}\boldsymbol{A}^\top(\boldsymbol{\Lambda}_k^{\mathrm{E},-j})^\top$. The calculation of $\nu_k^t$, $\nu_k^{t,-j}$ $\Bar{\nu}_k$, $\Bar{\nu}_k^{-j}$ are specified in \cite[Appendix G]{weng2025codingenforcedrobustsecureaggregation}.
}

When an eavesdropper (possibly the server) observes the partial sum $\boldsymbol{S}_k^t$, it should not be able to identify any participating client $j$, i.e., it should not confidently infer whether client $j$'s contribution $g_{k,j}(\Delta\boldsymbol{\Theta}_j^{t}+\boldsymbol{N}_j^t)$ is included in $\boldsymbol{S}_k^t$. 
In this case, privacy stems from: \textit{(i)} the random participation of client $j$ at client $k$, and \textit{(ii)} the aggregated privacy noise $\boldsymbol{M}_k^{t,-j}$. 
The challenges are threefold:  
\textit{(i)} The privacy noise $\boldsymbol{N}_m^t$ are correlated, $\nu_k^{t}$ and $\nu_k^{t,-j}$ do not vary monotonically with the number of participating devices; 
\textit{(ii)} The $\nu_k^{t}$ and $\nu_k^{t,-j}$ are r.v.s with respect to the network topology; 
\textit{(iii)} The message to be protected, i.e., the $j$-th contribution $g_{k,j}(\Delta\boldsymbol{\Theta}_j^{t}+\boldsymbol{N}_j^t)$, is unbounded. 
These challenges are addressed through the choice of the Bernstein parameter \cite{saha2024privacy} and the Chi-squared tail bound.


\begin{theorem}[{LDP in protecting identity in a partial sum}]\label{theo: relay LDP 1}
{For $\forall \Delta\boldsymbol{\Theta}_k^t, \Delta\boldsymbol{\Theta}_k^{t'}\in\mathbb{B}_D(R)$, 
$\forall\delta_0$, $r_1>0$ and $\delta'\in(0,1)$ such that $\mathrm{P_r}(\lvert \nu_k^{t,-j}-\Bar{\nu}_k^{-j}\rvert\geq r_1)\leq \delta'$,}\footnote{The choice of Bernstein parameter $r_1$, $ r_2$, $r_3$, $ r_4$ is discussed in \cite[Appendix H]{weng2025codingenforcedrobustsecureaggregation}.} the partial sum $\boldsymbol{S}_k^t$ is $(\epsilon_{k,j}^{(2)},(1-p_{k,j})(\delta'+\delta_j^{(2)}))$-differentially private w.p. at least $1- e^{-\frac{D}{2}(\delta_0-\ln{(1+\delta_0)})}$ in protecting the identity of any participating node $j$, where 
\begin{align}
    \epsilon_{k,j}^{(2)}=
    \begin{cases}
        \left[2\log\left( \frac{1.25}{\delta_j^{(2)}} \right)\right]^{\frac{1}{2}}\frac{\lvert g_{k,j} \rvert\left( R+ \lambda\sqrt{D(1+\delta_0)}\right)}{\sqrt{\Bar{\nu}_k^{-j}-r_1}}\;\;\; \mathrm{if}\;p_{k,j}\geq 0,\\
        \hspace{25mm}0\;\;\;\;\;\;\;\;\;\; \;\;\; \;   \hspace{15mm}\;\;\; \mathrm{if}\;p_{k,j}=0.
    \end{cases}
\end{align}
In other words, for any measurable set $\mathcal{S}$, it holds that 
\begin{align}
    \mathrm{P_r}(\boldsymbol{S}_k^{t,-j}\in \mathcal{S} )\leq e^{\epsilon_{k,j}^{(2)}}\mathrm{P_r}(\boldsymbol{S}_k^{t}\in \mathcal{S} )+(1-p_{k,j})(\delta'+\delta_j^{(2)}),
\end{align}
w.p. at least $1- e^{-\frac{D}{2}(\delta_0-\ln{(1+\delta_0)})}$.
\end{theorem}
\begin{proof}
Proofs are provided in \cite[Appendix E]{weng2025codingenforcedrobustsecureaggregation}.
\end{proof}
Apart from the participating identity, when the transmission from client $j$ is perturbed from $\Delta\boldsymbol{\Theta}_j^t=\boldsymbol{u}$ to $\Delta\boldsymbol{\Theta}_j^{t}=\boldsymbol{u}'$, the aggregated partial sums should not fluctuate significantly. Different from the previous case, the noise providing privacy protection here is $\boldsymbol{M}_k^t$, which includes the $j$-th secret key. 
\begin{theorem}[{LDP against perturbation in a partial sum}]\label{theo: relay LDP 2}
{For $\forall \Delta\boldsymbol{\Theta}_k^t, \Delta\boldsymbol{\Theta}_k^{t'}\in\mathbb{B}_D(R)$,  $r_2>0$ and $\delta'\in(0,1)$ such that $\mathrm{P_r}(\lvert\nu_k^t-\Bar{\nu}_k^t\rvert\geq r_2)\leq \delta'$,} the partial sum $\boldsymbol{S}_k^t$ is $\left(\epsilon_{k,j}^{(3)},(1-p_{k,j})(\delta'+\delta_j^{(3)})\right)$-differentially private in protecting the local data of any participating node $j$ from perturbations, where 
\begin{align}
    \epsilon_{k,j}^{(3)}=
    \begin{cases}
        2\cdot\left[2\log\left( \frac{1.25}{\delta_j^{(3)}} \right)\right]^{\frac{1}{2}}\frac{\lvert g_{k,j} \rvert R}{\sqrt{\Bar{\nu}_k-r_2}}\;\;\;\;\;  \mathrm{if}\;p_{k,j}\geq 0,\\
        \hspace{15mm}\;\;\;\;\;0\;\;\;\;\;\;  \hspace{15mm}\;\;\; \mathrm{if}\;p_{k,j}=0.
    \end{cases}
\end{align}
In other words, for any $\boldsymbol{u}, \boldsymbol{u'}\in\mathbb{B}_D(R)$, and any measurable set $\mathcal{S}$, it holds that 
\begin{align*}
    \mathrm{P_r}(\boldsymbol{S}_k^{t}\in \mathcal{S}\vert \Delta\boldsymbol{\Theta}_j^{t}=\boldsymbol{u} )
    &\leq e^{\epsilon_{k,j}^{(3)}}\mathrm{P_r}(\boldsymbol{S}_k^{t}\in \mathcal{S}\vert \Delta\boldsymbol{\Theta}_j^{t}=\boldsymbol{u}')\\
    &+(1-p_{k,j})(\delta'+\delta_j^{(3)}).
    \numberthis
\end{align*}
\end{theorem}
\begin{proof}
    Proofs are provided in\cite[Appendix F]{weng2025codingenforcedrobustsecureaggregation}.
\end{proof}
\subsection{LDP Under Global Update Failures} 
{This section examines the model privacy in global model recovery failures. Assume that the server receives partial sums from clients in $\mathcal{V}^t\triangleq \{k: \tau_k^t=1 \}$, but fails to reconstruct the global model using the combinator $ \boldsymbol{c}_{f'}$ (referred to as a $\boldsymbol{c}_{f'}$-Failure). }
The resulting aggregation result $\boldsymbol{F}_{f'}^t=\sum_{k\in\mathcal{V}^t} c_{f',k}\sum_{m\in\mathcal{U}_k^t} g_{k,m}(\Delta\boldsymbol{\Theta}_m^{t,I}+\boldsymbol{N}_m^t)$ still contains residual privacy noise. The effective noise in this aggregated result is given by $\boldsymbol{N}_{F_{f'}}^t=\sum_{k\in\mathcal{V}^t} c_{f',k}\sum_{m\in\mathcal{U}_k^t} g_{k,m}\boldsymbol{N}_m^t$. 
The variance of $\boldsymbol{N}_{F_{f'}}^t$ is $\nu_{F_{f'}}^t\boldsymbol{I}_D=\mathrm{var} (\boldsymbol{N}_{F_{f'}}^t)$, 
and its expectation is $\Bar{\nu}_{F_{f'}}^t=\mathbb{E}[\mathrm{var} (\boldsymbol{N}_{F_{f'}}^t)]$.  
Similarly, define $\boldsymbol{F}_{f'}^{t, \{k,-j\}}=\sum_{k\in\mathcal{V}^t} c_{f',k}\sum_{m\in\mathcal{U}_k^t\setminus\{j\}} g_{k,m}(\Delta\boldsymbol{\Theta}_m^{t,I}+\boldsymbol{N}_m^t)$ as the aggregation excluding client $j$'s contribution at relaying client $k$, and $\boldsymbol{N}_{F_{f'}}^{t,\{k,-j\}}=\sum_{k\in\mathcal{V}^t} c_{f',k}\sum_{m\in\mathcal{U}_k^t \setminus\{j\}} g_{k,m}\boldsymbol{N}_m^t$. 
Moreover, define $\nu_{F_{f'}}^{t,\{k,-j\}}=\mathrm{var} (\boldsymbol{N}_{F_{f'}}^{t,\{k,-j\}})$ and $ \Bar{\nu}_{F_{f'}}^{\{k,-j\}}=\mathbb{E}[\mathrm{var} (\boldsymbol{N}_{F_{f'}}^{t,\{k,-j\}})]$ . 
The $\nu_{F_{f'}}^t$, $\Bar{\nu}_{F_{f'}}^t$, $\nu_{F_{f'}}^{t,\{k,-j\}}$ and $ \Bar{\nu}_{F_{f'}}^{\{k,-j\}}$ are further specified in \cite[Appendix H]{weng2025codingenforcedrobustsecureaggregation}.

Let $\mathcal{V}_{f'}\triangleq \{k: c_{f',k}\neq 0\}$, and $\mathcal{V}_{f'}^t\triangleq \mathcal{V}_{f'}\cap \mathcal{V}^t$. 
From $\boldsymbol{F}_{f'}^t$, the server should not infer the participation of client $j$ at the relaying client $k$. This privacy is obtained by: $\textit{(i)}$ the privacy noise $\boldsymbol{N}_{F_{f'}}^{t,\{k,-j\}}$, which consists of:
(a) the privacy noise sent from other clients (excluding client $j$), 
$\mathcal{U}_k \setminus \{j\}$, to client $k$; and 
(b) the privacy noise sent from $\mathcal{U}_m$ to client $m\in \mathcal{V}_{f'}\setminus \{k\}$, corresponding to non-zero combination coefficients 
$c_{f',m}$; $\textit{(ii)}$  the link intermittency $\tau_{k,j}$ from client $j$ to client $k$; and $\textit{(iii)}$ the link intermittency $\tau_k$ from client $k$ to the server. 

If $\Delta\boldsymbol{\Theta}_j^t=\boldsymbol{u}$ is perturbed to $\Delta\boldsymbol{\Theta}_j^{t}=\boldsymbol{u}'$, the server should not observe a significant fluctuation in the resulting $\boldsymbol{F}_{f'}^t$. The analytical difficulty arises as the perturbation propagates through multiple relaying nodes. The server can observe the perturbation unless every relaying path fails. 
The perturbation of the $j$-th local data is protected by 
$\textit{(i)}$ the privacy noise $\boldsymbol{N}_{F_{f'}}^{t}$, which consists of
(a) the privacy noise sent from clients (including client $j$), 
$\mathcal{U}_k$, to client $k$; and 
(b) the privacy noise sent from $\mathcal{U}_m$ to client $m\in\mathcal{V}_{f'}\setminus \{k\}$, corresponding to non-zero combination coefficients 
$c_{f',m}$; 
$\textit{(ii)}$ the link intermittency $\tau_{m,j}$ from client $j$ to all clients $m \in \mathcal{V}_j$ that relay the $j$-th message; and 
$\textit{(iii)}$ the link intermittency $\tau_m$ from client $m \in \mathcal{V}_j$ to the server. 
\begin{lemma}
The server cannot observe the perturbation of the local data on client $j$ through $\boldsymbol{c}_{f'}$-combination, only if all relaying clients that successfully hear from client  $j$ and corresponds to a non-zero combination coefficient in $\boldsymbol{c}_{f'}$ fails to communicate with the server, i.e., for $\forall m\in \mathcal{V}_{j,f'}^t\triangleq \mathcal{V}_j^t\cap \mathcal{V}_{f'}$, $\tau_m=0$, where $\mathcal{V}_j^t\triangleq\{m: m\in  \mathcal{V}_j\}\cap\{m: \tau_{m,j}\neq 0\}$ denotes the set of neighboring clients who successfully hears from client $j$, and  $\mathcal{V}_{f'}\triangleq \{k: c_{f',k}\neq 0\}$ denotes the clients who correspond to a non-zero combination coefficients in $\boldsymbol{c}_{f'}$. As a result, the probability of this event is given by 
\begin{align}
    \tilde{p}_{j,f'} =\hspace{-1mm} \sum_{ \mathcal{V}_{j,f'}^t\in \{\mathcal{V}_j\cap \mathcal{V}_{f'}\} } \prod_{k_1\in \mathcal{V}_{j,f'}^t}(1-p_{k_1}) \cdot p_{k_1,j}
    \hspace{-1mm}\hspace{-3mm}\prod_{k_2\in \{\mathcal{V}_j\cap \mathcal{V}_{f'}\}\setminus\mathcal{V}_{j,f'}^t} \hspace{-3mm} p_{k_2}.
\end{align}

\end{lemma}


\begin{theorem}[{LDP in $\boldsymbol{c}_{f'}$-failure}]\label{theo: server LDP-failure}
In $\boldsymbol{c}_{f'}$-Failure, for any $r_3>0$ and $\delta'\in(0,1)$ such that $\mathrm{P_r}(\lvert\nu_{F_{f'}}^{t,\{k,-j\}}-\Bar{\nu}_{F_{f'}}^{t,\{k,-j\}}\rvert\geq r_3 )\leq \delta'$,
$\boldsymbol{F}_{f'}^t$ is $(\epsilon_{f',\{k,j\}}^{(4)},(1-p_k)(1-p_{k,j})(\delta_{f',\{k,j\}}^{(4)}
    +\delta' )$ differentially private w.p. at least $1- e^{-\frac{D}{2}(\delta_0-\ln{(1+\delta_0)})}$ in protecting the identity of any participating client $j$ at relay $k$, where 
  \begin{align}
    &\epsilon_{f',\{k,j\}}^{(4)}\hspace{-1mm}=\hspace{-1mm}\left[\hspace{-0.5mm}2\log\left(\hspace{-1mm} \frac{1.25}{\delta_{f',\{k,j\}}^{(4)}} \hspace{-1mm}\right)\hspace{-1.5mm}\right]^{\frac{1}{2}}\hspace{-2mm}\frac{\lvert c_{f',k}g_{k,j}\rvert (R+ \lambda\sqrt{D(1+\delta_0)})}{\sqrt{\Bar{\nu}_{F_{f'}}^{t,\{k,-j\}}-r_3}}.
\end{align}
In other words, for any measurable set $\mathcal{S}$, it holds that 
\begin{align}
    \mathrm{P_r}(\boldsymbol{F}_{f'}^{t,\{k,-j\}}\in \mathcal{S} )\leq &e^{\epsilon_{f',\{k,j\}}^{(4)}}\mathrm{P_r}(\boldsymbol{F}_{f'}^t\in \mathcal{S})\notag\\
    &\hspace{-5mm}+(1-p_k)(1-p_{k,j})(\delta_{f',\{k,j\}}^{(4)}
    +\delta' ).
\end{align}
Moreover, for any $r_4>0$ and $\delta'\in(0,1)$ such that $\mathrm{P_r}(\lvert\nu_{F_{f'}}^{t}-\Bar{\nu}_{F_{f'}}^{t}\rvert\geq r_4 )\leq \delta'$, for any $\boldsymbol{u}, \boldsymbol{u'}\in\mathbb{B}_D(R)$, it holds that 
\begin{align*}
    \mathrm{P_r}(\boldsymbol{F}_{f'}^t\in \mathcal{S}\vert \Delta\boldsymbol{\Theta}_j^{t}=\boldsymbol{u} )
    &\leq e^{\epsilon_{k,j,f'}^{(5)}}\mathrm{P_r}(\boldsymbol{F}_{f'}^t\in \mathcal{S}\vert \Delta\boldsymbol{\Theta}_j^{t}=\boldsymbol{u}')\\
    &+(1-\tilde{p}_{j,f'})(\delta'+\delta_j^{(5)}),
    \numberthis
\end{align*}
w.p. at least $1- e^{-\frac{D}{2}(\delta_0-\ln{(1+\delta_0)})}$, where 
\begin{align}
    &\epsilon_{k,j,f'}^{(5)}=2\left[2\log\left( \frac{1.25}{\delta_j^{(5)}} \right)\right]^{\frac{1}{2}}\frac{\sum_{k=1}^K\lvert c_{f',k} g_{k,j} \rvert R}{\sqrt{\Bar{\nu}^{t,f'}-r_4}}.
\end{align}
\end{theorem}
\begin{proof}
    Proofs are provided in \cite[Appendix G]{weng2025codingenforcedrobustsecureaggregation}.
\end{proof}
{At the server side, extending multi-source inference to condition on all observed partial sums is not compatible with the standard LDP formulation, and the joint distribution of the partial sums becomes overly complex for a tractable exact MIP analysis, so we leave this as future work.}

\begin{remark}[{Multi-round Privacy}]
{Multi-round privacy in SecCoGC is discussed from two complementary perspectives.}
\begin{itemize}
    \item {
Through its coding structure, SecCoGC ensures that the injected privacy noise is completely canceled only during successful reconstruction of the intended global model, which aggregates all local models. For any other linear combination of local models, the injected privacy noise remains preserved. 
Consequently, the protocol prevents unintended disclosure of accurate local models through partial global aggregation across multiple rounds \cite{so2023securing}. }

    \item {
The LDP guarantees established in Theorems~\ref{theo: p2p LDP}--\ref{theo: server LDP-failure} characterize the privacy of the randomized mechanisms employed in each training round. Assuming that the secret keys are generated independently across rounds, the cumulative privacy loss follows directly from the standard adaptive sequential composition theorem for LDP~\cite{dwork2014algorithmic}.
}
\end{itemize}

\end{remark}

\begin{remark}[{Privacy leakage in successful global recovery}]
\label{remark: Privacy leakage in global model}
{LDP provides a worst-case guarantee independent of the statistics of local inputs, thus cannot capture the information leakage in the successful global update. In contrast, $\mu$-local mutual information privacy ($\mu$-LMIP) \cite{9448019} explicitly accounts for the statistical relationship between local and global models, enabling the quantification of information leakage in the global updates.}

{Assuming independent local training and multiple sources of randomness (e.g., data sampling and data noise), by the Central Limit Theorem (CLT), we model the local updates as Gaussian random variables with $ \operatorname{Cov}(\Delta\boldsymbol{\Theta}_k^t) = \mathrm{Diag}(\boldsymbol{I}_D\zeta^2)$. 
The information leakage of a local update $\Delta\boldsymbol{\Theta}_k^t$ from the global update $\sum_{m=1}^{K} w_m \Delta\boldsymbol{\Theta}_m^t$ is quantified by $\mu$-LMIP, where} 
\begin{align}
    \mu=\frac{D}{2}\log\left(1+\frac{w_k^2}{\sum_{m\neq k} w_m^2}\right).
\end{align}
{When clients are equally important, i.e., $ w_k=\frac{1}{M}$, $\mu=\frac{D}{2}\log\left(1+\frac{1}{M-1}\right)$.} 
\end{remark}
\begin{proof}
    The proof follows directly from MI property. 
\end{proof}
\begin{remark}[Impact of aggregation weights on privacy]
From Remark~\ref{remark: Privacy leakage in global model}, the information leakage $\mu_3$ is an increasing function of the aggregation weight $w_k$. Consequently, clients with larger aggregation weights are more susceptible to information leakage from the global model.
\end{remark}
\begin{remark}[Vanishing leakage in large systems]
From Remark \ref{remark: Privacy leakage in global model}, when $w_k=\frac{1}{K}$, $\lim_{K\rightarrow\infty}\mu_3=0$. As the number of clients grows, the information leakage from the global model about any individual local model asymptotically vanishes.
\end{remark}

\vspace{-0.5em}
\section{Simulations}\label{sec:simulation}
\subsection{Experimental Setups}
We evaluate the following methods on MNIST and CINIC-10, across varying levels of data heterogeneity, privacy preservation, and network conditions.
\begin{enumerate}[label=(\roman*)]
\item SecCoGC, as described in Section \ref{sec: Secure CoGC}. 
\item Standard FL with perfect connectivity. This benchmark provides insights into the ideal performance for any FL system under the same data heterogeneity. 
\item Non-private standard FL under unreliable communication from clients to the server. {This baseline could also represent the best performance of the 2-round SA protocol proposed in \cite{zhao2022information}, as it can be viewed as the case without dropouts in the second-round communication. }
\item Private standard FL under unreliable communication, which employs the Gaussian mechanism.
\item Private diversity network codes (DNC)-based FL\cite{10802992}, which employs the Gaussian mechanism and enables full client-to-client communication among $K$ clients.
\end{enumerate} 
The update rule under partial participation (iii), (iv), (v) is set non-blind and follows \cite[(17)]{10802992}. In such a design, the effective step size is not reduced by the number of missing updates.

The MNIST \cite{lecun1998gradient} is an easy-level benchmark dataset with a clean and standardized format.
The CINIC-10 \cite{darlow2018cinic}, formed by both CIFAR-10 and downsampled ImageNet, is more difficult to train due to greater diversity and real-world noise.
To simulate varying levels of data heterogeneity, local datasets are assigned using the Dirichlet distribution. The concentration parameter $\Gamma$ controls the level of heterogeneity among clients, where smaller values of $\Gamma$ indicate a higher imbalance.

\begin{figure*}[htbp]
  \centering
  \begin{subfigure}[b]{0.24\textwidth}
    \centering
    \includegraphics[width=\linewidth]{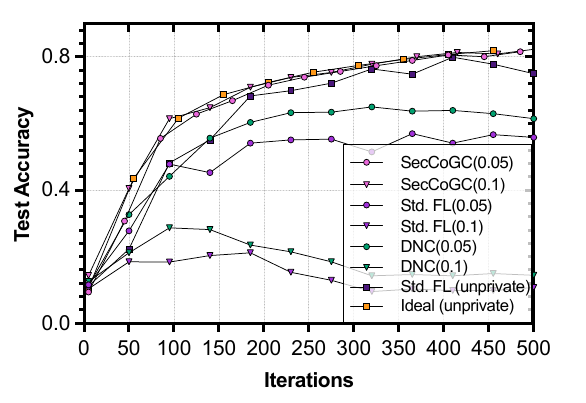}
    \caption{Symmetric network, $\Gamma=0.1$}
    \label{fig:mnist_data1_sym}
  \end{subfigure}
  \hfill
  \begin{subfigure}[b]{0.24\textwidth}
    \centering
    \includegraphics[width=\linewidth]{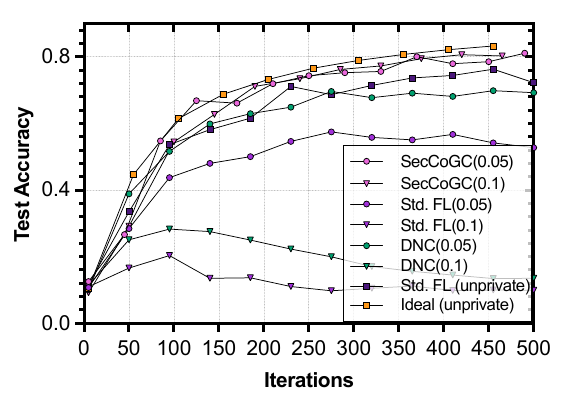}
    \caption{Asymmetric network, $\Gamma=0.1$}
    \label{fig:mnist_data1_ssym}
  \end{subfigure}
  \hfill
  \begin{subfigure}[b]{0.24\textwidth}
    \centering
    \includegraphics[width=\linewidth]{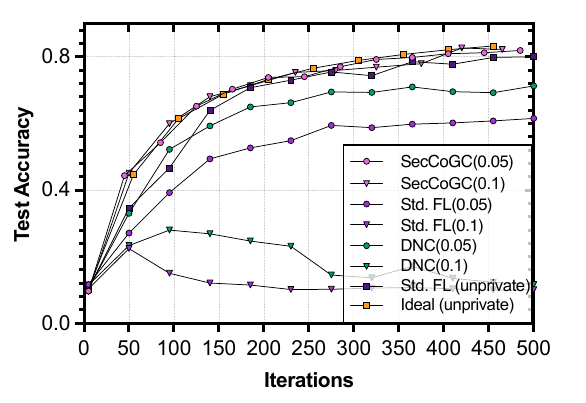}
    \caption{Symmetric network, $\Gamma=0.2$}
    \label{fig:mnist_data2_sym}
  \end{subfigure}
  \hfill
  \begin{subfigure}[b]{0.24\textwidth}
    \centering
    \includegraphics[width=\linewidth]{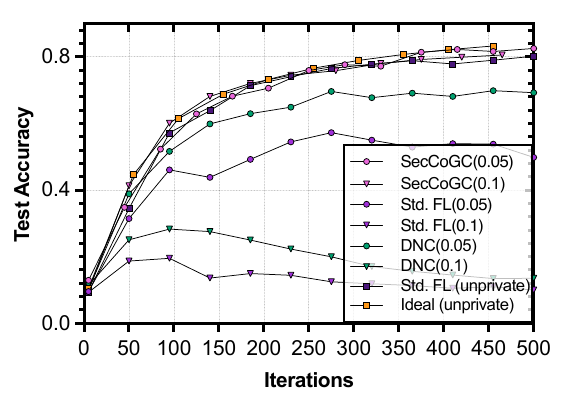}
    \caption{Asymmetric network, $\Gamma=0.2$}
    \label{fig:mnist_data2_asym}
  \end{subfigure}
  \caption{Performance comparison under varying privacy levels, network conditions, and data heterogeneity on the \textbf{MNIST} dataset.}
  \label{fig:mnist_all}
\end{figure*}
\begin{figure*}[htbp]
  \centering
  \begin{subfigure}[b]{0.24\textwidth}
    \centering
    \includegraphics[width=\linewidth]{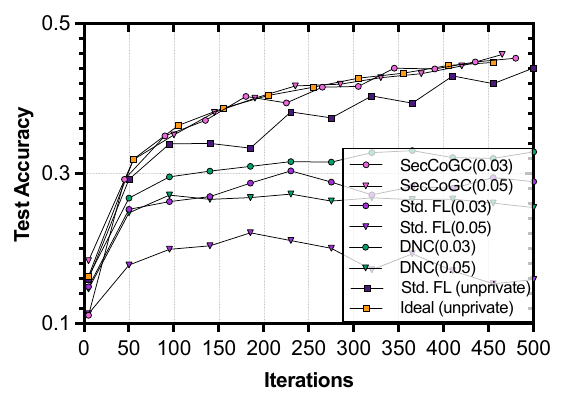}
    \caption{Symmetric network, $\Gamma=0.1$}
    \label{fig:cinic_data1_sym}
  \end{subfigure}
  \hfill
  \begin{subfigure}[b]{0.24\textwidth}
    \centering
    \includegraphics[width=\linewidth]{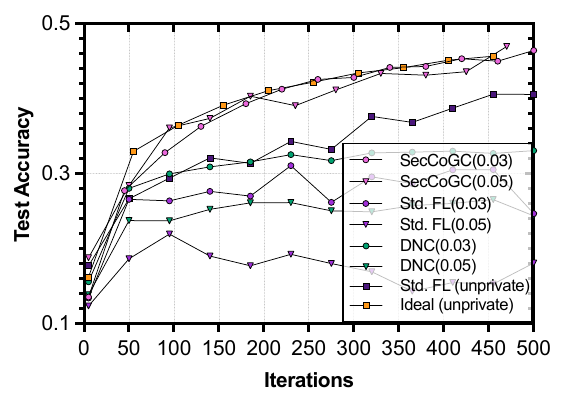}
    \caption{Asymmetric network, $\Gamma=0.1$}
    \label{fig:cinic_data1_ssym}
  \end{subfigure}
  \hfill
  \begin{subfigure}[b]{0.24\textwidth}
    \centering
    \includegraphics[width=\linewidth]{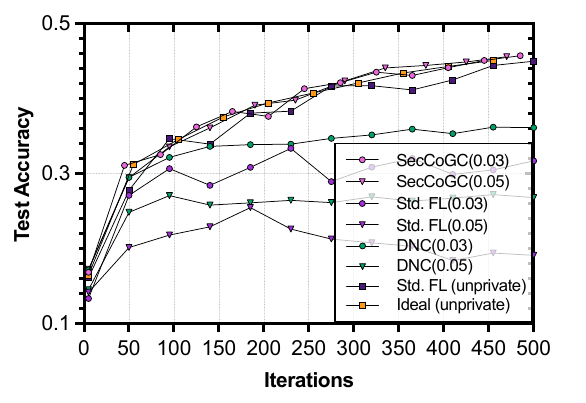}
    \caption{Symmetric network, $\Gamma=0.2$}
    \label{fig:cinic_data2_sym}
  \end{subfigure}
  \hfill
  \begin{subfigure}[b]{0.24\textwidth}
    \centering
    \includegraphics[width=\linewidth]{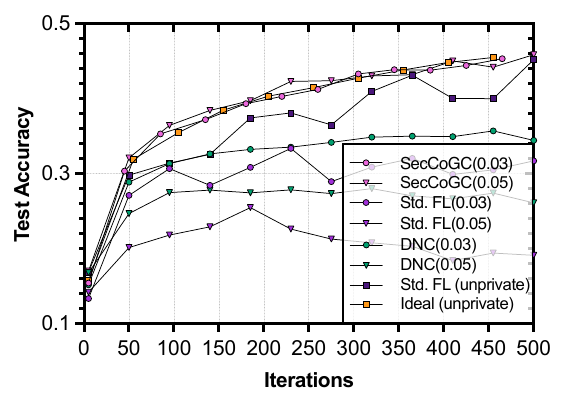}
    \caption{Asymmetric network, $\Gamma=0.2$}
    \label{fig:cinic_data2_asym}
  \end{subfigure}
  \caption{Performance comparison under varying privacy levels, network conditions, and data heterogeneity on the \textbf{CINIC} dataset.}
  \label{fig:cinic_all}
\end{figure*}

In the simulation, the number of clients is set to $K=10$. Each client is assigned an equal number of data samples. The total number of training rounds is set to $T=100$, and each client performs $I=5$ local training iterations per round.
The coding parameter $s$ is set to $7$ without further specification. 
The inter-client connectivity is configured with $p_{m,k}=0.9, \forall m\neq k$. In symmetric network settings, the local connectivity is set uniformly as $p_k=0.7$ for all clients. In asymmetric networks, each client is assigned a connectivity value $p_k\in\{0.500, 0.533, \cdots, 0.8\}$, for $k=\{1,\cdots, K\}$. For MNIST, the privacy level $\lambda$ is set to $0.05$ and $0.1$; for CINIC-10, it is set to $0.03$ and $0.05$.
Additional hyperparameters and neural network (NN) architectures are detailed in Table~\ref{tbl: cnn structures}.
\begin{table}[!ht]
\caption{
\footnotesize
NN architecture, hyperparameter specifications}
\label{tbl: cnn structures}
\resizebox{\linewidth}{!}{
\begin{tabular}{ccc}
\toprule
{\bf Datasets}& 
{\bf MNIST} & 
{\bf CINIC-10}\\
\toprule
Neural network &
CNN &
CNN \\
Model architecture$^*$ & 
\begin{tabular}{p{.18\textwidth}}
\centering
{\bf C}(1,10)
-- {\bf C}(10,20)
-- {\bf D}
-- {\bf L}(50)
-- {\bf L}(10)
\end{tabular}
&
\begin{tabular}{p{.18\textwidth}}
\centering
{\bf C}(3,32)
-- {\bf R}
-- {\bf M}
-- {\bf C}(32,32)
-- {\bf R}
-- {\bf M}
-- {\bf D}
-- {\bf L}(512)
-- {\bf R}
-- {\bf D}
-- {\bf L}(256)
-- {\bf R}
-- {\bf D}
-- {\bf L}(10)
\end{tabular}
\\
\midrule
Loss function &
Negative log-likelihood loss (NLLL) & Cross-entropy loss\\
\addlinespace[1ex]
Learning rate $\eta$
&
$\eta=0.002$ & $\eta=0.02$ \\ 
\addlinespace[1ex] 
\midrule
Batch size  &
\multicolumn{2}{c}{1024}\\
\addlinespace[1ex]
\bottomrule
\end{tabular}}
\vskip.2\baselineskip
\begin{tabular}{p{.47\textwidth}}
$^*$
\begin{footnotesize}    
{\bf C}(\# in-channel, \# out-channel): a 2D convolution layer (kernel size 3, stride 1, padding 1);
{\bf R}: ReLU activation function;
{\bf M}: a 2D max-pool layer (kernel size 2, stride 2);
{\bf L}: (\# outputs): a fully-connected linear layer;
{\bf D}: a dropout layer (probability 0.2).
\end{footnotesize}
\end{tabular}
\vspace{-5mm}
\end{table}

\subsection{Results and Discussions}
We report the average results of multiple runs. For a fair comparison, all results are truncated at the $100^\mathrm{th}$ training round. The simulation results are presented in Fig. \ref{fig:mnist_all} and Fig. \ref{fig:cinic_all}. These results demonstrate the strong robustness of SecCoGC against frequent communication failures between clients and the server, while preserving arbitrarily strong privacy guarantees. Detailed analyses are provided below.
\subsubsection{Fast convergence under privacy preservation}
When $\lambda$ is set to $0.05$ for MNIST and $0.03$ for CINIC-10, i.e., moderate levels of privacy noise, the private benchmark methods (including both private standard FL and private DNC) can work, but with significantly degraded performance. In addition to slower and less stable convergence, the final accuracy is also substantially reduced. This is because noise increases the amplitude of oscillations during optimization, and the larger step size prevents the final convergence point from accurately approaching the true global optimum. 
For this reason, even though DNC-based FL nearly ensures full recovery of all masked local models, masking still causes a considerable performance gap between ideal FL with perfect connectivity and DNC-based FL. As for the private standard FL, the perturbation induced by privacy-preserving noise, compounded by unreliable communication, disrupts the optimization trajectory of the masked global model. This results in a more pronounced divergence from the true global optimum, thereby leading to inferior performance compared to private DNC. With an increased privacy level ($\lambda = 0.1$ for MNIST and $\lambda = 0.05$ for CINIC-10), the utility degradation introduced by privacy noise becomes unacceptable. In this high-privacy regime, private standard FL consistently diverges across all configurations. Despite private DNC's near-complete removal of communication unreliability, the global model's accuracy remains significantly degraded, highlighting the intrinsic trade-off between strong privacy guarantees and model utility.
In contrast, our proposed SecCoGC consistently approaches ideal performance across all privacy levels. This is because SecCoGC reconstructs the global model exactly whenever recovery is possible. Nonetheless, we also observe slight fluctuations of SecCoGC compared to ideal FL, which is due to the increased variance resulting from the accumulation of local model updates, implied by Lemma \ref{lemma: local accum}.

\subsubsection{Robust SA under unreliable communication} On both MNIST and CINIC-10, the performance drop solely due to unreliable communication becomes more pronounced with increasing data imbalance and network asymmetry, as evidenced by the widening gap between ideal FL and unprivate standard FL. 
In contrast, our proposed SecCoGC naturally ensures the optimality over heterogeneous networks with varying data heterogeneity, owing to its binary decoding mechanism. 
It is worth noting that SecCoGC effectively mitigates the impact of unreliable communication and achieves exact recovery of the global model through its structured coding scheme. 
{This coding structure enforces the correct coefficient of each component in the aggregation, thereby ensuring that recovery is independent of any prior information.} 
Furthermore, SecCoGC leverages this coefficient-enforcing property to enable reliable secure aggregation, where user privacy is preserved, and ambiguity is ultimately removed from the final aggregation in successful global recovery, providing additional benefits that existing straggler-resilient schemes do not offer.

{
\subsubsection{Resilience to communication}
The resilience of SecCoGC to unreliable communication is determined by the underlying coding scheme and is unaffected by the integrated privacy mechanism. Since SecCoGC achieves the same global model recovery conditions and accuracy as our previously proposed Cooperative Gradient Coding (CoGC) framework \cite{10901785,11175173}, it inherits the same communication resilience properties.
A comprehensive evaluation of communication resilience under various coding parameters, SNR regimes, communication networks, data heterogeneity levels, and varying datasets has been presented in \cite{10901785,11175173}. As SecCoGC employs the same coding framework, these results directly apply. Therefore, the experiments in this section focus on assessing the robustness of the privacy mechanism and its impact on model accuracy in the presence of unreliable communications.
}

\section{Conclusion}\label{sec: conclusion}
In this work, we introduce SecCoGC, a coding-enforced robust secure aggregation (SA) method. {SecCoGC improves the reliability of zero-sum noise cancellation under unreliable communication, supports arbitrary privacy levels, and preserves the utility of the resulting global model. }
To conclude, a complete problem formulation of the real-field zero-sum noise is provided, along with the construction conditions and methods. 
A thorough performance analysis of the proposed SecCoGC in terms of privacy and convergence is provided.
Finally, the superiority of the proposed method is validated through numerical comparisons with state-of-the-art benchmarks over multiple datasets.


\balance
\bibliographystyle{IEEEtran.bst}
\bibliography{IEEEabrv,reference}

\appendices

\vspace{-0.5em}
\section{Table of Notations} \label{appx: notation}
\begin{table}[ht]
\centering
\caption{List of Notations}
\label{tab:notations}
\renewcommand{\arraystretch}{1.2}
\begin{tabular}{ll}
\hline
\textbf{Notation} & \textbf{Description} \\
\hline
$\boldsymbol{R}$ ($\boldsymbol{R}_{k,m}$) & Covariance matrix ($(k,m)$-th entry) \\
$\boldsymbol{N}_m^t$ ($\boldsymbol{n}_m$) & Secret key at  the $m$-th client (specific realization) \\
$\boldsymbol{Y}_k^{t}$ & Masked local update \\
$\boldsymbol{S}_k^t$ & The partial sum computed by client $k$ \\
$\boldsymbol{S}_k^{t, -j}$ & The partial sum excluding client $j$’s contribution \\
$\boldsymbol{M}_k^t$ & The aggregated privacy noise in $\boldsymbol{S}_k^t$ \\
$\nu_k^t$ & The variance of $\boldsymbol{M}_k^t$ \\
$\boldsymbol{M}_k^{t,-j}$ & The aggregated privacy noise in $\boldsymbol{S}_k^t$ excluding $\boldsymbol{N}_j^t$ \\
$\nu_k^{t,-j}$ & The variance of $\boldsymbol{M}_k^{t,-j}$ \\
 $\boldsymbol{\Lambda}_k^t$ &  Auxiliary vector \\
$\Bar{\nu}_k$ & Expectation of $\nu_k^t$\\
$\Bar{\nu}_k^{-j}$ &  Expectation of $\nu_k^{t,-j}$\\
$\boldsymbol{\Lambda}_k^{t,-k_1}$ & $\boldsymbol{\Lambda}_k^t$ with its $k_1$-th entry set to zero \\
$\mathcal{V}^t$ & The set of partial sums received by server \\
$\boldsymbol{F}_{f'}^t$ & The resulting aggregation result using combinator $\boldsymbol{c}_{f'}$\\
$\boldsymbol{N}_{F_{f'}}^t$ & The effective noise in this aggregated result $\boldsymbol{F}_{f'}^t$ \\
$\nu_{F_{f'}}^t\boldsymbol{I}_D$ & The variance of $\boldsymbol{N}_{F_{f'}}^t$ \\
$\Bar{\nu}_{F_{f'}}^t$ & The expectation of $\nu_{F_{f'}}^t$ \\
$\boldsymbol{F}_{f'}^{t, \{k,-j\}}$ & \makecell[l]{The aggregation excluding client $j$'s contribution \\ at relaying client $k$} \\
$\boldsymbol{N}_{F_{f'}}^{t,\{k,-j\}}$ & The aggregated noise in $\boldsymbol{F}_{f'}^{t, \{k,-j\}}$ \\
$\mathcal{V}_j^t$  &  \makecell[l]{The set of neighboring clients who successfully hears\\ from client $j$ } \\
$\mathcal{V}_{f'}$ & The set of non-trivial clients using combinator $\boldsymbol{c}_{f'}$  \\
$\mathcal{V}_{j,f'}^t$ & $\mathcal{V}_j^t\cap \mathcal{V}_{f'}$ \\
\hline
\end{tabular}
\end{table}
\vspace{-1em}
\section{Proofs of Theorem \ref{theo:Converge}} \label{Appx: theo:Converge}
Under Assumption \ref{assump:1}, it holds that
\begin{align*}
&\mathbb{E}\left[ \mathcal{L}(\boldsymbol{\Theta}^{t_s})\right]-\mathbb{E}\left[ \mathcal{L}(\boldsymbol{\Theta}^{t_s-1})\right]\\
&\leq \mathbb{E}\hspace{-0.5mm}\left[\left\langle \nabla\mathcal{L}(\boldsymbol{\Theta}^{t_s-1}) , \boldsymbol{\Theta}^{t_s}\hspace{-0.5mm}-\hspace{-0.5mm}\boldsymbol{\Theta}^{t_s-1}\right\rangle\right]
\hspace{-1mm}+\hspace{-1mm}\frac{G}{2} \mathbb{E}\hspace{-0.5mm}\left[ \lVert \boldsymbol{\Theta}^{t_s}\hspace{-0.5mm}-\hspace{-0.5mm}\boldsymbol{\Theta}^{t_s-1} \rVert^2 \right].
\numberthis
\label{eq:L-smooth}
\end{align*}

\begin{lemma}\label{lemma: first term}
Under Assumption \ref{assump:1}--\ref{assump:3}, it holds that 
\begin{align*}
    &\mathbb{E}\left[\left\langle \nabla\mathcal{L}(\boldsymbol{\Theta}^{t_s-1}), \boldsymbol{\Theta}^{t_s} -\boldsymbol{\Theta}^{t_s-1}\right\rangle\right]\\
    &\leq -\frac{1}{2}\eta\sum_{i_s=1}^{R_{t_s}I}a_{i_s}\mathbb{E}\left[\left\lVert \nabla\mathcal{L}(\boldsymbol{\Theta}^{t_s-1}) \right\lVert^2\right]\\
    &\hspace{5mm}+\frac{1}{2}\eta \frac{G^2}{K}\sum_{i_s=1}^{R_{t_s}I}a_{i_s} \sum_{k=1}^K\mathbb{E}\left[\left\lVert  
    \boldsymbol{\Theta}_k^{t_s-1}-\boldsymbol{\Theta}^{t_s,i_s-1}\right\lVert^2\right] \numberthis
\end{align*}
\end{lemma}
\begin{proof}
    The proof of Lemma \ref{lemma: first term} is provided in Appendix \ref{appx: lemma: first term}.
\end{proof}
\begin{lemma}\label{lemma: second term}
Under Assumption \ref{assump:1}--\ref{assump:3}, it holds that 
\begin{align*}
    &\frac{G}{2} \mathbb{E}\left[ \lVert \boldsymbol{\Theta}^{t_s}-\boldsymbol{\Theta}^{t_s-1} \rVert^2 \right]\\
    &\leq\frac{\eta^2G}{2} R_{t_s}^2\lVert\boldsymbol{a}\rVert_1^2 \left(\sigma^2+ \beta^2 \mathbb{E}\left[ \left\lVert \nabla \mathcal{L}(\boldsymbol{\Theta}^{t_s-1}) \right\rVert^2 \right]+\kappa^2  \right)\\
    &+\frac{\eta^2G^3}{2} \frac{1}{K}\sum_{k=1}^K R_{t_s}\lVert\boldsymbol{a}\rVert_1 \sum_{i_s=1}^{R_{t_s}I}a_{i_s}\mathbb{E}\left[\left\lVert \boldsymbol{\Theta}_k^{t_s, i_s-1} 
    -\boldsymbol{\Theta}^{t_s-1} \right\rVert^2 \right].
    \numberthis
\end{align*}
\end{lemma}
\begin{proof}
    The proof of Lemma \ref{lemma: second term} is provided in Appendix \ref{appx: lemma: second term}.
\end{proof}
Substitute Lemma \ref{lemma: first term} and \ref{lemma: second term} into \eqref{eq:L-smooth}, then substitute Lemma \ref{lemma: local accum}, we obtain the immediate result for one round,
\begin{align*}
    &\mathbb{E}\left[\left\langle \nabla\mathcal{L}(\boldsymbol{\Theta}^{t_s-1}), \boldsymbol{\Theta}^{t_s} -\boldsymbol{\Theta}^{t_s-1}\right\rangle\right]\\
    &\leq \left( -\frac{1}{2}\eta R_{t_s} \lVert \boldsymbol{a} \rVert_1 
    +\eta^2 G R_{t_s}^2\lVert \boldsymbol{a} \rVert_1^2\beta^2\right) \mathbb{E}\left[ \lVert \nabla\mathcal{L}(\boldsymbol{\Theta}^{t_s-1}) \rVert^2\right]\\
    &\hspace{5mm} + \eta^2 G R_{t_s}^2 \lVert \boldsymbol{a} \rVert_1^2 \sigma^2
    + \eta^2 G R_{t_s}^2 \lVert \boldsymbol{a} \rVert_1^2 \kappa^2\\
    &\hspace{5mm}+ \left( \eta^2 G^3 R_{t_s} \lVert \boldsymbol{a} \rVert_1 
    +\frac{1}{2} \eta G^2 \right) 
     \frac{2\eta^2R_{t_s}^3 \lVert \boldsymbol{a} \rVert_1^3}{1-2\eta^2 G^2 R_{t_s}^2 \lVert \boldsymbol{a} \rVert_1^2}\cdot\\
    &\hspace{5mm}\hspace{5mm}\frac{1}{K}\sum_{k=1}^K\mathbb{E}\left[ \lVert \nabla\mathcal{L}_k(\boldsymbol{\Theta}^{t_s-1}) \rVert^2\right],
    \numberthis
\end{align*}
Using Assumption \ref{assump:3}, this further becomes 
\begin{align*}
    &\mathbb{E}\left[\left\langle \nabla\mathcal{L}(\boldsymbol{\Theta}^{t_s-1}), \boldsymbol{\Theta}^{t_s} -\boldsymbol{\Theta}^{t_s-1}\right\rangle\right]\\
    &\leq \left( -\frac{1}{2}\eta R_{t_s} \lVert \boldsymbol{a} \rVert_1 
    +\eta^2 G R_{t_s}^2\lVert \boldsymbol{a} \rVert_1^2\beta^2\right) \mathbb{E}\left[ \lVert \nabla\mathcal{L}(\boldsymbol{\Theta}^{t_s-1}) \rVert^2\right]\\
    &\hspace{5mm} + \eta^2 G R_{t_s}^2 \lVert \boldsymbol{a} \rVert_1^2 \sigma^2
    + \eta^2 G R_{t_s}^2 \lVert \boldsymbol{a} \rVert_1^2 \kappa^2\\
    &\hspace{5mm}+ \left( \eta^2 G^3 R_{t_s} \lVert \boldsymbol{a} \rVert_1 
    +\frac{1}{2} \eta G^2 \right) 
     \frac{2\eta^2R_{t_s}^3 \lVert \boldsymbol{a} \rVert_1^3}{1-2\eta^2 G^2 R_{t_s}^2 \lVert \boldsymbol{a} \rVert_1^2}\cdot\\
    &\hspace{5mm}\hspace{5mm}\left(\beta^2\mathbb{E}\left[ \lVert \nabla\mathcal{L}(\boldsymbol{\Theta}^{t_s-1}) \rVert^2\right] +\kappa^2 \right),
    \numberthis
\end{align*}
\subsection{Averaging over Successful Rounds}
After minor rearrangement, and averaging the immediate result over all $T_s$ successful rounds, we get 
\begin{align*}
    &\frac{1}{T_s}\sum_{t_s=1}^{T_s}\bigg(  \frac{1}{2} R_{t_s} \lVert \boldsymbol{a} \rVert_1- \eta G R_{t_s}^2 \lVert \boldsymbol{a} \rVert_1^2 \beta^2-\\
    &\hspace{1mm}\frac{ 2\eta R_{t_s}^3 \lVert \boldsymbol{a} \rVert_1^3(\eta^2 G^3 R_{t_s} \lVert \boldsymbol{a} \rVert_1+\frac{1}{2}\eta G^2)\beta^2 }{1-2\eta^2 G^2 R_{t_s}^2 \lVert \boldsymbol{a} \rVert_1^2}  \bigg) \mathbb{E}\left[ \lVert \nabla\mathcal{L}(\boldsymbol{\Theta}^{t_s-1}) \rVert^2\right]\\
    &\leq \frac{1}{\eta T_s}(\mathcal{L}(\boldsymbol{\Theta}^{0})-\mathcal{L}^\star)+\frac{1}{2}\eta G \lVert \boldsymbol{a} \rVert_1^2 \sigma^2 \cdot \frac{1}{T_s}\sum_{t_s=1}^{T_s} R_{t_s}^2+\\
    &\frac{1}{T_s}\sum_{t_s=1}^{T_s}\hspace{-1mm}\bigg(\hspace{-1mm}\eta G \lVert \boldsymbol{a} \rVert_1^2 R_{t_s}^2\hspace{-1mm}+\hspace{-1mm} 
     \frac{2\eta R_{t_s}^3 \lVert \boldsymbol{a} \rVert_1^3(\eta^2 G^3 R_{t_s} \lVert \boldsymbol{a} \rVert_1 
    \hspace{-1mm}+\hspace{-1mm}\frac{1}{2} \eta G^2)}{1-2\eta^2 G^2 R_{t_s}^2 \lVert \boldsymbol{a} \rVert_1^2}\hspace{-1mm}\bigg) \kappa^2.
    \numberthis
    \label{eq: average-1}
\end{align*}
If the learning rate is chosen as $\eta=\frac{1}{L}\sqrt{\frac{K}{T}}$, by neglecting the non-dominant terms in the coefficients, \eqref{eq: average-1} becomes
\begin{align*}
    &\bigg( \frac{1}{T_s}\sum_{t_s=1}^{T_s} R_{t_s} \bigg) \cdot \frac{1}{2} \lVert \boldsymbol{a} \rVert_1 \cdot   \mathbb{E}\left[ \lVert \nabla\mathcal{L}(\boldsymbol{\Theta}^{t_s-1}) \rVert^2\right]\\
    &\leq \frac{G}{\sqrt{\frac{K}{T}}\cdot T_s}(\mathcal{L}(\boldsymbol{\Theta}^{0})-\mathcal{L}^\star)+\frac{1}{2}\sqrt{\frac{K}{T}} \lVert \boldsymbol{a} \rVert_1^2 \sigma^2 \bigg(\frac{1}{T_s}\sum_{t_s=1}^{T_s} R_{t_s}^2\bigg)\\
    &\hspace{5mm}+\sqrt{\frac{K}{T}} \lVert \boldsymbol{a} \rVert_1^2  \kappa^2\bigg(\frac{1}{T_s}\sum_{t_s=1}^{T_s} R_{t_s}^2\bigg).
    \numberthis
    \label{eq: average-2}
\end{align*}
\subsection{Relaxation with High Probability Guarantee}
By \eqref{eq: bound_Ts}, \eqref{eq: average-2} can be further bounded as follows with probability $0.9974$,  
\begin{align*}
    &\bigg( \frac{1}{T_s}\sum_{t_s=1}^{T_s} R_{t_s} \bigg) \cdot \frac{1}{2} \lVert \boldsymbol{a} \rVert_1 \cdot   \mathbb{E}\left[ \lVert \nabla\mathcal{L}(\boldsymbol{\Theta}^{t_s-1}) \rVert^2\right]\\
    &\leq\frac{G(\mathcal{L}(\boldsymbol{\Theta}^{0})-\mathcal{L}^\star)}{\sqrt{\frac{K}{T}}\left( \frac{T}{e_R}-3\sqrt{\frac{\nu_R T}{e_R^3}}  \right)}\\
    &+\frac{ \lVert \boldsymbol{a} \rVert_1^2}{2}\sqrt{\frac{K}{T}} \left(\hspace{-1mm} e_{R^2}+3\sqrt{\frac{\nu_{R^2}}{\frac{T}{e_R}-3\sqrt{\frac{\nu_R T}{e_R^3}}}} \right)\hspace{-1mm}
    \left( \sigma^2+ \kappa^2 \right).
    \numberthis
\end{align*}
Substitute $e_R, \nu_R, e_{R^2}, \nu_{R^2}$, we get 
\begin{align*}
&\frac{1}{T_s}\hspace{-1mm}\sum_{t_s=1}^{T_s} \mathbb{E}\left[ \lVert \Delta \mathcal{L}(\boldsymbol{\Theta}^{t_s-1})\rVert^2\right]\hspace{-0.5mm}\leq\hspace{-0.5mm} \frac{1}{T_s}\hspace{-1mm}\sum_{t_s=1}^{T_s}\hspace{-1mm} R_{t_s} \mathbb{E}\left[ \lVert \Delta \mathcal{L}(\boldsymbol{\Theta}^{t_s-1})\rVert^2\right]\\
&\leq \frac{2G(\mathcal{L}(\boldsymbol{\Theta}^{0})-\mathcal{L}^\star)}{\lVert \boldsymbol{a}\rVert_1\left((1-P_O)\sqrt{KT}-3\sqrt{P_O(1-P_O)}\right)}\\
&\hspace{5mm}+2\lVert \boldsymbol{a}\rVert_1\sqrt{\frac{K}{T}} (\sigma^2+2\kappa^2) \cdot \Bigg(\frac{1-P_O}{P_O} \mathrm{Li}_{-2}(P_O)\\
&\hspace{5mm}+3\sqrt{\frac{\frac{1-P_O}{P_O}\mathrm{Li}_{-4}(P_O)-\left(\frac{1-P_O}{P_O}\mathrm{Li}_{-2}(P_O) \right)^2}{(1-P_O)T-3\sqrt{P_O(1P_O)T}}} \Bigg). 
\numberthis 
\end{align*}
By retaining only the dominant terms and substituting $\mathrm{Li}_{-2}(P_O)=\frac{P_O(1+P_O)}{(1-P_O)^3}$, the proof of Theorem~\ref{theo:Converge} is complete.

\section{Proofs of Auxiliary Lemmas}
\subsection{Proof of Lemma \ref{lemma: first term}}
\label{appx: lemma: first term}
For the first term, we have
\begin{subequations}
\begin{align*}
    &\mathbb{E}\left[\left\langle \nabla\mathcal{L}(\boldsymbol{\Theta}^{t_s-1}), \boldsymbol{\Theta}^{t_s} -\boldsymbol{\Theta}^{t_s-1}\right\rangle\right]\\
    &=\mathbb{E}\hspace{-1mm}\left[\hspace{-1mm}\left\langle\hspace{-1mm} \nabla\mathcal{L}(\boldsymbol{\Theta}^{t_s-1}),-\eta\frac{1}{K}\sum_{k=1}^K\sum_{i_s=1}^{R_{t_s}I}a_{i_s}\nabla \mathcal{L}_k(\boldsymbol{\Theta}_k^{t_s, i_s-1}, \boldsymbol{\xi}_k^{t_s, i_s})\hspace{-1mm}\right\rangle\hspace{-1mm}\right]\\
    &=-\eta\hspace{-1mm}\sum_{i_s=1}^{R_{t_s}I}\hspace{-1mm}a_{i_s}\mathbb{E}\hspace{-1mm}\left[\hspace{-1mm}\left\langle\hspace{-1mm} \nabla\mathcal{L}(\boldsymbol{\Theta}^{t_s-1}),\frac{1}{K}\hspace{-1mm}\sum_{k=1}^K\nabla \mathcal{L}_k(\boldsymbol{\Theta}_k^{t_s, i_s-1})\hspace{-1mm}\right\rangle\hspace{-1mm}\right]
    \numberthis\label{eq:term1-1}\\
    &\leq -\frac{1}{2}\eta\sum_{i_s=1}^{R_{t_s}I}a_{i_s}\mathbb{E}\left[\left\lVert \nabla\mathcal{L}(\boldsymbol{\Theta}^{t_s-1}) \right\lVert^2\right]\\
    &\hspace{5mm}+\frac{1}{2}\eta\hspace{-1mm}\sum_{i_s=1}^{R_{t_s}I}a_{i_s} \frac{1}{K}\sum_{k=1}^K\mathbb{E}\hspace{-1mm}\left[\left\lVert  
    \nabla\mathcal{L}(\boldsymbol{\Theta}_k^{t_s-1})\hspace{-1mm}-\hspace{-1mm}\nabla\mathcal{L}_k(\boldsymbol{\Theta}^{t_s,i_s-1})\right\lVert^2\right]\\
    &\hspace{5mm}-\frac{1}{2}\eta\sum_{i_s=1}^{R_{t_s}I}a_{i_s}\mathbb{E}\left[\left\lVert \frac{1}{K}\sum_{k=1}^K\nabla\mathcal{L}_k(\boldsymbol{\Theta}^{t_s,i_s-1}) \right\lVert^2\right]
    \numberthis\label{eq:term1-2}\\
    &\leq -\frac{1}{2}\eta\sum_{i_s=1}^{R_{t_s}I}a_{i_s}\mathbb{E}\left[\left\lVert \nabla\mathcal{L}(\boldsymbol{\Theta}^{t_s-1}) \right\lVert^2\right]\\
    &\hspace{5mm}+\frac{1}{2}\eta \frac{G^2}{K}\sum_{i_s=1}^{R_{t_s}I}a_{i_s} \sum_{k=1}^K\mathbb{E}\left[\left\lVert  
    \boldsymbol{\Theta}_k^{t_s-1}-\boldsymbol{\Theta}^{t_s,i_s-1}\right\lVert^2\right]
    \numberthis \label{eq:term1-3},
\end{align*}
\end{subequations}
where \eqref{eq:term1-1} follows Assumption \ref{assump:2}, \eqref{eq:term1-2} is due to the convexity of $\ell_2$-norm, and \eqref{eq:term1-3} is due to Assumption \ref{assump:1}.
\subsection{Proof of Lemma \ref{lemma: second term}} 
\label{appx: lemma: second term}
For the second term, it holds that
\begin{subequations}
\begin{align*}
&\frac{G}{2} \mathbb{E}\left[ \lVert \boldsymbol{\Theta}^{t_s}-\boldsymbol{\Theta}^{t_s-1} \rVert^2 \right]\\
&=\frac{G}{2} \mathbb{E}\left[ \left\lVert -\eta\frac{1}{K}\sum_{k=1}^K\sum_{i_s=1}^{R_{t_s}I}a_{i_s}\nabla \mathcal{L}_k(\boldsymbol{\Theta}_k^{t_s, i_s-1}, \boldsymbol{\xi}_k^{t_s, i_s}) \right\rVert^2 \right]\\
&\leq \frac{\eta^2G}{2} \frac{1}{K}\sum_{k=1}^K\mathbb{E}\left[ \left\lVert \sum_{i_s=1}^{R_{t_s}I}a_{i_s}\nabla \mathcal{L}_k(\boldsymbol{\Theta}_k^{t_s, i_s-1}, \boldsymbol{\xi}_k^{t_s, i_s}) \right\rVert^2 \right] \numberthis\label{eq:term2-1}\\
&\leq \frac{\eta^2G}{2} \frac{1}{K}\sum_{k=1}^K R_{t_s}\lVert\boldsymbol{a}\rVert_1 \sum_{i_s=1}^{R_{t_s}I}a_{i_s} \mathbb{E}\hspace{-0.5mm}\left[ \left\lVert \nabla \mathcal{L}_k(\boldsymbol{\Theta}_k^{t_s, i_s-1}, \boldsymbol{\xi}_k^{t_s, i_s}) \right\rVert^2 \right]\numberthis\label{eq:term2-2}\\
&= \frac{\eta^2G}{2} \frac{1}{K}\sum_{k=1}^K R_{t_s}\lVert\boldsymbol{a}\rVert_1 \sum_{i_s=1}^{R_{t_s}I}a_{i_s} \Bigg(\mathbb{E}\bigg[ \bigg\lVert \nabla \mathcal{L}_k(\boldsymbol{\Theta}_k^{t_s, i_s-1}, \boldsymbol{\xi}_k^{t_s, i_s})\\
&\hspace{5mm}-\nabla \mathcal{L}_k(\boldsymbol{\Theta}_k^{t_s, i_s-1}) \bigg\rVert^2 \bigg]
+\mathbb{E}\left[ \left\lVert \nabla \mathcal{L}_k(\boldsymbol{\Theta}_k^{t_s, i_s-1}) \right\rVert^2 \right]\Bigg)\numberthis\label{eq:term2-3}\\
&\leq \frac{\eta^2G}{2} \frac{1}{K}\sum_{k=1}^K R_{t_s}\lVert\boldsymbol{a}\rVert_1 \sum_{i_s=1}^{R_{t_s}I}a_{i_s} \Bigg(\sigma^2+ \mathbb{E}\bigg[\bigg\lVert \nabla \mathcal{L}_k(\boldsymbol{\Theta}_k^{t_s, i_s-1})\\
&\hspace{5mm}-\nabla \mathcal{L}_k(\boldsymbol{\Theta}^{t_s-1}) \bigg\rVert^2 \bigg]
+\mathbb{E}\left[ \left\lVert \nabla \mathcal{L}_k(\boldsymbol{\Theta}^{t_s-1}) \right\rVert^2 \right]\Bigg)\numberthis\label{eq:term2-4}\\
&\leq \frac{\eta^2G}{2} \frac{1}{K}\sum_{k=1}^K R_{t_s}^2\lVert\boldsymbol{a}\rVert_1^2 \sigma^2
+ \frac{\eta^2G}{2} \frac{1}{K}\sum_{k=1}^K R_{t_s}\lVert\boldsymbol{a}\rVert_1 \sum_{i_s=1}^{R_{t_s}I}a_{i_s}\cdot\\
&\bigg( G^2\mathbb{E}\bigg[\bigg\lVert \boldsymbol{\Theta}_k^{t_s, i_s-1} 
-\boldsymbol{\Theta}^{t_s-1} \bigg\rVert^2 \bigg]
+\mathbb{E}\left[ \left\lVert \nabla \mathcal{L}_k(\boldsymbol{\Theta}^{t_s-1}) \right\rVert^2 \right]\Bigg)\numberthis\label{eq:term2-5}\\
&\leq\frac{\eta^2G}{2} R_{t_s}^2\lVert\boldsymbol{a}\rVert_1^2 \left(\sigma^2+ \beta^2 \mathbb{E}\left[ \left\lVert \nabla \mathcal{L}(\boldsymbol{\Theta}^{t_s-1}) \right\rVert^2 \right]+\kappa^2  \right)\\
&+\frac{\eta^2G^3}{2} \frac{1}{K}\sum_{k=1}^K R_{t_s}\lVert\boldsymbol{a}\rVert_1 \sum_{i_s=1}^{R_{t_s}I}a_{i_s}\mathbb{E}\left[\left\lVert \boldsymbol{\Theta}_k^{t_s, i_s-1} 
    -\boldsymbol{\Theta}^{t_s-1} \right\rVert^2 \right],
\numberthis\label{eq:term2-6}
\end{align*}
\end{subequations}
where \eqref{eq:term2-1} and \eqref{eq:term2-2} use Jensen's Inequality,  \eqref{eq:term2-3} is due to unbiased estimation, \eqref{eq:term2-4} follows Assumption \ref{assump:2}, \eqref{eq:term2-5} follows Assumption \ref{assump:1}, and \eqref{eq:term2-6} follows Assumption \ref{assump:3}.



\section{Proof of Theorem \ref{theo: p2p LDP}} \label{Appx: theo: p2p LDP}
\begin{definition}[$(\epsilon, \delta)$-LDP] 
The randomized mechanism $\mathcal{M}: \mathcal{O}\rightarrow \mathbb{R}^D$ is $(\epsilon, \delta)$ locally differentially private if for all dataset instances $\forall \boldsymbol{\Theta}, \boldsymbol{\Theta}'\in \mathcal{O}$, and any measurable set $\mathcal{S}\subseteq\mathrm{Range}(\mathcal{M})$, it holds that 
\begin{align}
    \mathrm{P_r}(\mathcal{M}(\boldsymbol{\Theta})\in \mathcal{S})\leq e^{\epsilon} \mathrm{P_r}(\mathcal{M}(\boldsymbol{\Theta}')\in \mathcal{S})+\delta.
\end{align}
\end{definition}
\begin{definition}[Guassian Mechanism\cite{dwork2014algorithmic}] 
If a node wishes to release a function $f(\boldsymbol{\Theta})$ of an input $\boldsymbol{\Theta}$ subject to $(\epsilon, \delta)$-LDP. The Gaussian release mechanism is defined as
\begin{align}
    \mathcal{M}(\boldsymbol{\Theta})\triangleq f(\boldsymbol{\Theta})+\mathcal{N}(0,\varepsilon^2\boldsymbol{I}).    
\end{align}
If the $\ell_2$-sensitivity of the function is bounded by $\Delta_{f}$, i.e., $\lVert f(\boldsymbol{\Theta})-f(\boldsymbol{\Theta}')\rVert\leq \Delta_{f}$, then $\forall \boldsymbol{\Theta},\boldsymbol{\Theta}'$, $\forall\delta\in(0,1]$, the Guassian mechanism satisfies $(\epsilon, \delta)$-LDP, where 
\begin{align}
    \epsilon=\frac{\Delta_{f}}{\varepsilon}\left[2\log\left( \frac{1.25}{\delta} \right)\right]^{\frac{1}{2}}. 
\end{align}
\end{definition}

Despite the standard Gaussian mechanism $\boldsymbol{N}_k^t\sim\mathcal{N}(\mathbf{0}, \lambda^2\boldsymbol{I}_D)$ is applied to the peer-to-peer transmission from client $k$ to client $m$, the correlation between the secret key $\boldsymbol{N}_m^t$ at client $m$ and $\boldsymbol{N}_k^t$ can reveal some information about the unknown secret key. This correlation will weaken privacy to some extent. The variance reflecting the uncertainty reduction decreases from $\lambda^2$ to $\lambda^2 - \frac{1}{ \lambda^2}\boldsymbol{R}_{m,k}\boldsymbol{R}_{k,m}$. By utilizing the correlation between $\boldsymbol{N}_m^t$ and $\boldsymbol{N}_k^t$, the transmission is equivalent to applying the Gaussian mechanism $
\mathcal{N}\!\left(
\frac{\boldsymbol{n}_m}{ \lambda^2}\boldsymbol{R}_{m,k} ,\;
 (\lambda^2 - \frac{1}{ \lambda^2}\boldsymbol{R}_{m,k}\boldsymbol{R}_{k,m})\cdot\boldsymbol{I}_D
\right)$. The expected value $\frac{\boldsymbol{n}_m}{ \lambda^2}\boldsymbol{R}_{m,k}$ does not affect the privacy loss, hence it is further equivalent to applying the Gaussian mechanism $
\mathcal{N}\!\left(\mathbf{0} ,\;
 (\lambda^2 - \frac{1}{ \lambda^2}\boldsymbol{R}_{m,k}\boldsymbol{R}_{k,m})\cdot\boldsymbol{I}_D
\right)$.


When client $m$ fails to decode message from client $k$, i.e., $\tau_{m,k}=0$, we have $\boldsymbol{Y}_k^{t}=0$ regardless of the values of $\Delta\boldsymbol{\Theta}_k^t, \Delta\boldsymbol{\Theta}_k^{t'}\in\mathbb{B}_D(R)$, it yields perfect privacy, i.e., 
\begin{align*}
     &\mathrm{P_r}(\boldsymbol{Y}_k^{t}\in \mathcal{S} \vert \Delta\boldsymbol{\Theta}_k^t\in\mathcal{O}_k, \tau_{m,k}=0)\\
    &\hspace{2cm}=\mathrm{P_r}(\boldsymbol{Y}_k^{t}\in \mathcal{S} \vert \Delta\boldsymbol{\Theta}_k^{t'}\in\mathcal{O}_k, \tau_{m,k}=0),
    \numberthis
    \label{eq: pr_broken_link}
\end{align*}
If $\tau_{m,k}=1$, client $m$ successfully decode $\boldsymbol{Y}_k^{t}$, the standard Gaussian mechanism applies, with the $\ell_2$-sensitivity given by 
\begin{align*}
    \sup_{\Delta\boldsymbol{\Theta}_k^t, \Delta\boldsymbol{\Theta}_k^{t'}\in\mathbb{B}_D(R)} \lVert \Delta\boldsymbol{\Theta}_k^t- \Delta\boldsymbol{\Theta}_k^{t'}\rVert\leq 2R.
\end{align*}
Thus, it holds that
\begin{align*}
     &\mathrm{P_r}(\boldsymbol{Y}_k^{t}\in \mathcal{S} \vert \Delta\boldsymbol{\Theta}_k^t\in\mathcal{O}_k, \tau_{m,k}=1)\\
    &\leq e^{\epsilon_{m,k}^{(1)}}\mathrm{P_r}(\boldsymbol{Y}_k^{t}\in \mathcal{S} \vert \Delta\boldsymbol{\Theta}_k^{t'}\in\mathcal{O}_k, \tau_{m,k}=1)+\delta_k^{(1)},
    \numberthis
\end{align*}
where 
\begin{align}
    \epsilon_{m, k}^{(1)}= \frac{2R}{\sqrt{\lambda^2 - \frac{1}{ \lambda^2}\boldsymbol{R}_{m,k}\boldsymbol{R}_{k,m}}}\left[2\log\left( \frac{1.25}{\delta_{m,k}^{(1)}} \right)\right]^{\frac{1}{2}}. 
\end{align}
Following similar steps in \cite[(73)]{saha2024privacy}, we have 
\begin{subequations}
\begin{align}
     &\mathrm{P_r}(\boldsymbol{Y}_k^{t}\in \mathcal{S} \vert \Delta\boldsymbol{\Theta}_k^t\in\mathcal{O}_k)\notag\\
    &=p_{m,k}\mathrm{P_r}(\boldsymbol{Y}_k^{t}\in \mathcal{S} \vert \Delta\boldsymbol{\Theta}_k^t\in\mathcal{O}_k, \tau_{m,k}=0)\notag\\     
    &\hspace{5mm}+(1-p_{m,k})\mathrm{P_r}(\boldsymbol{Y}_k^{t}\in \mathcal{S} \vert \Delta\boldsymbol{\Theta}_k^t\in\mathcal{O}_k, \tau_{m,k}=1)\\
    &\leq p_{m,k} e^{\epsilon_{m,k}^{(1)}} \mathrm{P_r}(\boldsymbol{Y}_k^{t}\in \mathcal{S} \vert \Delta\boldsymbol{\Theta}_k^t\in\mathcal{O}_k, \tau_{m,k}=0) \notag\\
    &\hspace{5mm}+(1-p_{m,k}) e^{\epsilon_{m,k}^{(1)}} \mathrm{P_r}(\boldsymbol{Y}_k^{t}\in \mathcal{S} \vert \Delta\boldsymbol{\Theta}_k^{t'}\in\mathcal{O}_k, \tau_{m,k}=1)  \notag\\
    &\hspace{5mm}+(1-p_{m,k})\delta_k^{(1)} \label{eq: pr_all_connect}\\
    &=e^{\epsilon_{m,k}^{(1)}}\mathrm{P_r}(\boldsymbol{Y}_k^{t}\in \mathcal{S} \vert \Delta\boldsymbol{\Theta}_k^{t'}\in\mathcal{O}_k)+(1-p_{m,k})\delta_k^{(1)},
    \numberthis
\end{align}
\end{subequations}
where \eqref{eq: pr_all_connect} is due to \eqref{eq: pr_broken_link} and that $e^{\epsilon_{m,k}^{(1)}}\geq 1$. 
\section{Proof of Theorem \ref{theo: relay LDP 1}} \label{Appx: theo: relay LDP 1}
Now let us proceed to LDP at the relaying client $k$. The relaying client $k$ successfully collects the masked local model updates from clients in $\mathcal{U}_k^t$ (including itself), computes their partial sum $\boldsymbol{S}_k^{t}$, and transmits $\boldsymbol{S}_k^{t}$ to the server. During the transmission, an eavesdropper should not be able to distinguish the participation of client $j$ by observing $\boldsymbol{S}_k^{t}$. 

If client $j\notin \mathcal{U}_k^t$, then $\boldsymbol{S}_k^{t,-j}=\boldsymbol{S}_k^t$. So for any measurable set $\mathcal{S}$, it holds that 
\begin{align*}
    \mathrm{P_r}(\boldsymbol{S}_k^{t,-j}\hspace{-1mm}\in\hspace{-0.5mm} \mathcal{S})
    \hspace{-0.5mm}
    =&\mathrm{P_r}(j\in\mathcal{U}_k^t)\mathrm{P_r}(\boldsymbol{S}_k^{t,-j}\hspace{-0.5mm}\in \hspace{-1mm}\mathcal{S}\vert j\in\mathcal{U}_k^t)\\
    &+\mathrm{P_r}(j\notin\mathcal{U}_k^t)\mathrm{P_r}(\boldsymbol{S}_k^{t}\in \hspace{-1mm}\mathcal{S}\vert j\notin\mathcal{U}_k^t). 
    \numberthis
    \label{eq: 67}
\end{align*}
Consider the network realizations reflected by $\mathcal{U}_k^t$, such that corresponding variance $\nu_k^{t,-j}$ lies within the event $\mathcal{E}\triangleq\{ \lvert \nu_k^{t,-j}-\Bar{\nu}_k^{-j}\rvert\leq r_1 \}$, where $r_1$ is chosen such that $\mathrm{P_r}(\mathcal{E}^\mathrm{c})<\delta'$, $\delta'\in (0,1)$, then 
\begin{align*}
    \hspace{-0.5mm}\mathrm{P_r}(\boldsymbol{S}_k^{t,-j}\hspace{-0.5mm}\in \hspace{-1mm}\mathcal{S}\vert j\in\mathcal{U}_k^t)
    =&\mathrm{P_r}(\mathcal{E})\mathrm{P_r}(\boldsymbol{S}_k^{t,-j}\hspace{-0.5mm}\in \hspace{-1mm}\mathcal{S}\vert j\in\mathcal{U}_k^t,\mathcal{E})\\
    &\hspace{-12mm}+\mathrm{P_r}(\mathcal{E}^\mathrm{c})\mathrm{P_r}(\boldsymbol{S}_k^{t,-j}\hspace{-0.5mm}\in \hspace{-1mm}\mathcal{S}\vert j\in\mathcal{U}_k^t,\mathcal{E}^\mathrm{c})\\
    &\hspace{-15mm}\leq\delta'+\mathrm{P_r}(\mathcal{E})\mathrm{P_r}(\boldsymbol{S}_k^{t,-j}\hspace{-0.5mm}\in \hspace{-1mm}\mathcal{S}\vert j\in\mathcal{U}_k^t,\mathcal{E}). 
    \numberthis
    \label{eq: 68}
\end{align*} 
Moreover, by counting all network realizations $\mathcal{A}$ lying within the event $\mathcal{E}$, we have 
\begin{align}
    \mathrm{P_r}(\boldsymbol{S}_k^{t,-j}\hspace{-0.5mm}\in \hspace{-1mm}\mathcal{S}\vert j\in\mathcal{U}_k^t,\mathcal{E})
    =\hspace{-7mm}\sum_{\substack{\mathcal{A}:\mathcal{A}\in \mathcal{U}_k, j,k\in\mathcal{A}\\ \lvert \nu_k^{t,-j}-\Bar{\nu}_k^{-j}\rvert\leq r_1}} \hspace{-7mm}\mathrm{P_r}(\boldsymbol{S}_k^{t,-j}\hspace{-0.5mm}\in \hspace{-1mm}\mathcal{S}\vert i\in\mathcal{A})  \mathrm{P_r}(\mathcal{A}). 
    \label{eq: relay_ldp_Ut}
\end{align}
For each specific realization $\mathcal{A}$, the form of its privacy loss is given by 
\begin{align*}
&\log\left( \frac{\mathrm{P_r}(\boldsymbol{S}_k^{t,-j}=\boldsymbol{z}\vert i\in\mathcal{A})}{\mathrm{P_r}(\boldsymbol{S}_k^{t}=\boldsymbol{z}\vert i\in\mathcal{A})} \right)\\
=&\log\left( \frac{\exp{\left(-\frac{\lVert\boldsymbol{z}-\sum_{m\in\mathcal{A}\setminus\{j\}}(\Delta\boldsymbol{\Theta}_m^t+\boldsymbol{N}_m^t)\rVert^2}{2\nu_k^{t,-j}}\right)}}{\exp{\left(-\frac{\lVert\boldsymbol{z}-\sum_{m\in\mathcal{A}}(\Delta\boldsymbol{\Theta}_m^t+\boldsymbol{N}_m^t)\rVert^2}{2\nu_k^{t,-j}}\right)}} \right)\\
=&-\frac{1}{2\nu_k^{t,-j}} \Bigg(\left\lVert\boldsymbol{z}-\sum_{m\in\mathcal{A}\setminus\{j\}}(\Delta\boldsymbol{\Theta}_m^t+\boldsymbol{N}_m^t)\right\rVert^2\\
&\hspace{30mm}-\left\lVert\boldsymbol{z}-\sum_{m\in\mathcal{A}}(\Delta\boldsymbol{\Theta}_m^t+\boldsymbol{N}_m^t)\right\rVert^2 \Bigg).
\numberthis
\label{eq: privacy_loss_relay}
\end{align*}
The form of \eqref{eq: privacy_loss_relay} is identical to the privacy loss of the Gaussian mechanism with Gaussian noise $\nu_k^{t,-j}$. The $\ell_2$-sensitivity is bounded by 
\begin{align*}
    &\sup_{\hspace{-3mm}\substack{\mathcal{A}:\mathcal{A}\in \mathcal{U}_k, j,k\in\mathcal{A}\\ \lvert \nu_k^{t,-j}-\Bar{\nu}_k^{-j}\rvert\leq r_1}} \hspace{-1mm}\left\lVert \sum_{m\in\mathcal{A}} \hspace{-1mm} g_{k,m} (\Delta\boldsymbol{\Theta}_m^t\hspace{-1mm}+\hspace{-1mm}\boldsymbol{N}_m^t)\hspace{-0.5mm}-
    \hspace{-5mm}\sum_{m\in\mathcal{A}\setminus\{j\}} \hspace{-4mm}g_{k,m}(\Delta\boldsymbol{\Theta}_m^t\hspace{-1mm}+\hspace{-1mm}\boldsymbol{N}_m^t)\right\rVert\\
    &= \left\lVert  g_{k,j} (\Delta\boldsymbol{\Theta}_j^t+\boldsymbol{N}_j^t)\right\rVert
    \leq \left\lVert  g_{k,j} \Delta\boldsymbol{\Theta}_j^t\right\rVert
    +\left\lVert  g_{k,j}\boldsymbol{N}_j^t\right\rVert\\
    &\leq \lvert g_{k,j} \rvert R+\left\lVert  g_{k,j}\boldsymbol{N}_j^t\right\rVert. 
    \label{eq: l2_sensitivity_relay1}
    \numberthis
\end{align*}
Since that $g_{k,j}\boldsymbol{N}_j^t\sim\mathcal{N}(\boldsymbol{0}, \lambda^2\boldsymbol{I}_D)$, the second term in \eqref{eq: l2_sensitivity_relay1} can be bounded by Chi-squared tail bound, i.e., $\forall \delta_0$,
\begin{align}
    \mathrm{P_r}\left(\left\lVert  g_{k,j}\boldsymbol{N}_j^t\right\rVert\hspace{-1mm}\leq\hspace{-1mm} \lvert  g_{k,j}\rvert\lambda\sqrt{D(1+\delta_0)}\right)\hspace{-0.5mm}\geq\hspace{-0.5mm} 1\hspace{-0.5mm}-\hspace{-0.5mm} e^{-\frac{D}{2}(\delta_0-\ln{(1+\delta_0)})}.
    \label{eq: chi_tail}
\end{align}
Substituting \eqref{eq: chi_tail} into \eqref{eq: l2_sensitivity_relay1}, we can bound the $\ell_2$-sensitivity by
\begin{align*}
    & \lvert g_{k,j} \rvert R+\left\lVert  g_{k,j}\boldsymbol{N}_j^t\right\rVert
    \leq \lvert g_{k,j} \rvert\left( R+\lambda\sqrt{D(1+\delta_0)}\right), 
    \label{eq: l2_sensitivity_relay2}
    \numberthis
\end{align*}
with probability (w.p.) at least $1- e^{-\frac{D}{2}(\delta_0-\ln{(1+\delta_0)})}$. 

Consequently, each specific realization $\mathcal{A}$ is $(\epsilon_{k,j}^{\mathcal{A},2},\delta_j^{(2)})$-differentially private w.p. at least $1- e^{-\frac{D}{2}(\delta_0-\ln{(1+\delta_0)})}$, i.e., 
\begin{align}
\mathrm{P_r}(\boldsymbol{S}_k^{t,-j}\hspace{-0.5mm}\in \hspace{-1mm}\mathcal{S}\vert i\in\mathcal{A})
\leq e^{\epsilon_{k,j}^{\mathcal{A},2}} \mathrm{P_r}(\boldsymbol{S}_k^{t}\hspace{-0.5mm}\in \hspace{-1mm}\mathcal{S}\vert i\in\mathcal{A})\hspace{-0.5mm}+\hspace{-0.5mm}\delta_j^{(2)},
\label{eq: relay_ldp_A}
\end{align} 
where $\epsilon_{k,j}^{\mathcal{A},2}=\hspace{-1mm}\left[2\log\left( \frac{1.25}{\delta_j^{(2)}} \right)\right]^{\frac{1}{2}}\hspace{-1mm}\frac{\lvert g_{k,j} \rvert\left( R+ \lambda\sqrt{D(1+\delta_0)}\right)}{\sqrt{\nu_k^{t,-j}}}\hspace{-1mm}\triangleq\frac{c_1}{\sqrt{\nu_k^{t,-j}}}$. 

Substitute \eqref{eq: relay_ldp_A} into \eqref{eq: relay_ldp_Ut}, we get 
\begin{subequations}
\begin{align}
     &\mathrm{P_r}(\boldsymbol{S}_k^{t,-j}\hspace{-0.5mm}\in \hspace{-1mm}\mathcal{S}\vert j\in\mathcal{U}_k^t,\mathcal{E})\\
     &\leq \hspace{-5mm}\sum_{\substack{\mathcal{A}:\mathcal{A}\in \mathcal{U}_k, j,k\in\mathcal{A}\\ \lvert \nu_k^{t,-j}-\Bar{\nu}_k^{-j}\rvert\leq r}} \hspace{-5mm} \mathrm{P_r}(\mathcal{A})\left( e^{\frac{c_1}{\sqrt{\nu_k^{t,-j}}}} \mathrm{P_r}(\boldsymbol{S}_k^{t}\hspace{-0.5mm}\in \hspace{-1mm}\mathcal{S}\vert i\in\mathcal{A})\hspace{-0.5mm}+\hspace{-0.5mm}\delta_j^{(2)}\right) \label{eq: 71b}\\
     &\leq e^{\frac{c_1}{\sqrt{\Bar{\nu}_k^{-j}-r_1}}} \cdot \hspace{-5mm} \sum_{\substack{\mathcal{A}:\mathcal{A}\in \mathcal{U}_k, j,k\in\mathcal{A}\\ \lvert \nu_k^{t,-j}-\Bar{\nu}_k^{-j}\rvert\leq r}} \hspace{-5mm} \mathrm{P_r}(\mathcal{A})\mathrm{P_r}(\boldsymbol{S}_k^{t}\hspace{-0.5mm}\in \hspace{-1mm}\mathcal{S}\vert i\in\mathcal{A})\hspace{-0.5mm}+\hspace{-0.5mm}\delta_j^{(2)}\label{eq: 71c}\\
    &\leq e^{\frac{c_1}{\sqrt{\Bar{\nu}_k^{-j}-r_1}}} 
    \cdot\mathrm{P_r}(\boldsymbol{S}_k^{t}\hspace{-0.5mm}\in \hspace{-1mm}\mathcal{S}\vert j\in\mathcal{U}_k^t,\mathcal{E})\hspace{-0.5mm}+\hspace{-0.5mm}\delta_j^{(2)}\label{eq: 71d},  
\end{align}
\end{subequations}
where \eqref{eq: 71c} holds since $\nu_k^{t,-j}>\Bar{\nu}_k^{-j}-r_1$, \eqref{eq: 71d} is similar to \eqref{eq: relay_ldp_Ut}. 

Substitute \eqref{eq: 71d} into \eqref{eq: 68}, we obtain
\begin{align*}
    &\hspace{-0.5mm}\mathrm{P_r}(\boldsymbol{S}_k^{t,-j}\hspace{-0.5mm}\in \hspace{-1mm}\mathcal{S}\vert j\in\mathcal{U}_k^t)\\
    &\leq\delta'+\mathrm{P_r}(\mathcal{E})\left( 
     e^{\frac{c_1}{\sqrt{\Bar{\nu}_k^{-j}-r_1}}} 
    \cdot\mathrm{P_r}(\boldsymbol{S}_k^{t}\hspace{-0.5mm}\in \hspace{-1mm}\mathcal{S}\vert j\in\mathcal{U}_k^t,\mathcal{E})\hspace{-0.5mm}+\hspace{-0.5mm}\delta_j^{(2)} \right)\\
    &\leq\delta'+ 
     e^{\frac{c_1}{\sqrt{\Bar{\nu}_k^{-j}-r_1}}} 
    \cdot\mathrm{P_r}(\boldsymbol{S}_k^{t}\hspace{-0.5mm}\in \hspace{-1mm}\mathcal{S}\vert j\in\mathcal{U}_k^t)\hspace{-0.5mm}+\hspace{-0.5mm}\delta_j^{(2)} . 
    \numberthis
    \label{eq:72}
\end{align*}
Furthermore, it holds that 
\begin{align*}
    &\hspace{-0.5mm}\mathrm{P_r}(\boldsymbol{S}_k^{t,-j}\hspace{-0.5mm}\in \hspace{-1mm}\mathcal{S}\vert j\notin\mathcal{U}_k^t)
    \leq
     e^{\frac{c_1}{\sqrt{\Bar{\nu}_k^{-j}-r_1}}} \mathrm{P_r}(\boldsymbol{S}_k^{t}\hspace{-0.5mm}\in \hspace{-1mm}\mathcal{S}\vert j\notin\mathcal{U}_k^t),
    \numberthis
    \label{eq:73}
\end{align*}
since that $e^{\frac{c_1}{\sqrt{\Bar{\nu}_k^{-j}-r_1}}}\geq 1$. 

Substitute both \eqref{eq:72} and \eqref{eq:73} into \eqref{eq: 67}, we get
\begin{align}
    \mathrm{P_r}(\boldsymbol{S}_k^{t,-j}\hspace{-1mm}\in\hspace{-0.5mm} \mathcal{S})
    \hspace{-0.5mm}\leq \hspace{-0.5mm}e^{\frac{c_1}{\sqrt{\Bar{\nu}_k^{-j}-r_1}}}  \mathrm{P_r}(\boldsymbol{S}_k^{t}\hspace{-1mm}\in\hspace{-0.5mm} \mathcal{S})\hspace{-0.5mm}+\hspace{-0.5mm}(1\hspace{-0.5mm}-\hspace{-0.5mm}p_{k,j})(\delta'\hspace{-1mm}+\hspace{-0.5mm}\delta_j^{(2)}). 
\end{align}
The proof is completed. 

\section{Proof of Theorem \ref{theo: relay LDP 2}}\label{Appx: theo: relay LDP 2}
For $\forall \boldsymbol{u},\boldsymbol{u}'\in \mathbb{B}_D(R)$, similar to (\ref{eq: 67}), it holds that 
\begin{align*}
    \mathrm{P_r}(\boldsymbol{S}_k^{t}&\in \mathcal{S} \vert \Delta\boldsymbol{\Theta}_j^{t}=\boldsymbol{u})\\
    =&\mathrm{P_r}(j\in\mathcal{U}_k^t)\mathrm{P_r}(\boldsymbol{S}_k^{t}\in \mathcal{S}\vert \Delta\boldsymbol{\Theta}_j^{t}=\boldsymbol{u},  j\in\mathcal{U}_k^t)\\
    &+\mathrm{P_r}(j\notin\mathcal{U}_k^t)\mathrm{P_r}(\boldsymbol{S}_k^{t}\in \mathcal{S}\vert \Delta\boldsymbol{\Theta}_j^{t}=\boldsymbol{u}. j\notin\mathcal{U}_k^t).
    \numberthis
    \label{eq: 74}
\end{align*}
Moreover, similar to (\ref{eq: 68}), it holds that
\begin{align*}
   \mathrm{P_r}(&\boldsymbol{S}_k^{t}\in\mathcal{S}\vert \Delta\boldsymbol{\Theta}_j^{t}=\boldsymbol{u},  j\in\mathcal{U}_k^t)\\
    &\leq\delta'+\mathrm{P_r}(\mathcal{E})\mathrm{P_r}(\boldsymbol{S}_k^{t}\hspace{-0.5mm}\in \hspace{-1mm}\mathcal{S}\vert \Delta\boldsymbol{\Theta}_j^{t}=\boldsymbol{u},  j\in\mathcal{U}_k^t,\mathcal{E}),  
    \numberthis
\end{align*}
and similar to \eqref{eq: relay_ldp_A}, it holds that 
\begin{align*}
&\mathrm{P_r}(\boldsymbol{S}_k^{t}\hspace{-0.5mm}\in \hspace{-1mm}\mathcal{S}\vert \Delta\boldsymbol{\Theta}_j^{t}=\boldsymbol{u},  i\in\mathcal{A})\\
&\leq e^{\epsilon_{k,j}^{\mathcal{A},3}} \mathrm{P_r}(\boldsymbol{S}_k^{t}\hspace{-0.5mm}\in \hspace{-1mm}\mathcal{S}\vert \Delta\boldsymbol{\Theta}_j^{t}=\boldsymbol{u}',  i\in\mathcal{A})+\hspace{-0.5mm}\delta_j^{(3)}.
\numberthis
\label{eq: relay_ldp_perturb_A}
\end{align*} 
The $\ell_2$-sensitivity is given by 
\begin{align*}
    &\hspace{5mm}\sup_{\hspace{-5mm}\substack{\mathcal{A}:\mathcal{A}\in \mathcal{U}_k, j,k\in\mathcal{A}\\ \lvert \nu_k^{t,-j}-\Bar{\nu}_k^{-j}\rvert\leq r_2}} \hspace{0mm}\left\lVert \boldsymbol{S}_k^{t,-j}+\boldsymbol{u}+\boldsymbol{N}_j^t - \boldsymbol{S}_k^{t,-j}-\boldsymbol{u}'-\boldsymbol{N}_j^t \right\rVert\\
    &\hspace{10mm}\leq 2 \lvert g_{k,j} \rvert R,
    \numberthis
\end{align*}
which gives $\epsilon_{k,j}^{\mathcal{A},3}= \left[2\log\left( \frac{1.25}{\delta_j^{(3)}} \right)\right]^{\frac{1}{2}}\frac{2 \lvert g_{k,j} \rvert R}{\sqrt{\nu_k^t}}\triangleq\frac{c_2}{\sqrt{\nu_k^t}}$. 

Similar to \eqref{eq:73}, we have
\begin{align*}
    &\hspace{-0.5mm}\mathrm{P_r}(\boldsymbol{S}_k^{t}\in \mathcal{S}\vert \Delta\boldsymbol{\Theta}_j^{t}=\boldsymbol{u},  j\in\mathcal{U}_k^t)\\
    &\leq\delta'+ 
     e^{\frac{c_2}{\sqrt{\Bar{\nu}_k-r_2}}} 
    \mathrm{P_r}(\boldsymbol{S}_k^{t}\in \mathcal{S}\vert \Delta\boldsymbol{\Theta}_j^{t}=\boldsymbol{u}',  j\in\mathcal{U}_k^t)\hspace{-0.5mm}+\hspace{-0.5mm}\delta_j^{(3)} . 
    \numberthis
\end{align*}
By substituting \eqref{eq:73} into \eqref{eq: 74}, we obtain the final expression,
\begin{align*}
    \mathrm{P_r}(\boldsymbol{S}_k^{t}\in \mathcal{S}\vert \Delta\boldsymbol{\Theta}_j^{t}
    =\boldsymbol{u} )
    \leq& e^{\frac{c_2}{\sqrt{\Bar{\nu}_k-r_2}}}  \mathrm{P_r}(\boldsymbol{S}_k^{t}\in \mathcal{S}\vert \Delta\boldsymbol{\Theta}_j^{t}=\boldsymbol{u}' )\\
    &+(1-p_{k,j})(\delta'+\delta_j^{(3)}), 
    \numberthis
\end{align*}
thereby completing the proof.
\section{Proof of Theorem \ref{theo: server LDP-failure}}\label{Appx: theo: server LDP-failure}
When the combinator $\boldsymbol{c}_{f'}$ fails to reconstruct the global model, the combination result is noisy, and the server should not learn the participation of client $j$ at the relaying client $k\in\mathcal{V}_j$. That is, from the aggregated result according to $f'$-th combinator, $\boldsymbol{F}_{f'}=\sum_{k\in\mathcal{V}^t} c_{f',k}\sum_{m\in\mathcal{U}_k^t} g_{k,m}(\Delta\boldsymbol{\Theta}_m^{t,I}+\boldsymbol{N}_m^t)$ should not precisely reveal the identity of the $j$-th participating client at relay $k$. Let $\boldsymbol{F}_{f'}^{t,\{k,-j\}}$ denote the combined results excluding the $j$-th contribution via relay $k$. 
It holds that
\begin{align*}
&\mathrm{P_r}(\boldsymbol{F}_{f'}^{t,\{k,-j\}}\hspace{-1mm}\in\mathcal{S})=\\
&\mathrm{P_r}(\hspace{-0.5mm}\{j\in \hspace{-0.5mm}\mathcal{U}_k^t\}\hspace{-0.5mm}\cap\hspace{-0.5mm} \{k\in \hspace{-0.5mm}\mathcal{V}^t\}\hspace{-0.5mm})\mathrm{P_r}(\boldsymbol{F}_{f'}^{t,\{k,-j\}}\hspace{-1.5mm}\in\hspace{-0.5mm}\mathcal{S}\vert \{j\in\hspace{-0.5mm} \mathcal{U}_k^t\}\hspace{-0.5mm}\cap\hspace{-0.5mm} \{k\in\hspace{-0.5mm} \mathcal{V}^t\}\hspace{-0.5mm})\\
&+\hspace{-0.5mm}\mathrm{P_r}(\hspace{-0.5mm}\{j\notin \mathcal{U}_k^t\}\hspace{-0.5mm}\cup \hspace{-0.5mm}\{k\notin \mathcal{V}^t\}\hspace{-0.5mm})\mathrm{P_r}(\boldsymbol{F}_{f'}^t\hspace{-1mm}\in\hspace{-0.5mm}\mathcal{S}\vert \{j\notin \mathcal{U}_k^t\}\hspace{-0.5mm}\cup\hspace{-0.5mm}\{k\notin \hspace{-0.5mm}\mathcal{V}^t\}\hspace{-0.5mm}), 
\numberthis
\label{eq: multi_selay_1}
\end{align*}
where \eqref{eq: multi_selay_1} follows that the scenarios where client $j$ fails to convey its local model update $\Delta\boldsymbol{\Theta}_j^{t}$ via relaying client $k$ either results from the event $\{j\notin \mathcal{U}_k^t\}$, where $j$ fails to share with client $m$; or $\{k\notin \mathcal{V}^t\}$, where $k$ fails to connect with the server.  

Consider the events such that resulting variance $\nu_k^{t,-j}$ lies within the event $\mathcal{E}\triangleq\{ \lvert\nu_{F_{f'}}^{t,\{k,-j\}}-\Bar{\nu}_{F_{f'}}^{t,\{k,-j\}}\rvert\leq r_3 \}$, where $r_3$ is chosen such that $\mathrm{P_r}(\mathcal{E}^\mathrm{c})<\delta'$, $\delta'\in (0,1)$, then 
\begin{align*}
    &\mathrm{P_r}(\boldsymbol{F}_{f'}^{t,\{k,-j\}}\in\mathcal{S}\vert \{j\in \mathcal{U}_k^t\}\cap \{k\in \mathcal{V}^t\})\\
    =&\mathrm{P_r}(\boldsymbol{F}_{f'}^{t,\{k,-j\}}\in\mathcal{S}\vert \{j\in \mathcal{U}_k^t\}\cap \{k\in \mathcal{V}^t\},\mathcal{E})\mathrm{P_r}(\mathcal{E})\\
    &+\mathrm{P_r}(\boldsymbol{F}_{f'}^{t,\{k,-j\}}\in\mathcal{S}\vert \{j\in \mathcal{U}_k^t\}\cap \{k\in \mathcal{V}^t,\mathcal{E}^c\})\mathrm{P_r}(\mathcal{E}^c)\\
    \leq&\mathrm{P_r}(\boldsymbol{F}_{f'}^{t,\{k,-j\}}\hspace{-1mm}\in\hspace{-0.5mm}\mathcal{S}\vert \{j\hspace{-1mm}\in \mathcal{U}_k^t\}\hspace{-0.5mm}\cap\hspace{-0.5mm} \{k\in \hspace{-1mm}\mathcal{V}^t\},\mathcal{E})\mathrm{P_r}(\mathcal{E})+\delta'. 
    \numberthis
    \label{eq: multi_selay_allE}
\end{align*}
For each realization of $\mathcal{U}_k^t=\mathcal{A}$ and $\mathcal{V}^t=\mathcal{B}$, similar to \eqref{eq: privacy_loss_relay}, the privacy loss is identical to that of the Gaussian mechanism, with variance $\nu_{F_{f'}}^{t,\{k,-j\}}$. The $\ell_2$-sensitivity is bounded by
\begin{align*}
&\hspace{4mm}\sup_{\hspace{-2mm}\substack{\mathcal{A},\mathcal{B}:\mathcal{A}\in \mathcal{U}_k, j,k\in\mathcal{A},\\
    \mathcal{B}\in[K], k\in \mathcal{B}\\ 
    \lvert\nu_{F_{f'}}^{t,\{k,-j\}}-\Bar{\nu}_{F_{f'}}^{t,\{k,-j\}}\rvert\leq r_3 }} \hspace{0mm}\left\lVert c_{f',k}g_{k,j}\Delta\boldsymbol{\Theta}_j^{t}-c_{f',k}g_{k,j}\boldsymbol{N}_j^t \right\rVert\\
    &\hspace{15mm}\leq\lvert c_{f',k}g_{k,j}\rvert (R+ \lVert\boldsymbol{N}_j^t\rVert)\\
    &\hspace{15mm}\leq\lvert c_{f',k}g_{k,j}\rvert (R+ \lambda\sqrt{D(1+\delta_0)}),
    \numberthis
    \label{eq: multi_selay_privacy_loss}
\end{align*}
with probability $1- e^{-\frac{D}{2}(\delta_0-\ln{(1+\delta_0)})}$, where Chi-Square tail bound is applied in \eqref{eq: multi_selay_privacy_loss}. 
Consequently, for each realization of $\mathcal{U}_k^t=\mathcal{A}$ and $\mathcal{V}^t=\mathcal{B}$, $\boldsymbol{F}_{f'}^{t,\{k,-j\}}$ is $(\epsilon_{f',\{k,j\}}^{\mathcal{A},4}, \delta_{f',\{k,j\}}^{(4)})$-differentially private,
where $\epsilon_{f',\{k,j\}}^{\mathcal{A},4}=\hspace{-1mm}\left[2\log\left( \frac{1.25}{\delta_{f',\{k,j\}}^{(4)}} \right)\right]^{\frac{1}{2}}\hspace{-1mm}\frac{\lvert c_{f',k}g_{k,j}\rvert (R+ \lambda\sqrt{D(1+\delta_0)})}{\sqrt{\nu_{F_{f'}}^{t,\{k,-j\}}}}\hspace{-1mm}\triangleq\frac{c_3}{\sqrt{\nu_{F_{f'}}^{t,\{k,-j\}}}}$. Thus, it holds that 
\begin{subequations}
\begin{align}
     &\mathrm{P_r}(\boldsymbol{F}_{f'}^{t,\{k,-j\}}\in\mathcal{S}\vert \{j\in \mathcal{U}_k^t\}\cap \{k\in \mathcal{V}^t\},\mathcal{E})\\
     &\leq \hspace{-5mm}\sum_{\hspace{-2mm}\substack{\mathcal{A},\mathcal{B}:\mathcal{A}\in \mathcal{U}_k, j,k\in\mathcal{A},\\
    \mathcal{B}\in[K], k\in \mathcal{B}\\ 
    \lvert\nu_{F_{f'}}^{t,\{k,-j\}}-\Bar{\nu}_{F_{f'}}^{t,\{k,-j\}}\rvert\leq r_3 }} \hspace{-5mm} \mathrm{P_r}(\{j\in \mathcal{A}\}\cap \{k\in \mathcal{B}\})\Bigg( e^{\frac{c_1}{\sqrt{\nu_{F_{f'}}^{t,\{k,-j\}}}}} \cdot\notag\\
    &\hspace{10mm}\vspace{-3mm}\mathrm{P_r}(\boldsymbol{F}_{f'}^t\in\mathcal{S}\vert \{j\in \mathcal{A}\}\cap \{k\in \mathcal{B}\})\hspace{-0.5mm}+\hspace{-0.5mm}\delta_{f',\{k,j\}}^{(4)}\Bigg) \label{eq: 90b}\\
     &\leq e^{\frac{c_3}{\sqrt{\Bar{\nu}_{F_{f'}}^{t,\{k,-j\}}-r_3}}} \cdot \hspace{-5mm} \sum_{\hspace{-2mm}\substack{\mathcal{A},\mathcal{B}:\mathcal{A}\in \mathcal{U}_k, j,k\in\mathcal{A},\\
    \mathcal{B}\in[K], k\in \mathcal{B}\\ 
    \lvert\nu_{F_{f'}}^{t,\{k,-j\}}-\Bar{\nu}_{F_{f'}}^{t,\{k,-j\}}\rvert\leq r_3 }}  \hspace{-5mm} \mathrm{P_r}(\{j\in \mathcal{A}\}\cap \{k\in \mathcal{B}\})\cdot\notag\\
    &\hspace{10mm}\mathrm{P_r}(\boldsymbol{F}_{f'}^t\in\mathcal{S}\vert \{j\in \mathcal{A}\}\cap \{k\in \mathcal{B}\})\hspace{-0.5mm}+\hspace{-0.5mm}\delta_{f',\{k,j\}}^{(4)}\label{eq: 90c}\\
    &\leq e^{\frac{c_3}{\sqrt{\Bar{\nu}_{F_{f'}}^{t,\{k,-j\}}-r_3}}} 
    \mathrm{P_r}(\boldsymbol{F}_{f'}^t\in\mathcal{S}\vert \{j\in \mathcal{U}_k^t\}\cap \{k\in \mathcal{V}^t\},\mathcal{E})\hspace{-0.5mm}\notag\\
    &\hspace{60mm}+\hspace{-0.5mm}\delta_{f',\{k,j\}}^{(4)}\label{eq: 90d},  
\end{align}
\end{subequations}
where \eqref{eq: 90c} holds since $\nu_{F_{f'}}^{t,\{k,-j\}}>\Bar{\nu}_{F_{f'}}^{t,\{k,-j\}}-r_3$, \eqref{eq: 90d} is similar to \eqref{eq: relay_ldp_Ut}. 

Furthermore, according to \eqref{eq: multi_selay_allE}, we have
\begin{subequations}
\begin{align}
    &\mathrm{P_r}(\boldsymbol{F}_{f'}^{t,\{k,-j\}}\in\mathcal{S}\vert \{j\in \mathcal{U}_k^t\}\cap \{k\in \mathcal{V}^t\})\notag\\
    \leq& \mathrm{P_r}(\mathcal{E})\bigg( e^{\frac{c_3}{\sqrt{\Bar{\nu}_{F_{f'}}^{t,\{k,-j\}}-r_3}}} \mathrm{P_r}(\boldsymbol{F}_{f'}^t\in\mathcal{S}\vert \{j\in \mathcal{U}_k^t\}\cap \{k\in \mathcal{V}^t\},\mathcal{E})\notag\\
    &\hspace{45mm}+\delta_{f',\{k,j\}}^{(4)}\bigg)
    +\delta'\\
    \leq&  e^{\frac{c_3}{\sqrt{\Bar{\nu}_{F_{f'}}^{t,\{k,-j\}}-r_3}}} \mathrm{P_r}(\boldsymbol{F}_{f'}^t\in\mathcal{S}\vert \{j\in \mathcal{U}_k^t\}\cap \{k\in \mathcal{V}^t\}) \notag\\
    &\hspace{50mm}+\delta_{f',\{k,j\}}^{(4)}
    +\delta'.
    \label{eq: multi_selay in UV}
\end{align}
\end{subequations}
In addition, we have 
\begin{align}
    &\mathrm{P_r}(\boldsymbol{F}_{f'}^t\in\mathcal{S}\vert \{j\notin \mathcal{U}_k^t\}\cup \{k\notin \mathcal{V}^t\})\notag\\
    &\leq e^{\frac{c_3}{\sqrt{\Bar{\nu}_{F_{f'}}^{t,\{k,-j\}}-r_3}}} \mathrm{P_r}(\boldsymbol{F}_{f'}^t\in\mathcal{S}\vert \{j\notin \mathcal{U}_k^t\}\cup \{k\notin \mathcal{V}^t\}).
    \label{eq: multi_selay notin UV}
\end{align}
Substitute \eqref{eq: multi_selay in UV} and \eqref{eq: multi_selay notin UV} into \eqref{eq: multi_selay_1}, we get
\begin{align*}
    &\mathrm{P_r}(\boldsymbol{F}_{f'}^{t,\{k,-j\}}\in\mathcal{S})\leq e^{\frac{c_3}{\sqrt{\Bar{\nu}_{F_{f'}}^{t,\{k,-j\}}-r_3}}} \mathrm{P_r}(\boldsymbol{F}_{f'}^t\in\mathcal{S})\\
    &\hspace{10mm}+\mathrm{P_r}(\{j\in \mathcal{U}_k^t\}\cap \{k\in \mathcal{V}^t\})
    \left(\delta_{f',\{k,j\}}^{(4)}
    +\delta' \right),
    \numberthis
\end{align*}
where $\mathrm{P_r}(\{j\in \mathcal{U}_k^t\}\cap \{k\in \mathcal{V}^t\})=(1-p_k)(1-p_{k,j})$. 
The proof is thereby completed. 

When $\Delta\boldsymbol{\Theta}_j^{t}=u$ is perturbed to $\Delta\boldsymbol{\Theta}_j^{t}=\boldsymbol{u}'$. The $\boldsymbol{F}_{f'}^t$ should not fluctuate significantly.
Notably, any relaying neighbor of client $j$ (including client $j$),  say client $k$, can convey the perturbation to $\boldsymbol{F}_{f'}^t$ if $c_{f'k}\neq 0$. 
The perturbation does not affect $\boldsymbol{F}_{f'}^t$ only if all relaying neighbors of client $j$ that successfully receive from $j$ either fail to connect to the server or correspond to a zero coefficient in $\boldsymbol{c}_{f'}$.
Denote the event that the perturbation of $\Delta\boldsymbol{\Theta}_j^{t}$ can cause a perturbation of $\boldsymbol{F}_{f'}^t$ as event $\mathcal{J}$, then $\mathcal{J}=\{\exists k: k\in\mathcal{V}_{f'}, k\in\mathcal{V}^t, j\in\mathcal{U}_k^t \}$.    

For $\forall \boldsymbol{u},\boldsymbol{u}'\in \mathbb{B}_D(R)$, it holds that 
\begin{align*}
    \mathrm{P_r}(\boldsymbol{F}_{f'}^t\in \mathcal{S} \vert \Delta\boldsymbol{\Theta}_j^{t}=\boldsymbol{u})
    &=\mathrm{P_r}(\mathcal{J})
    \mathrm{P_r}(\boldsymbol{F}_{f'}^t\in \mathcal{S} \vert \Delta\boldsymbol{\Theta}_j^{t}=\boldsymbol{u}, \mathcal{J})\\
    &\hspace{-15mm}+\mathrm{P_r}(\mathcal{J}^c)
    \mathrm{P_r}(\boldsymbol{F}_{f'}^t\in \mathcal{S} \vert \Delta\boldsymbol{\Theta}_j^{t}=\boldsymbol{u}', \mathcal{J}^c). 
    \numberthis
    \label{eq: 74}
\end{align*}
Consider the events such that resulting variance $\nu_{F_{f'}}^{t}$ lies within the event $\mathcal{E}\triangleq\{ \lvert\nu_{F_{f'}}^{t}-\Bar{\nu}_{F_{f'}}^{t}\rvert\leq r_4 \}$, where $r_4$ is chosen such that $\mathrm{P_r}(\mathcal{E}^\mathrm{c})<\delta'$, $\delta'\in (0,1)$,
then
\begin{subequations}
\begin{align}
     \mathrm{P_r}(\boldsymbol{F}_{f'}^t \hspace{-1mm}\in \hspace{-1mm} \mathcal{S} \vert \Delta\boldsymbol{\Theta}_j^{t}\hspace{-1mm}=\boldsymbol{u}, \mathcal{J})
     &=\mathrm{P_r}(\mathcal{E})\mathrm{P_r}(\boldsymbol{F}_{f'}^t\hspace{-1mm}\in\hspace{-1mm} \mathcal{S} \vert \Delta\boldsymbol{\Theta}_j^{t}\hspace{-1mm}=\boldsymbol{u}, \mathcal{J},\mathcal{E}) \notag\\
     &\hspace{-16mm}+\mathrm{P_r}(\mathcal{E}^c)\mathrm{P_r}(\boldsymbol{F}_{f'}^t\in \mathcal{S} \vert \Delta\boldsymbol{\Theta}_j^{t}=\boldsymbol{u}, \mathcal{J},\mathcal{E}^c)\\
     &\hspace{-20mm}\leq \delta'+\mathrm{P_r}(\mathcal{E})\mathrm{P_r}(\boldsymbol{F}_{f'}^t\hspace{-1mm}\in\hspace{-1mm} \mathcal{S} \vert \Delta\boldsymbol{\Theta}_j^{t}\hspace{-1mm}=\boldsymbol{u}, \mathcal{J},\mathcal{E})
\end{align}
\end{subequations}
For each realization of $\mathcal{U}_k^t=\mathcal{A}$ and $\mathcal{V}^t=\mathcal{B}$, similar to \eqref{eq: privacy_loss_relay}, the privacy loss is identical to that of the Gaussian mechanism, with variance $\nu_{F_{f'}}^{t}$. The $\ell_2$-sensitivity is bounded by
\begin{align*}
&\hspace{4mm}\sup_{\hspace{-2mm}\substack{\mathcal{A},\mathcal{B}: \mathcal{B}\in[K], \mathcal{B}\cap\mathcal{V}_{f'}=\emptyset,\\
     k\in \mathcal{B}, \mathcal{A}\in \mathcal{U}_k, j,k\in\mathcal{A},\\ 
    \lvert\nu_{F_{f'}}^{t}-\Bar{\nu}_{F_{f'}}^{t}\rvert\leq r_4 }} \hspace{0mm}
    \left\lVert \sum_{k\in\mathcal{B}} c_{f',k}g_{k,j}(\boldsymbol{u}-\boldsymbol{u}') \right\rVert\\
    &\hspace{15mm}\leq\sum_{k=1}^K \lvert c_{f',k}g_{k,j}\rvert \cdot 2R. 
    \numberthis
    \label{eq: multi_selay_privacy_perturb}
\end{align*}
Consequently, for each realization of $\mathcal{U}_k^t=\mathcal{A}$ and $\mathcal{V}^t=\mathcal{B}$, $\boldsymbol{F}_{f'}^{t}$ is $(\epsilon_{f',j}^{\mathcal{A},\mathcal{B},5}, \delta_{f',j}^{(5)})$-differentially private,
where $\epsilon_{f',j}^{\mathcal{A},\mathcal{B},5}=\hspace{-1mm}\left[2\log\left( \frac{1.25}{\delta_{f',j}^{(5)}} \right)\right]^{\frac{1}{2}}\hspace{-1mm}\frac{2R\sum_{k=1}^K \lvert c_{f',k}g_{k,j}\rvert}{\sqrt{\nu_{F_{f'}}^{t}}}\hspace{-1mm}\triangleq\frac{c_4}{\sqrt{\nu_{F_{f'}}^{t}}}$. 

Thus, it holds that 
\begin{subequations}
\begin{align}
    &\mathrm{P_r}(\boldsymbol{F}_{f'}^t\hspace{-1mm}\in\hspace{-1mm} \mathcal{S} \vert \Delta\boldsymbol{\Theta}_j^{t}\hspace{-1mm}=\boldsymbol{u}, \mathcal{J},\mathcal{E})\notag\\
    &=\sum_{\hspace{-2mm}\substack{\mathcal{A},\mathcal{B}: \mathcal{B}\in[K], \mathcal{B}\cap\mathcal{V}_{f'}=\emptyset,\\
     k\in \mathcal{B}, \mathcal{A}\in \mathcal{U}_k, j,k\in\mathcal{A},\\ 
    \lvert\nu_{F_{f'}}^{t}-\Bar{\nu}_{F_{f'}}^{t}\rvert\leq r_4 }}
    \mathrm{P_r}(\mathcal{A},\mathcal{B}) 
    \mathrm{P_r}(\boldsymbol{F}_{f'}^t\hspace{-1mm}\in\hspace{-1mm} \mathcal{S} \vert \Delta\boldsymbol{\Theta}_j^{t}\hspace{-1mm}=\boldsymbol{u}, \mathcal{J}(\mathcal{A},\mathcal{B}),\mathcal{E})\\
    &\leq \hspace{0mm}\sum_{\hspace{-10mm}\substack{\mathcal{A},\mathcal{B}: \mathcal{B}\in[K], \mathcal{B}\cap\mathcal{V}_{f'}=\emptyset,\\
     k\in \mathcal{B}, \mathcal{A}\in \mathcal{U}_k, j,k\in\mathcal{A},\\ 
    \lvert\nu_{F_{f'}}^{t}-\Bar{\nu}_{F_{f'}}^{t}\rvert\leq r_4 }}
    \mathrm{P_r}(\mathcal{A},\mathcal{B})
    \big(e^{\epsilon_{f',j}^{\mathcal{A},\mathcal{B},5}}
    \mathrm{P_r}(\boldsymbol{F}_{f'}^t\hspace{-1mm}\in\hspace{-1mm} \mathcal{S} \vert \Delta\boldsymbol{\Theta}_j^{t}\hspace{-1mm}=\boldsymbol{u}', \mathcal{J}(\mathcal{A},\mathcal{B}),\mathcal{E}) \notag\\
    &\hspace{60mm}+\delta_{f',j}^{(5)}\big)\\
    &\leq e^{\frac{c_4}{\sqrt{\Bar{\nu}_{F_{f'}}^{t}-r_4}}} \mathrm{P_r}(\boldsymbol{F}_{f'}^t\hspace{-1mm}\in\hspace{-1mm} \mathcal{S} \vert \Delta\boldsymbol{\Theta}_j^{t}\hspace{-1mm}=\boldsymbol{u}', \mathcal{J},\mathcal{E})+\delta_{f',j}^{(5)}.
\end{align}
\end{subequations}
Similar to \eqref{eq: multi_selay_allE} and \eqref{eq: multi_selay in UV}, it holds that
\begin{subequations}
\begin{align}
    &\mathrm{P_r}(\boldsymbol{F}_{f'}^t\hspace{-1mm}\in\hspace{-1mm} \mathcal{S} \vert \Delta\boldsymbol{\Theta}_j^{t}\hspace{-1mm}=\boldsymbol{u}, \mathcal{J})\notag\\
    &\leq \delta'+\mathrm{P_r}(\mathcal{E})\left(e^{\frac{c_4}{\sqrt{\Bar{\nu}_{F_{f'}}^{t}-r_4}}} \mathrm{P_r}(\boldsymbol{F}_{f'}^t\hspace{-1mm}\in\hspace{-1mm} \mathcal{S} \vert \Delta\boldsymbol{\Theta}_j^{t}\hspace{-1mm}=\boldsymbol{u}', \mathcal{J},\mathcal{E})+\delta_{f',j}^{(5)} \right)\\
    &\leq e^{\frac{c_4}{\sqrt{\Bar{\nu}_{F_{f'}}^{t}-r_4}}} \mathrm{P_r}(\boldsymbol{F}_{f'}^t\hspace{-1mm}\in\hspace{-1mm} \mathcal{S} \vert \Delta\boldsymbol{\Theta}_j^{t}\hspace{-1mm}=\boldsymbol{u}', \mathcal{J})+(\delta_{f',j}^{(5)}+\delta')
\end{align}
\end{subequations}
Additionally, it holds that 
\begin{align}
    &\mathrm{P_r}(\boldsymbol{F}_{f'}^t\hspace{-1mm}\in\hspace{-1mm} \mathcal{S} \vert \Delta\boldsymbol{\Theta}_j^{t}\hspace{-1mm}=\boldsymbol{u}', \mathcal{J}^c)\notag \\
    &\leq e^{\frac{c_4}{\sqrt{\Bar{\nu}_{F_{f'}}^{t}-r_4}}} \mathrm{P_r}(\boldsymbol{F}_{f'}^t\hspace{-1mm}\in\hspace{-1mm} \mathcal{S} \vert \Delta\boldsymbol{\Theta}_j^{t}\hspace{-1mm}=\boldsymbol{u}', \mathcal{J}^c)
\end{align}
As a result, 
\begin{align}
    &\mathrm{P_r}(\boldsymbol{F}_{f'}^t\hspace{-1mm}\in\hspace{-1mm} \mathcal{S} \vert \Delta\boldsymbol{\Theta}_j^{t}\hspace{-1mm}=\boldsymbol{u})\notag \\
    &\leq e^{\frac{c_4}{\sqrt{\Bar{\nu}_{F_{f'}}^{t}-r_4}}}
    \mathrm{P_r}(\boldsymbol{F}_{f'}^t\hspace{-1mm}\in\hspace{-1mm} \mathcal{S} \vert \Delta\boldsymbol{\Theta}_j^{t}\hspace{-1mm}=\boldsymbol{u}')+\mathrm{P_r}(\mathcal{J})(\delta_{f',j}^{(5)}+\delta'),
\end{align}
where $\mathrm{P_r}(\mathcal{J})=1-\tilde{p}_{j,f'}$, and $\tilde{p}_{j,f'}$ denotes the probability of client $j$'s contribution not being conveyed by those participating in $\boldsymbol{c}_{f'}$-combination. 

\section{Choices of Bernstein Parameters}\label{Appendix: Bernstein Parameter}
\begin{lemma}[Bernstein’s Inequality]
Let $X$ be a zero-mean real-valued random variable, bounded by $M$, almost surely. Then, for $\forall r\geq 0$, we have
\begin{align}
      \mathrm{P_r}(X\geq r)\leq \exp\left(-\frac{r^2}{\mathbb{E}[X^2]+\frac{2}{3}Mr}\right),  
\end{align}
or
\begin{align}
      \mathrm{P_r}(\lvert X\rvert\geq r)\leq 2\exp\left(-\frac{r^2}{\mathbb{E}[X^2]+\frac{2}{3}Mr}\right).  
\end{align}
\end{lemma}
By choosing $\delta'=2\exp\left(-\frac{r^2}{\mathbb{E}[X^2]+\frac{2}{3}Mr}\right)$, this is a quadratic in $r$ and $r$ can be solved accordingly, i.e.,
\begin{align}
    r=\ln\left(\frac{2}{\delta'}\right)\cdot\left( \frac{1}{3}M+\sqrt{\frac{1}{9}M^2+\frac{\mathbb{E}[X^2]}{\ln\left(\frac{2}{\delta'}\right)}} \right). 
\end{align}

\subsection{Choice of $r_1$ and $r_2$}
Let $X_{k,2}=\nu_k^t-\Bar{\nu}_k$. Then, $\mathbb{E}[X_{k,2}]=0$, and $\max\{X_{k,2}\}\leq \max\{\nu_k^t\}\leq \max_k\{\nu_k^t\}\triangleq M_2$. Due to Bernoulli r.v.s, $M_2$ is computable, through e.g., brute-force search.
Moreover, 
\begin{align}
\mathbb{E}[X_{k,2}^2]=\mathbb{E}[(\nu_k^t-\Bar{\nu}_k)^2]=\mathbb{E}[(\nu_k^t)^2]-\Bar{\nu}_k^2, 
\label{eq: var_X_k}
\end{align}
where $\Bar{\nu}_k$ and $\mathbb{E}[(\nu_k^t)^2]$ are given in \eqref{eq: calc_bar_v_k} and \eqref{eq: E[v_k^t_square]}, respectively. 
Thereby, $r_2$ is chosen as
\begin{align}
    r_2=\ln\left(\frac{2}{\delta'}\right)\cdot\left( \frac{1}{3}M_2+\sqrt{\frac{1}{9}M_2^2+\frac{\mathbb{E}[X_{k,2}^2]}{\ln\left(\frac{2}{\delta'}\right)}} \right). 
\end{align}

Similarly, let $X_{k,1}=\nu_k^{t,-j}-\Bar{\nu}_k^{-j}$. Then, $\mathbb{E}[X_{k,1}]=0$, and $\max\{X_{k,1}\}\leq \max\{\nu_k^{t,-j}\}\leq \max_k\{\nu_k^{t,-j}\}\triangleq M_1$. By setting $\tau_{k,j}=0$, $M_1$ is computable, through e.g., brute-force search.
Moreover, 
\begin{align}
\mathbb{E}[X_{k,1}^2]=\mathbb{E}[(\nu_k^{t,-j}-\Bar{\nu}_k^{-j})^2]=\mathbb{E}[(\nu_k^{t,-j})^2]-(\Bar{\nu}_k^{-j})^2, 
\label{eq: var_X_k}
\end{align}
where $(\Bar{\nu}_k^{-j})^2$ is given in \eqref{eq: calc_bar_v_k_-j} and $\mathbb{E}[(\nu_k^{t,-j})^2]$ can be computed by setting $\tau_{k,j}=0$ in \eqref{eq: E[v_k^t_square]}. 
Thereby, $r_1$ is chosen as
\begin{align}
    r_1=\ln\left(\frac{2}{\delta'}\right)\cdot\left( \frac{1}{3}M_1+\sqrt{\frac{1}{9}M_1^2+\frac{\mathbb{E}[X_{k,1}^2]}{\ln\left(\frac{2}{\delta'}\right)}} \right). 
\end{align}

\subsection{Choice of $r_3$ and $r_4$}
Let $X_{k,4}=\nu_{F_{f'}}^t-\Bar{\nu}_{F_{f'}}$. Then, $\mathbb{E}[X_{k,4}]=0$, and $\max\{X_{k,4}\}\leq \max\{\nu_{F_{f'}}^t\}\leq \max_{f'}\{\nu_{F_{f'}}^t\}\triangleq M_4$. Due to the Bernoulli r.v.s, $M_4$ is computable, through e.g., brute-force search.
Moreover, 
\begin{align}
\mathbb{E}[X_{k,4}^2]=\mathbb{E}[(\nu_{F_{f'}}^t-\Bar{\nu}_{F_{f'}})^2]=\mathbb{E}[(\nu_{F_{f'}}^t)^2]-\Bar{\nu}_{F_{f'}}^2, 
\label{eq: var_X_k}
\end{align}
where $\Bar{\nu}_{F_{f'}}$ and $\mathbb{E}[(\nu_{F_{f'}}^t)^2]$ are given in \eqref{eq: Bar_nu_f'} and \eqref{eq: E[v_k^t_square]}, respectively. 
Thereby, $r_4$ is chosen as
\begin{align}
    r_4=\ln\left(\frac{2}{\delta'}\right)\cdot\left( \frac{1}{3}M_4+\sqrt{\frac{1}{9}M_4^2+\frac{\mathbb{E}[X_{k,4}^2]}{\ln\left(\frac{2}{\delta'}\right)}} \right). 
\end{align}

Similarly, let $X_{k,3}=\nu_{F_{f'}}^{t,\{k,-j\}}-\Bar{\nu}_{F_{f'}}^{\{k,-j\}}$. Then, $\mathbb{E}[X_{k, 3}]=0$, and $\max\{X_{k,3}\}\leq \max\{\nu_{F_{f'}}^{t,\{k,-j\}}\}\leq \max_{f'}\{\nu_{F_{f'}}^{t,\{k,-j\}}\}\triangleq M_3$. By setting $\tau_{k,j}=0$, $M_3$ is computable, through e.g., brute-force search.
Moreover, 
\begin{align}
\mathbb{E}[X_{k,3}^2]&=\mathbb{E}[(\nu_{F_{f'}}^{t,\{k,-j\}}-\Bar{\nu}_{F_{f'}}^{\{k,-j\}})^2] \notag\\
&=\mathbb{E}[(\nu_{F_{f'}}^{t,\{k,-j\}})^2]-(\Bar{\nu}_{F_{f'}}^{\{k,-j\}})^2, 
\label{eq: var_X_k}
\end{align}
where $\Bar{\nu}_{F_{f'}}^{\{k,-j\}}$ and $\mathbb{E}[(\nu_{F_{f'}}^{t,\{k,-j\}})^2]$ can be computed be setting $\tau_{k,j}=0$ in \eqref{eq: Bar_nu_f'} and \eqref{eq: E_nu_f_k_-j}, respectively. 
Thereby, $r_3$ is chosen as
\begin{align}
    r_3=\ln\left(\frac{2}{\delta'}\right)\cdot\left( \frac{1}{3}M_3+\sqrt{\frac{1}{9}M_3^2+\frac{\mathbb{E}[X_{k,3}^2]}{\ln\left(\frac{2}{\delta'}\right)}} \right). 
\end{align}


\section{Calculation of Privacy Noise Variance}\label{appx: variance}
\subsection{Calculation of $\Bar{\nu}_k$ and $\mathbb{E}[(\nu_k^t)^2]$}

Let us begin by computing the variance of the aggregated secret keys $\boldsymbol{M}_k^t$ at relaying client $k$. It can be verified that $\boldsymbol{M}_k^t=\sum_{m\in\mathcal{U}_k^t} g_{k,m}\boldsymbol{N}_m^t=\sum_{m=1}^K\boldsymbol{\Lambda}_k^t(m)\boldsymbol{N}_k^t$, so $\mathrm{var}(\boldsymbol{M}_k^t)=\nu_k^t \boldsymbol{I}_D=\lVert \boldsymbol{\Lambda}_k^t\boldsymbol{A}\rVert^2\boldsymbol{I}_D$. 
Similarly, we define $\boldsymbol{M}_{k}^{t,-j}=\sum_{m\in\mathcal{U}_k^t\setminus\{j\}} g_{k,m}\boldsymbol{N}_m^t=\sum_{m\neq j}\boldsymbol{\Lambda}_k^t(m)\boldsymbol{N}_k^t$, and we have 
 $\mathrm{var}(\boldsymbol{M}_{k}^{t,-j})=\nu_{k}^{t,-j} \boldsymbol{I}_D=\lVert \boldsymbol{\Lambda}_k^{t,-j}\boldsymbol{A}\rVert^2\boldsymbol{I}_D$. 
The $\boldsymbol{\Lambda}_k^t$ depends on $\{\tau_{k,m}\}_{m=1}^K$, so the variance $\nu_k^t$ of the effective noise $\boldsymbol{M}_k^{t}$ is also a random variable concerning the network topology. It should be noted that due to correlation, the variance of $\boldsymbol{M}_k^t$ does not necessarily increase monotonously with the number of received updates $\vert \mathcal{U}_k^t\rvert$ at client $k$. The expected variance $\Bar{\nu}_k$ is given by 
\begin{align}
    \Bar{\nu}_k=\mathbb{E}_{\mathcal{U}_k^t}[\nu_k^t]=\mathbb{E}\left[\lVert \boldsymbol{\Lambda}_k^t\boldsymbol{A}\rVert^2\right].
\end{align}
Furthermore, we have
\begin{align*}
    &\mathbb{E}\left[\lVert \boldsymbol{\Lambda}_k^t\boldsymbol{A}\rVert^2\right]
    =\mathbb{E}\left[ \boldsymbol{\Lambda}_k^t\boldsymbol{A}\boldsymbol{A}^\top(\boldsymbol{\Lambda}_k^{t})^\top \right]\\
    &=\operatorname{Tr}\left( \operatorname{Cov}[\boldsymbol{\Lambda}_k^t] \boldsymbol{A}\boldsymbol{A}^\top \right)+\mathbb{E}\left[ \boldsymbol{\Lambda}_k^t\boldsymbol{A}\boldsymbol{A}^\top(\boldsymbol{\Lambda}_k^{t})^\top\right]\\
    &=\lambda^2\sum_{m=1}^K(1-p_{k,m})g_{m,k}^2+\boldsymbol{\Lambda}_k^{\mathrm{E}}\boldsymbol{A}\boldsymbol{A}^\top(\boldsymbol{\Lambda}_k^{\mathrm{E}})^\top,
    \numberthis
    \label{eq: calc_bar_v_k}
\end{align*}
where $\boldsymbol{\Lambda}_k^{\mathrm{E}}=\mathbb{E}[\boldsymbol{\Lambda}_k^t]=[(1-p_{k,1})g_{k,1},\cdots,(1-p_{k,K})g_{k,K}]$.

Similarly, we also have 
\begin{align}
    \Bar{\nu}_k^{-j}=\mathbb{E}_{\mathcal{U}_k^t}[\nu_k^{t,-j}]=\mathbb{E}\left[\lVert \boldsymbol{\Lambda}_k^{t,-j}\boldsymbol{A}\rVert^2\right]\boldsymbol{I}_D,
\end{align}
and 
\begin{align*}
    &\mathbb{E}\left[\lVert \boldsymbol{\Lambda}_k^{t,-j}\boldsymbol{A}\rVert^2\right]\\
    &=\lambda^2\sum_{m=1}^K(1-p_{k,m})g_{m,k}^2+\boldsymbol{\Lambda}_k^{\mathrm{E},-j}\boldsymbol{A}\boldsymbol{A}^\top(\boldsymbol{\Lambda}_k^{\mathrm{E},-j})^\top,
    \numberthis
     \label{eq: calc_bar_v_k_-j}
\end{align*}
where $\boldsymbol{\Lambda}_k^{\mathrm{E},-j}$ is identical to $\boldsymbol{\Lambda}_k^{\mathrm{E}}$ except the $j$-th element is set to $0$.

It can be verified that $\nu_k^t$ can also be written in the form 
$\nu_k^t=\sum_{k_1,k_2\in\mathcal{U}_k} C^{(k)}_{k_1,k_2} \tau^t_{k,k_1}\tau^t_{k,k_2}$,  where $C^{(k)}_{k_1,k_2}=g_{k,k_1}[\boldsymbol{A}\boldsymbol{A}^\top]_{k_1k_1}g_{k,k_2}$. Following this, we have
\begin{subequations}
\begin{align}
\left(\nu_k^t\right)^2&=\hspace{-2mm}\sum_{k_1,k_2,k_3,k_4\in\mathcal{U}_k} \hspace{-2mm} C^{(k)}_{k_1,k_2} C^{(k)}_{k_3,k_4} \tau^t_{k,k_1}\tau^t_{k,k_2}\tau^t_{k,k_3}\tau^t_{k,k_4},\\
\mathbb{E}\left[ \left(\nu_k^t\right)^2 \right]&=\hspace{-2mm}\sum_{k_1,k_2,k_3,k_4\in\mathcal{U}_k} \hspace{-2mm} C^{(k)}_{k_1,k_2} C^{(k)}_{k_3,k_4} \mathbb{E}\left[\tau^t_{k,k_1}\tau^t_{k,k_2}\tau^t_{k,k_3}\tau^t_{k,k_4}
\right], 
\label{eq: E[v_k^t_square]}
\end{align}
\end{subequations}
where $\mathbb{E}\left[\tau^t_{k,k_1}\tau^t_{k,k_2}\tau^t_{k,k_3}\tau^t_{k,k_4}
\right]$ is given by 
\begin{align}
\mathbb{E}\left[\tau^t_{k,k_1}\tau^t_{k,k_2}\tau^t_{k,k_3}\tau^t_{k,k_4}
\right]=\prod_{u\in\{k_1\}\cup\{k_2\}\cup\{k_3\}\cup\{k_4\}}(1-p_{k,u}), 
\label{eq: link1234}
\end{align}
since $\tau^t_{k,k_1}$ and $\tau^t_{k,k_2}$ are independent Bernoulli r.v., $(\tau^t_{k,k_1})^2=\tau^t_{k,k_1}$,  
and $\{k_1\}\cup\{k_2\}\cup\{k_3\}\cup\{k_4\}$ ensures the uniqueness of the links in the product.

\subsection{Calculation of $\Bar{\nu}_{F_{f'}}$ and $\mathbb{E}[(\nu_{F_{f'}}^t)^2]$}
The aggregated secret key according to $\boldsymbol{c}_{f'}$ is 
\begin{subequations}
\begin{align}
\boldsymbol{N}_{F_{f'}}^t&=\sum_{k\in\mathcal{V}^t} c_{f',k}\sum_{m\in\mathcal{U}_k^t} g_{k,m}\boldsymbol{N}_m^t\\
&=\sum_{k=1}^K c_{f',k} \tau_k^t\sum_{m=1}^K \tau_{k,m}^t g_{k,m}\boldsymbol{N}_m^t\\
&=\sum_{k=1}^K c_{f',k} \tau_k^t\sum_{m=1}^K \tau_{k,m}^t g_{k,m}\sum_{l=1}^L \alpha_{m,l} \boldsymbol{Z}_l^t\\
&=\sum_{l=1}^L \sum_{k=1}^K c_{f',k} \tau_k^t\sum_{m=1}^K \tau_{k,m}^t g_{k,m}\alpha_{m,l} \boldsymbol{Z}_l^t.
\end{align}
\end{subequations}
Thus, $\mathrm{var}(\boldsymbol{N}_{F_{f'}}^t)=\nu_{F_{f'}}^t\boldsymbol{I}_D$, where
\begin{subequations}
\begin{align}
   \nu_{F_{f'}}^t&=\sum_{l=1}^L \left(\sum_{k=1}^K c_{f',k} \tau_k^t\sum_{m=1}^K \tau_{k,m}^t g_{k,m}\alpha_{m,l}\right)^2 \label{eq: var_f' form 1}\\
    &=\sum_{l=1}^L \left(\sum_{k=1}^K c_{f',k} \tau_{k}^t \boldsymbol{\Lambda}_{k}^t \boldsymbol{\alpha}_l\right)^2\\
    &\hspace{-3mm}=\sum_{l=1}^L \boldsymbol{\alpha}_l^\top \left(\sum_{k_1,k_2} c_{f',k_1} c_{f',k_2} \tau_{k_1}^t\tau_{k_2}^t \boldsymbol{\Lambda}_{k_1}^t \boldsymbol{\Lambda}_{k_2}^t\right) \boldsymbol{\alpha}_l. 
\end{align}
\end{subequations}
Thus, we have 
\begin{align}
    \Bar{\nu}_{F_{f'}}=\mathbb{E}\left[\nu_{F_{f'}}^t\right]
    =\sum_{l=1}^L \boldsymbol{\alpha}_l^\top \left(\sum_{k_1,k_2} c_{f',k_1} c_{f',k_2} \tau_{k_1}^t\tau_{k_2}^t \Sigma_k\right) \boldsymbol{\alpha}_l, 
    \label{eq: Bar_nu_f'}
\end{align}
where $\Sigma_k=\mathbb{E}\left[\boldsymbol{\Lambda}_{k_1}^t \boldsymbol{\Lambda}_{k_2}^t\right]$ and is given by
\begin{align}
    [\Sigma_k]_{ij}=\begin{cases}
        g_{k,i}^2(1-p_{k,i}),\;\;\;\;\;\text{if}\; i=j\\
        \left[(\boldsymbol{\Lambda}_k^{\mathrm{E}})^\top \boldsymbol{\Lambda}_k^{\mathrm{E}}\right]_{ij}\;\;\;\;\;\text{if}\; i\neq j
    \end{cases}. 
\end{align}

Next, let us calculate $\mathbb{E}\left[(\nu_{F_{f'}}^t)^2\right]$. The \eqref{eq: var_f' form 1} can be rewritten as
\begin{subequations}
\begin{align}
    &\nu_{F_{f'}}^t=\sum_{l=1}^L \left(\sum_{k_1,m_1} c_{f',k_1} g_{k_1,m_1}
    \tau_{k_1}^t \tau_{k_1,m_1}^t \alpha_{m_1,l} \right)^2\\
    &=\sum_{l=1}^L \sum_{k_1,m_1} \sum_{k_2,m_2} c_{f',k_1}c_{f',k_2} g_{k_1,m_1} g_{k_2,m_2} \alpha_{m_1,l} \alpha_{m_2,l} \notag \\
    &\hspace{4cm}\cdot\tau_{k_1}^t \tau_{k_1,m_1}^t  \tau_{k_2}^t\tau_{k_2,m_2}^t. 
    \label{eq: var_f' form 2}
\end{align}
\end{subequations}
Let $A_{k_1,m_1}^{f'}=c_{f',k_1} g_{k_1,m_1}$, $t(k_1,m_1)=\tau_{k_1}^t \tau_{k_1,m_1}^t$, and $H_{m_1,m_2}=\sum_{l=1}^L \alpha_{m_1,l} \alpha_{m_2,l}$ \eqref{eq: var_f' form 2} can be simplified to 
\begin{align}
   \nu_{F_{f'}}^t=\sum_{k_1,m_1} \sum_{k_2,m_2} A_{k_1,m_1}^{f'} A_{k_2,m_2}^{f'} H_{m_1,m_2} t(k_1,m_1) t(k_2,m_2).
\end{align}
Thus,
\begin{align}
    &(\nu_{F_{f'}}^t)^2=\hspace{-2mm}\sum_{k_1,m_1} \sum_{k_2,m_2} \sum_{k_3,m_3} \sum_{k_4,m_4} A_{k_1,m_1}^{f'} A_{k_2,m_2}^{f'} 
    A_{k_3,m_3}^{f'} A_{k_4,m_4}^{f'}\notag\\
    &\hspace{0.5cm} \cdot H_{m_1,m_2}H_{m_3,m_4} 
    t(k_1,m_1) t(k_2,m_2)
    t(k_3,m_3) t(k_4,m_4). 
\end{align}
Thus, 
\begin{align} &\mathbb{E}\left[(\nu_{F_{f'}}^t)^2\right]\hspace{-1mm}=\hspace{-2mm}\sum_{k_1,m_1} \hspace{-0.5mm}\sum_{k_2,m_2} \hspace{-0.5mm}\sum_{k_3,m_3} \hspace{-0.5mm}\sum_{k_4,m_4} A_{k_1,m_1}^{f'} A_{k_2,m_2}^{f'} 
A_{k_3,m_3}^{f'} A_{k_4,m_4}^{f'}\notag\\
&\hspace{0.2cm} \cdot H_{m_1,m_2}H_{m_3,m_4} 
    \mathbb{E}\left[t(k_1,m_1) t(k_2,m_2)
    t(k_3,m_3) t(k_4,m_4)\right], 
\label{eq: E_nu_f_k_-j}
\end{align}
where 
\begin{align} 
    &\mathbb{E}\left[t(k_1,m_1) t(k_2,m_2)
    t(k_3,m_3) t(k_4,m_4)\right]\notag\\
    &=\prod_{\substack{u\in\{k_1\}\cup\{k_2\}
    \\ \cup\{k_3\}\cup\{k_4\}}}(1-p_{u}) \prod_{\substack{(u,v)\in\{(k_1,m_1)\}\cup\{(k_2,m_2)\}\\
    \cup\{(k_3,m_3)\}\cup\{(k_4,m_4)\}}}(1-p_{u,v}),
\end{align}
since $\mathbb{E}\left[t(k_1,m_1)\right]=(1-p_{k_1})(1-p_{k_1,m_1})$, $t(k_1,m_1)=t(k_2,m_2)$ for $k_1=k_2$, $ m_1=m_2$, and $\cup$ ensures the uniqueness of the set elements.


\end{document}

%% file: SecCoGC.tex
\begin{tikzpicture}[
    client/.style={draw, minimum width=1.5cm, minimum height=0.8cm},
    arrow/.style={>=stealth'},
    PS/.style={draw,minimum height=0.8cm,text width=4cm, text centered}
]

\node[PS] (top) at (0,2) {Any two: $\boldsymbol{\Theta}_1+\boldsymbol{\Theta}_2+\boldsymbol{\Theta}_3$};

\node[client] (c1) at (-2.5,0) {$\mathcal{D}_1$};
\node[client] (c2) at (0,0) {$\mathcal{D}_2$};
\node[client] (c3) at (2.5,0) {$\mathcal{D}_3$};

\draw[->] (c1.north) -- ([xshift=-0.7cm]top.south)node[above left, pos=0.2, xshift=0.3cm] {\scriptsize{
\shortstack{$\frac{1}{2}\boldsymbol{\Theta}_1+\boldsymbol{\Theta}_2$\\
$+\frac{1}{2}\boldsymbol{N}_1+\boldsymbol{N}_2 $}
}};
\draw[->] (c2.north) -- (top.south)node[above right, pos=0.1] {\scriptsize{\shortstack{$\boldsymbol{\Theta}_2-\boldsymbol{\Theta}_3$\\
$+\boldsymbol{N}_2-\boldsymbol{N}_3$}}};
\draw[->] (c3.north) -- ([xshift=0.7cm]top.south) node[above right, pos=0.2, xshift=-0.3cm] {\scriptsize{\shortstack{$\frac{1}{2}\boldsymbol{\Theta}_1+\boldsymbol{\Theta}_3$\\
$+\frac{1}{2}\boldsymbol{N}_1+\boldsymbol{N}_3$}}};

\draw[dotted, ->] 
    ([xshift=-0.3cm]c1.south) -- ++(0,-0.8) -- ++(5.6,0)node[below, pos=0.5, xshift=-0cm] {$\boldsymbol{\Theta}_1+\boldsymbol{N}_1$} -- ++(0,0.8);
    
\draw[dotted, ->] 
    ([xshift=-0.3cm]c2.south) -- ++(0,-0.6) -- ++(-1.9,0) node[above, pos=0.5, xshift=-0cm] {$\boldsymbol{\Theta}_2+\boldsymbol{N}_2$} -- ++(0,0.6);

\draw[dotted, ->] 
    ([xshift=-0.3cm]c3.south) -- ++(0,-0.6) -- ++(-1.9,0) node[above, pos=0.5, xshift=-0cm] {$\boldsymbol{\Theta}_3+\boldsymbol{N}_3$} -- ++(0,0.6); 

\end{tikzpicture}